\documentclass[%
 aip,
 amsmath,amssymb,
 reprint,%
]{revtex4-1}

\usepackage{graphicx}
\usepackage{tabularx}
\usepackage{dcolumn}
\usepackage{bm}
\usepackage{float}
\usepackage[utf8]{inputenc}
\usepackage[T1]{fontenc}
\usepackage{mathptmx}
\usepackage{amsmath}
\usepackage{etoolbox}
\usepackage{xcolor}
\usepackage[shortlabels]{enumitem}
\usepackage{hyperref}
\usepackage{tikz}

\newcommand{\sophie}[1]{}%{{\color{blue}[Sophie:] #1}}
\newcommand{\mhc}[1]{}%{{\color{violet}[M: #1]}}
\newcommand{\todo}[1]{}%{{\itshape #1}}

\newcommand{\x}{\mathbf{x}}

\newcommand{\s}{\mathbf{s}}
\newcommand{\z}{\mathbf{z}}
\newcommand{\Z}{\mathbf{Z}}
\newcommand{\y}{\mathbf{y}}
\newcommand{\w}{\mathbf{w}}

\newcommand{\utot}{u^{\rm tot}}
\newcommand{\Utot}{U^{\rm tot}}
\newcommand{\Uc}{U^c}
\newcommand{\Ue}{U}
\renewcommand{\c}{\mathbf{c}}
\renewcommand{\v}{\mathbf{v}}
\renewcommand{\j}{\mathbf{j}}
\newcommand{\R}{\mathbb{R}}
\newcommand{\E}{\mathbb{E}}
\newcommand{\half}{\frac{1}{2}}
\newcommand{\dd}[2]{\frac{d#1}{d#2}}  % derivative
\newcommand{\pp}[2]{\frac{\partial #1}{\partial #2}}  % partial derivative
\newcommand{\grad}{\mbox{grad}\,}
\renewcommand{\div}{\mbox{div}\,}
\newcommand{\gradM}{\mbox{grad}_{\!\!\mathcal M}\,}
\newcommand{\divM}{\mbox{div}_{\!\!\mathcal M}\,}
\newcommand{\dsigmaM}{d\sigma_{\!\!\mathcal M}}
\newcommand{\PM}{P_M}   %{P^{(M)}}
\newcommand{\QM}{Q_M}   %{P^{(M)}}
\DeclareMathOperator*{\argmin}{argmin}
\DeclareMathOperator*{\col}{col}

\makeatletter
\def\@email#1#2{%
 \endgroup
 \patchcmd{\titleblock@produce}
  {\frontmatter@RRAPformat}
  {\frontmatter@RRAPformat{\produce@RRAP{*#1\href{mailto:#2}{#2}}}\frontmatter@RRAPformat}
  {}{}
}%
\makeatother
\begin{document}

\preprint{AIP/123-QED}

\title[Constraints for Brownian dynamics]{Brownian dynamics with soft constraints in soft matter systems}

\author{S. Marbach}
\email{sophie.marbach@cnrs.fr}
\affiliation{CNRS, Sorbonne Université, Physicochimie des Electrolytes et Nanosystèmes Interfaciaux, F-75005 Paris, France}
\author{A. Carter}
\affiliation{CNRS, Sorbonne Université, Physicochimie des Electrolytes et Nanosystèmes Interfaciaux, F-75005 Paris, France}
%\affiliation{Sorbonne Université, Physicochimie des Electrolytes et Nanosystèmes Interfaciaux, F-75005 Paris, France}
\author{M. Holmes-Cerfon}
\affiliation{Department of Mathematics, University of British Columbia, 1984 Mathematics Rd, V6T 1Z2, Vancouver, BC, Canada}
\email{holmescerfon@math.ubc.ca}

\date{\today}

\begin{abstract}
Stiff forces, which bind objects together or  otherwise confine motion, are found widely in soft-matter systems -- colloids with short range attractions, ligand-receptor contacts, particles in optical traps, fibres that resist stretching, etc. 
To assess the long-term effect of these stiff forces on dynamics and structure, it is useful to consider the limit where they are treated as \textit{constraints}, so the system evolves strictly within allowed configurations. 
Efforts to derive equations involving both constraints, and the stochastic motion appropriate at the scales of soft matter, began around 50 years ago, yet, we are still lacking a straightforward way to extract the projected equations and apply them in modern formulations of mesoscale dynamics.
Here, we address this gap with two key contributions: (1) a practical summary of the  constrained Brownian dynamics equations with ``soft'' constraints, i.e. constraints imposed by stiff forces, which is illustrated through several representative examples, taking care to highlight the nontrivial effects of the constraints;  and (2) a novel derivation using singular perturbation theory, establishing the validity of these equations over timescales exceeding the relaxation of stiffly constrained degrees of freedom. We further extend our approach to ``soft soft'' constraints, where mobility varies on lengthscales comparable to the restraining forces—a scenario typical for particles in fluids experiencing hydrodynamic interactions.
We hope our results will be useful for soft matter research, as a robust toolkit for studying tethered or confined systems.
\end{abstract}

\maketitle

\section{Introduction}

Many problems in soft-matter physics involve stiff forces whose dynamical role is mainly to \textit{restrain} certain degrees of freedom, \textit{i.e.} that hinder access to certain configurations. 
Common examples include the backbone bonds of a polymer chain, which keep particles at a nearly fixed distance~\cite{leibler1991dynamics,schallamach1963theory}, colloidal particles interacting with a  short-ranged attractive depletion force \cite{meng2010free,lekkerkerker2024colloids},  or colloids moving in optical traps~\cite{grier1997optical,franosch2009persistent,franosch2011resonances}. 
Such forces are also prevalent in ligand-receptor interactions,  such as DNA pairs between synthetic colloids~\cite{mirkin1996dna,feng2013specificity,wang2015crystallization,rogers2016using,cui2022comprehensive,gehrels2022programming,macfarlane2011nanoparticle,jana2019translational,varilly2012general,melio2024soft}, protein linkers on microscale white blood cells used to stick to blood vessel walls~\cite{alon2002rolling,ley2007getting,korn2008dynamic}, and spike proteins on viruses used to adhere to mucus~\cite{mammen1998polyvalent,sakai2017influenza,sakai2018unique,muller2019mobility}. Continuous objects such as inextensible fibres may also be thought of as being subject to restraining forces on the length of the fibre \cite{maxian2021integral}. 

In this context it can be useful -- for at least two reasons -- to consider the limit where these restraints are treated as \textit{constraints}, \textit{i.e.} where the system is strictly restricted to certain configurations. One reason why is numerical: when a restraining force is strong and short-ranged, the governing equations are stiff, and hence tedious to simulate -- they require small timesteps to resolve the restraining forces, but they must be simulated over long timescales to capture dynamics of interest and average over many random configurations.  Replacing restraints with constraints, allows taking longer timesteps, at the expense of more computation per timestep. Such an approach began with the SHAKE~\cite{krautler2001fast} and RATTLE \cite{andersen1983rattle} algorithms of molecular dynamics, and continues to this day with more sophisticated computational algorithms and that are adapted to Brownian dynamics \cite{Leimkuhler.2016,maxian2023bending,maxian2024simulation,funkenbusch2024approaches,delmotte2025modeling}. Another reason why is conceptual: treating restraints as constraints can allow for simpler, lower-dimensional models, which can sometimes be more easily analyzed to learn about behavior~\cite{marbach2022mass,marbach2022nanocaterpillar,bressloff2013stochastic,mckinley2012asymptotic,lee2018modeling,fogelson2019transport,fogelson2018enhanced}. Such a treatment also highlights the nontrivial effects that restraints can have on the macroscopic dynamics, beyond simply constraining motion to a specific region of space.

A critical consideration when using constraints in mesoscale systems, is that such systems are subject to thermal fluctuations. When the stiff forces interact with the thermal fluctuations, they produce significant effects on the dynamics, that remain even in the limit when the forces become infinitely stiff. 
In particular, constraints alter both the macroscopic drift, and the mobility tensor. % which appears in a Brownian dynamics formulation of the equations of motion. 
This has led to the so-called ``paradox'' of \textit{soft} versus \textit{hard} constraints, that directly projecting the dynamics onto the constraint manifold (hard constraints) gives an apparently different result from considering the limit of an infinitely stiff restraining force (soft constraints) \cite{Kampen.1981}. 
This issue is not restricted to thermal systems: even for systems obeying deterministic evolution equations, such as Hamiltonian\cite{rubin1957motion} or quantum\cite{maraner1995complete,froese2001realizing} systems,  the effective dynamics induced by a stiff potential can differ from those obtained from classical constrained Lagrangian mechanics\cite{landau,arnold}, in intricate ways that depend on the particular system and even on the initial conditions. 

% hamiltonian: can oscillate around constraint surface. must account for weak limit of oscillations. see froese.
%
%Maraner: "Quantum mechanics is a field theory, after all, and the wavefunction of the system keeps on exploring the whole configuration space even if it is squeezed on the constraint’s surface. "

Clearly, from a soft matter modeling perspective, constraints should derive from stiff forces and thus, the \textit{soft constraints} are generally appealing. 
Yet, deriving the correct form of the overdamped Langevin equations for a soft constraint is nontrivial.

Efforts to derive softly constrained dynamics in noisy physical systems began around 50 years ago with Fixman \cite{fixman1978simulation}, continuing with Hinch \cite{hinch1994Brownian}, Ottinger \cite{ottinger1994brownian}, and summarized in the book chapter of Morse \cite{morse2003Theory}. These authors take different approaches for deriving the constrained equations, %though generally they are based on thermodynamic principles, and sometimes they use Lagrange multipliers e.g. in the Langevin equation\cite{hinch1994Brownian} or in a Poisson bracket formulation\cite{ottinger2015preservation}. Notably, 
and notably, they sometimes arrive at subtlely different conclusions for the constrained equations. Morse~\cite{morse2003Theory} has summarized the past approaches of Refs.~\onlinecite{fixman1978simulation,hinch1994Brownian,ottinger1994brownian}, and combined them in a unifying approach which points to some incorrect assumptions in the derivations of Refs.~\onlinecite{fixman1978simulation,hinch1994Brownian}. Yet, while thorough, the treatment of Ref.~\onlinecite{morse2003Theory} does not provide a straightforward way to extract the equations of motion and to apply to modern formulations of Brownian dynamics.

There have been simultaneous developments in the mathematical literature considering stochastic evolution equations subject to strong or rapidly oscillating forces. Kurtz \cite{kurtz1973limit,kurtz2005averaging} and Papanicolaou \cite{papanicolaou1975asymptotic,papanicolaou1976some} considered perturbation expansions of the governing operator semigroups, and
laid the mathematical framework for rigorously proving convergence results. These ideas were further developed to consider convergence of stochastic differential equations forced onto a manifold by a strong drift\cite{katzenberger1991,funaki1993,fatkullin2010reduced,lelievre2012langevin}. %, starting with the foundational paper by Katzenberger\cite{katzenberger1991} which was built upon in several directions \cite{funaki1993,fatkullin2010reduced,lelievre2012langevin}. 
Notably, most of this literature does not consider systems with spatially-dependent diffusion coefficients, which is essential for connecting the results to physical systems at the mesoscale. 

%More recent efforts have started with general stochastic differential equations and proven the convergence to a limiting equation, but 
%have rarely been connected to physical systems at the mesoscale \cite{fatkullin2010reduced,lelievre2012langevin}. 

In this article, we adopt a modern approach to deriving the softly constrained equations: we start with Brownian dynamics equations, which include stiff restraining forces, and asymptotically compute the ``soft limit'' where the restraining forces become infinitely stiff. Our approach relies on singular perturbation theory, which doesn't require making any assumptions about the thermodynamic properties of the system nor the precise form of the Lagrange multipliers (which as we discuss is subtle), but rather, is a straightforward mathematical procedure which gives the correct limiting equations given accurate assumptions on the scales involved. This derivation, which is applied here for the first time to constrained Brownian dynamics with multiplicative noise, recovers the equations in Ref.~\onlinecite{morse2003Theory}. Importantly, singular perturbation theory allows us to go further and to derive the softly constrained equations in the case where the mobility tensor varies on the same scale as the restraining potential, a situation we call \emph{soft soft} constraints. Such would be the case, for example, with colloids held in close proximity or near a boundary, since hydrodynamic lubrication forces between surfaces vary rapidly with the distance between them~\cite{sprinkle2020driven, franosch2009persistent,franosch2011resonances}. 
A further benefit of this approach is that it highlights the validity regime for the softly constrained equations: they are valid for timescales that are long enough compared to the timescales to explore the restrained degrees of freedom. Hence, it allows us to determine when constrained equations will be effective to capture the systems dynamics even when forces do not appear to be all that stiff.

We apply our theory to several physically-motivated examples, including systems which exhibit hydrodynamic interactions %(a colloid constrained to diffuse near a wall or two colloids constrained in lower-dimensional optical traps) 
and systems with bonds between particles.  %(a ligand-receptor pair diffusing together, a particle tethered to a wall with a ligand-receptor contact, and a 4-particle assembly in a square). 
Our examples highlight the nontrivial and diversity of effects that constraints can induce on macroscopic dynamics in soft-matter systems. We also include several non-physical, but low-dimensional examples to show the geometric origins behind the changes in the effective drift and effective mobility tensors. Examples of both kinds are hard to find in the literature, and indeed, one aim of this article is to provide a concise summary of the softly-constrained equations of Brownian dynamics and a pedagogical illustration of how to apply them to concrete examples.

\vspace{3mm}

This article arose out of conversations with Aleksandar Donev, who aimed to build fast solvers for Brownian dynamics, and who wished to understand what is the correct mobility tensor to use for constrained Brownian dynamics when particles are in close contact: should one project it first, and then average? Or average and then project? And is there a specific type of projection that should be used in such situations? While we regret that Dr. Donev did not live to see the resolution to this question, we are honoured to contribute post-humously to his research program by  showing that \textit{project-then-average} is the correct approach. Parts of this article are inspired by his unfinished notes on the topic.  

\vspace{3mm}

The manuscript is organized as follows. In Sec.~\ref{sec:setup} we introduce notation and summarize the softly constrained Brownian dynamics equations. % general notations and summarize our results, in particular the softly constrained equations that one can use to simulate/analyze restrained soft matter systems. 
In Sec.~\ref{sec:consequences} we illustrate how softly constrained systems can exhibit rather subtle dynamical phenomena in 4 typical ``soft matter'' examples, applying the derived formulas to: (A) a particle confined by gravity near a wall; (B) two particles diffusing next to one another when constrained by traps; (C) a particle tethered to another or to a surface; and (D) a square assembly of particles. In Sec.~\ref{sec:toymodels}, we derive,  for a few toy models, how restrained dynamics give rise to subtle constrained dynamics on long timescales and show how to recover constrained results applying the formulas of Sec.~\ref{sec:setup}. We also include numerical validation on some of the toy models. In Sec.~\ref{sec:derivation}, we derive the softly constrained  dynamics of an arbitrary system with stiff restraints and interpret the results. Finally, we show an alternative derivation of the constrained equations, using Lagrange multipliers, in Sec.~\ref{sec:lagrange}. Compared to singular perturbation, this technique lacks robustness, as the final result is sensitive to the assumed form of the Lagrange multipliers, which is not obvious in the presence of noise. %directly in the Brownian dynamics equations. however, we also show that the result is sensitive to the assumed form of the Lagrange multipliers -- some seemingly natural forms, such as It\^{o} noise or Stratonovich noise, give incorrect results. 
%%%%%%%%%%%%%%%%%%%%%%%%%%%%%%%%%%%%%%%%%%%%%%%
%%%%%   Section: Setup and Main results             %%%%%
%%%%%%%%%%%%%%%%%%%%%%%%%%%%%%%%%%%%%%%%%%%%%%%
\section{Setup and main results}
\label{sec:setup}

In this section we summarize the setup and our main results, which will be derived and motivated by examples in subsequent sections. Our hope is that a reader who wishes to use constrained dynamics without necessarily following the derivations, may use this section as a reference point.

\begin{figure}
    \centering
    \includegraphics[width=0.99\linewidth]{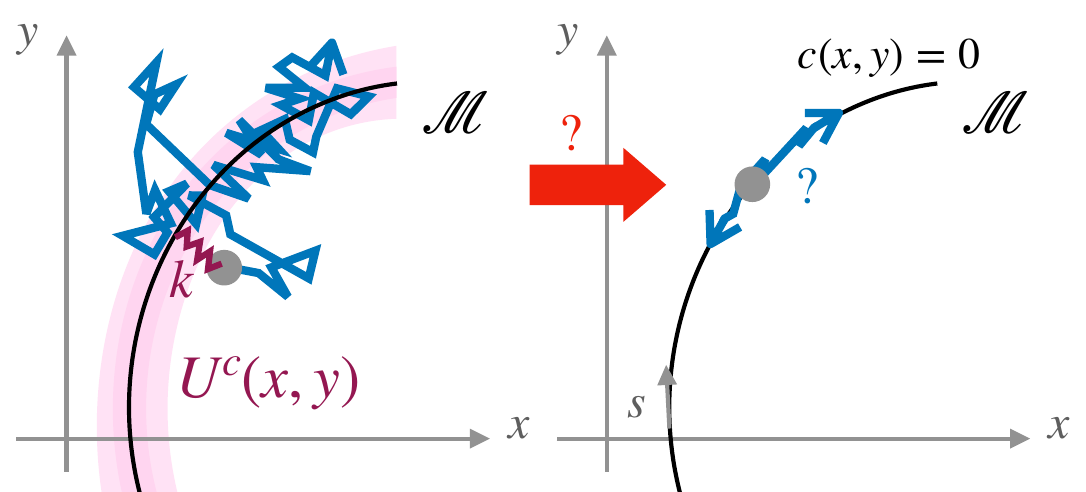}
    \caption{\textbf{Schematic of a softly constrained particle to a curved line in 2D.} Here $\mathcal{M}$ is the confined space over which the constraint $c(x,y)=0$, $\Uc(x,y)$ the confining potential around $\mathcal{M}$, and $k$ a typical spring constant associated with the confining potential. }
    \label{fig:schematic}
\end{figure}

\paragraph{Starting point: overdamped equations with a soft confining potential.}
Objects or particles with scales of nanometers or micrometers are well described by overdamped Langevin equations. We thus start with an $n$-dimensional overdamped equation of the form 
\begin{equation}\label{eq:OD}
    \dd{\x}{t} = -M\nabla \Utot + k_BT\nabla\cdot  M + \sqrt{2k_BT}M^{1/2} \eta(t).
\end{equation}
Here $\x(t)\in \R^n$ is the configuration of the process at time $t$, $\Utot(\x)$ is the potential energy, $\eta(t)\in \R^n$ is an $n$-dimensional white noise, $T$ is temperature, $k_B$ is Boltzmann's constant, and $M(\x)$ is the mobility tensor, a symmetric positive definite $n\times n$ matrix. We write $M^{1/2}$ to represent any square root of $M$, \textit{i.e.} any matrix such that $M^{1/2}(M^{1/2})^T = M$. 
The mobility tensor is related to both the diffusion tensor $D=k_BTM$, and the friction tensor $\Gamma = M^{-1}$. The stationary distribution for these dynamics (written as a density with respect to the usual Euclidean measure $d\x$) is the Boltzmann distribution 
\begin{equation}\label{eq:pi0}
\pi(\x) = \frac{e^{-\beta \Utot(\x)}}{Z},
\end{equation}
where $\beta=(k_BT)^{-1}$, and $Z$ is a normalizing constant.

We wish to model the situation where the potential strongly confines the system near some region of space, or \emph{manifold}, which we call $\mathcal M$ (fig.~\ref{fig:schematic}). The manifold $\mathcal M$ is defined as the level set of a collection of \emph{constraints}, \textit{i.e.} of functions $\c=(c_1,\ldots,c_m)^T$ where $m$ is the number of constraints, such that
\begin{equation}\label{eq:Mmanifold}
\mathcal M = \{\x\in \R^n: c_i(\x)=0,i=1,\ldots,m\}.
\end{equation}
%We call $\c(\x)$ the \emph{constraints}.
%Note that the constraints here are functions only of the particle coordinates and not of particle velocities. 
We give some examples of constraints in subsequent sections, however, one example could be a distance constraint between particles $\x_i,\x_j\in \R^2$ or $\R^3$, which would have the form $c(\x) = |\x_i-\x_j| - l_0$, where $l_0$ is a fixed length. Note, that although it does not seem essential at this stage, the constraints have to be homogeneous to distances; this will enter in the assumed form of the constraining potential. 

To model these constraints, we suppose the overall potential has the form
%\begin{equation}\label{eq:U}
$\Utot(\x) = \Uc(\x) + U(\x)$, where $U$ is an ``external'' contribution and $\Uc$ strongly confines the system near $\mathcal M$. The external potential $U$ must vary slowly compared with the scale of the confinement. One example (though not the only possibility) of a strongly-confining potential would be
\begin{equation}\label{eq:Uc}
\Uc(\x) = \half\sum_{i=1}^m k_i c_i^2(\x) .
\end{equation}
Here the $k_i$ can be interpreted as spring constants, and they may depend on $\x$ though are assumed to vary slowly compared to the scale of confinement, which is $\sim \sqrt{k_B T/k}$ where $k = \min_i|k_i|$.

We remark that there is some ambiguity in the relation between $k_i(\x)$ and $c_i(\x)$, since, given  dimensionless functions $\alpha_i(\x)>0$, the transformation $k_i(\x)
\to k_i(\x)/\alpha_i(\x)$, $c_i(\x) \to \alpha_i(\x) c_i(\x)$, leads to the same level sets $\mathcal M$ as in \eqref{eq:Mmanifold}, and the same confining potential \eqref{eq:Uc}. We choose to allow variations in both $k_i$ and $c_i$, both for generality, and to allow for some flexibility in modelling: in some problems (e.g. Section \ref{sec:2dflatline}) it may be natural to think of the spring constant as varying with space, as this avoids an unnatural change of spatial coordinates. Fortunately, the final constrained equations are equivalent no matter what the choice of $\alpha_i$.  

When $k$ is sufficiently large, the confining potential $\Uc$  has a narrow, deep minimum along $\mathcal M$, and the dynamics are constrained near $\mathcal{M}$. This corresponds to \emph{soft} constraints, in that small deviations relative to $\mathcal{M}$ are possible.
%(Note that $\Uc$ could have higher-order contributions in $c_i$, but we are only interested in the leading-order part.) 
%We may also have another, ``external'' contribution to the potential, $\Ue$, which varies  slowly compared to the scale of confinement, which is $\sim \sqrt{k_B T/\max_i|k_i|}$. %\sophie{that's the same for $\Ue$ but depending on how we write this it would be enough to say slow compared to the scale of confinement which is $\sqrt{k_B T/k}$ where $k \sim max|k_i|$}. 
%The overall potential has the form 
%\begin{equation}\label{eq:U}
%$\Utot(\x) = \Uc(\x) + U(\x)$.
%\end{equation}

To make progress on modeling, we consider the limit of a strongly confining potential, when the length scales of interest are much larger than the scale of the confinement, and we investigate the limiting equations constrained to $\mathcal{M}$. %when $k$ is large, and investigate the limiting equations constrained to $\mathcal{M}$. %In this limit the dynamics are exactly constrained to $\mathcal M$, and there is no more need for the confining potential $\Uc$ as its effect is included in the limiting dynamics. 
Physically, we will show that this limit is valid on timescales $\tau$ that are long enough such that diffusive motion on this timescale occurs over lengthscales that are much larger than the confining length, $\sqrt{D_0\tau} \gg \sqrt{k_B T/k}$ and so $\tau \gg k_BT/D_0k$ where $D_0$ is the typical amplitude of the diffusion coefficients. 
The limiting dynamics can be written using two ways: one is extrinsically, in the original Cartesian variables $\x$, and the other is intrinsically, using a collection of variables which parameterize the manifold. We recapitulate the form of the limiting dynamics in Table~\ref{tab:formulas} and briefly describe them here.

\paragraph{A few mathematical notations}

To express the limiting dynamics we need to define the Jacobian matrix 
\begin{equation}\label{eq:C}
C\equiv\nabla \c = \begin{pmatrix}
(\nabla c_1)^T \\
... \\
(\nabla c_m)^T
\end{pmatrix} \in \R^{m\times n}.
\end{equation}
We adopt the convention that the gradient of a scalar-valued function (as in $\nabla c_1$ or $\nabla U$) is a column vector, however for 
a  Jacobian (as in $\nabla \c$), gradients of each of the involved functions are assumed to be row vectors. We write $|C| = |CC^T|^{1/2}$ to mean the pseudodeterminant of $C$. 
In the above notation, all quantities depend on $\x$, i.e. $C=C(\x)$, however we usually suppress this in the notation. Note that since we supposed $c(\x)$ is homogeneous to a distance, the components of $C$  and its pseudodeterminant are unitless.

We must assume, for our results below to be valid, that $m\leq n$ and that $C$ has full rank, a condition which is equivalent to asking that there are no extra zero-modes in the system. This implies that $\mathcal M$ is a $d=(n-m)$-dimensional manifold. 
Note that this assumption may be violated even by distance constraints, which leads to nontrivial effects on the thermodynamics and dynamics of the system \cite{meng2010free,mao2015mechanical,kallus2017,mannattil2022Thermal}.  

\renewcommand{\arraystretch}{1.5}
\begin{table*}
    \centering
    \begin{tabular}{c|c}
    \hline 
    \multicolumn{2}{c}{Projection operators, conventions and general notations} \\
    \hline 
        Constraint matrix (with constraints $\c(\x) = 0$) & $C(\x) = \begin{pmatrix}
(\nabla c_1)^T \\
... \\
(\nabla c_m)^T  
\end{pmatrix}$ \\
        Projection on the constraint manifold & $P =  I - C^T(CC^T)^{-1}C $ \\
        Friction matrix & $\Gamma$ \\
        Projected friction matrix & $\Gamma_P = P \Gamma P$ \\
        Mobility matrix & $M = \Gamma^{-1}$ \\
        Projection oriented by the mobility matrix & $P_M =  I - MC^T(CMC^T)^{-1}C $ \\
        Projected mobility matrix & $M_P = P_M M = P_M MP_M =  \Gamma_P^{\dagger}$ \\
        Projected mobility square root & $\sigma_P$ such that $\sigma_P \sigma_P^T =M_P = \Gamma_P^{\dagger}$ \\
        Effective potential & $U^{\rm eff}(\x) = U(\x)-k_BT \log \kappa(\x)/\kappa^m_0$ \\
        Constraint potential prefactor term & If Eq.~\eqref{eq:Uc}, then $\kappa(\x) = \prod_{i=1}^m \sqrt{\frac{2\pi k_BT}{k_i(\x)}}$; else Eq.~\eqref{eq:kappa}.
 \\
 \hline 
    \multicolumn{2}{c}{Effective equations of motion for \textit{\textbf{soft constraints}} (infinitely stiff spring)} \\
    \hline 
 Equilibrium distribution & $\pi_{\rm constrained}(\x) =  \frac{1}{Z}  e^{-\beta U^{\rm eff}(\x)}\delta(\c(\x))$ \\
        Dynamics with $M_P$ (equivalently $\Gamma_P^{\dagger}$) & $\frac{dx_k}{dt} = - M_{P,ki}  \partial_i U^{\rm eff}(\x)  + k_B T \partial_iM_{P,ki} + \sqrt{2k_B T} \sigma_{P,ki} \eta_i(t)$ \\
        Alternative form & $\frac{dx_k}{dt} = - M_{P,ki} \partial_i \left( U^{\rm eff}(\x)+ k_B T \log |C| \right) +k_B T  P_{ij} \partial_j M_{P,ik}   + \sqrt{2k_BT} \sigma_{P,ki} \eta_i(t)$ \\
        Fokker-Planck equation &    $\partial_t p = \partial_i \left(  M_{P,ij} (\partial_j U^{\rm eff}(\x) ) p \right) + \partial_i  \left(  M_{P,ij}  \partial_j p \right)$  \\
        %Backward Kolmorov equation & 
%$\mathcal{L} f = - \left( M_P_{ij} \partial_j \mathcal{U}_{\rm eff}(x) \right) \partial_i f + k_B T \partial_j (M_P_{ij} . \partial_i f)$ \\
 \hline 
    \multicolumn{2}{c}{Effective equations of motion for \textbf{\textit{soft soft constraints}} (finite spring stiffness) } \\
    \hline 
 Same as soft but with $M_P \rightarrow \overline{M_P}$  & $\overline{M_P}(\x) = \int M_P(\x) e^{\frac{-\Uc(\x)}{k_BT}}d\c/\int e^{\frac{-\Uc(\x)}{k_BT}}d\c$ \\
        %Backward Kolmorov equation & 
%$\mathcal{L} f = - \left( M_P_{ij} \partial_j \mathcal{U}_{\rm eff}(x) \right) \partial_i f + k_B T \partial_j (M_P_{ij} . \partial_i f)$ \\
\hline 
    \multicolumn{2}{c}{Equations of motion for \textit{\textbf{hard constraints}} (rigid joints)} \\
    \hline 
        Equilibrium distribution & $\pi_{\rm constrained}(\x) = \frac{1}{Z} \frac{1}{|C|}e^{-\beta U^{\rm eff}(\x)}\delta(\c(\x))  = \frac{1}{Z} e^{-\beta (U^{\rm eff}(\x) - k_BT \log|C|)}\delta(\c(\x))$\\
        Dynamics with $M_P$ (or with $\Gamma_P^{\dagger}$...)  & $\frac{dx_k}{dt} = - M_{P,ki}  \partial_i \left(U^{\rm eff}(\x) - k_B T \log |C| \right) + k_B T \partial_iM_{P,ki} + \sqrt{2k_B T} \sigma_{P,ki} \eta_i(t)$ \\
        Alternative form  & $\frac{dx_k}{dt} = - M_{P,ki}  \partial_i U^{\rm eff}(\x) +k_B T  P_{ij} \partial_j M_{P,ik}   + \sqrt{2k_BT} \sigma_{P,ki} \eta_i(t)$ \\
        Fokker-Planck equation &    $\partial_t p = \partial_i \left(  M_{P,ij} ( \partial_j \left(U^{\rm eff}(\x) - k_B T \log |C| \right) ) p \right) + \partial_i  \left(  M_{P,ij}  \partial_j p \right)$ 
    \end{tabular}
    \caption{General formulas to obtain projected, constrained, dynamics. All the stochastic dynamics are to be read in the It\^{o} sense. We use summation over repeated indices. All operations are written in the usual Euclidean space or extrinsic coordinates.%\sophie{in the $\Gamma_P$ equations I am pretty sure in the first term the derivative of the potential is meant over all spatial dimensions} \sophie{@MIranda, can we add something associated with the soft-soft constraints? And how do I do it for non-uniform constraint? (does that latter case matter?)}
   The equivalence between Eq.~\eqref{eq:ODP0} and the ``alternative form'' provided in the table is shown in Appendix~\ref{app:equivalenceMobFriction}. %Notice $U^{\rm eff}(\x) = U(\x) - k_B T \log \kappa(\x)$, where $\kappa(\x)$ essentially describes how the strength of the constraints change in space. 
    }
    \label{tab:formulas}
\end{table*}

\paragraph{Extrinsic form of the equations of motion on the constrained space.} 

In the usual Euclidean space, \textit{i.e.} in \textit{extrinsic} variables, the soft limit of Eq.~\eqref{eq:OD} is
\begin{equation}\label{eq:ODP0}
    \dd{\x}{t} = -M_P\nabla U^{\rm eff} + k_BT\nabla\cdot  M_P + \sqrt{2k_BT}M^{1/2}_P \eta(t).
\end{equation}
This is directly similar to Eq.~\eqref{eq:OD} except it involves a projected mobility $M_P$, which in general is different from $M$, and an effective potential 
\begin{equation}
U^{\rm eff}(\x) = U(\x)-k_BT \log \kappa(\x)/\kappa^m_0.
\label{eq:Ueffsoft}
\end{equation}
This is modified from the external potential by a term $\kappa(\x)$, which equals, for a confining potential of the form \eqref{eq:Uc}, 
\begin{equation}\label{eq:kappa0}
\kappa(\x) = \prod_{i=1}^m \sqrt{\frac{2\pi k_BT}{k_i(\x)}}.%\int \exp \left( - {\frac{\Uc(\x)}{k_BT}} \right) d\c
\end{equation}
The constant $\kappa_0$ is any reference lengthscale, and is included in \eqref{eq:Ueffsoft} for dimensional consistency, but it is not involved in any calculation as the potential $U^{\rm eff}(\x)$ is defined up to a constant. 
Note that the term $\kappa(\x)$ is not needed in the effective potential if the $k_i$ are constants. In fact, one could absorb any variations in $k_i$, into the definitions of the constraint functions $c_i$, provided the constraints stay proportional to lengths and the constraint manifold $\mathcal{M}$ is not modified. A more general expression for $\kappa$ is given in our derivation, see Eq.~\eqref{eq:kappa}. %\mhc{valid if Uc has form given in Eq 4, add comment that if the potential is not harmonic, obtained in a different way -- ref to formula later}
%for $\x\in \mathcal M$. \sophie{I don't know what the above means at all unfortunately... I mean I have no clue how to calculate this practically...} \sophie{I also do not have the same equations in the table...}
%The notation above means we integrate over the fibres normal to the manifold, which are given by variations in $\c$. This average only makes sense for $\x\in \mathcal M$. 

The projected mobility tensor is
\begin{equation}\label{eq:MP}
M_P =P_MM \quad \text{with } P_M = I-MC^T(CMC^T)^{-1}C.
\end{equation}
It is such that $P_MMP_M = P_MM$, as we verify in Sec~\ref{sec:PMidentity}, hence, it is symmetric.
The matrix $P_M(\x)$ is a projection matrix with a specific geoemtrical property: it projects onto the %space spanned by $\{\M t_i\}_{i=1}^d$, where $\{t_i\}_{i=1}^d$ forms a basis for the 
space which is the linear mapping of $M(\x)$ with the level sets of $\{c_i(\x)\}$, and it is orthogonal in an inner product space weighted by $M^{-1}(\x)$, as we discuss in Sec~\ref{sec:projection}. 
Because the projection matrix is defined geometrically, it does not depend on the specific form of the constraints when evaluated \emph{on the manifold $\mathcal M$} -- however, it \emph{does} depend on the specific form of the constraints as it moves away from the manifold. Notably, the divergence term  $\nabla \cdot M_P$ in \eqref{eq:ODP0} gives a different drift, depending on the functional form of the constraints, as we discuss at the end of this section. 
Hence, this projected mobility is what encodes the \emph{physics} of the constraints -- this is how the specific forces imposing the constraints, enter into the effective dynamics. %\mhc{added some text here}
%Hence, this projected mobility is where the specific form of the constraints is encoded in the dynamical equations. 

%\sophie{addsomething that says that you can also do the alternative way...}
%We show later that this is an orthogonal projection of the original mobility $M$, in a weighted inner product space. 
%is a projection of the original mobility tensor onto the space defined by $\mbox{span}\{Mt_i\}_{i=1}^d$, where $\{t_i\}_{i=1}^d$ is a basis for the tangent space to $\mathcal M$ at $\x$, and which is such that the projection is orthogonal with respect to an inner product with weight $M^{-1}$. \sophie{I don't know how important it is to say that in the setup but I don't think so}
Note that the projected mobility tensor may be equivalently, and sometimes more easily, obtained as the inverse of the projected friction, $M_P = \Gamma_P^{\dagger}$ (see Appendix~\ref{app:equivalenceMobFriction} for a proof), where $\dagger$ denotes the Moore-Penrose pseudo inverse. The projected friction is directly obtained as 
\begin{equation}\label{eq:GammaP}
    \Gamma_P = P \Gamma P, \quad\text{where} \quad P = I - C^T(CC^T)^{-1}C
\end{equation} 
is the projection operator onto the tangent space to $\mathcal M$, which is orthogonal with respect to the usual Euclidean inner product. The projected friction is obtained using the perhaps more natural projection operator $P$, because, the physical object that can be projected are forces, which are linearly related to the friction matrix $\Gamma$. In analytical calculations, $\Gamma_P$ can be easier to calculate than $M_P$ as we show via several examples in Sec.~\ref{sec:consequences} and Sec.~\ref{sec:toymodels}; however in numerical calculations where the mobility $M$ is easier to obtain, it can be easier to work with $M_P$, as it avoids a matrix inverse. We discuss the differences between projected mobility and projected friction in Sec.~\ref{sec:projection}.

For ``soft-soft'' constraints where the 
 original mobility varies rapidly, i.e.  on the same scale as the confining potential,  the effective mobility should instead be replaced by an averaged mobility $\overline{M_P}(\x)$ defined as
 \begin{equation}\label{eq:MPvary}
\overline{M_P}(\x) = \frac{1}{\int e^{\frac{-\Uc(\x)}{k_BT}}d\c}\int M_P(\x) e^{\frac{-\Uc(\x)}{k_BT}}d\c.
\end{equation}
The notation above means we integrate over the $c_i$, assuming these are variables; for example $\int e^{-\half k_1c_1^2(\x)/k_BT}dc_1 = \sqrt{2\pi/ k_1}$. 
That is, the mobility should be replaced by the equilibrium average of $M_P$ with respect to $\Uc$, averaged over the fibres normal to $\mathcal M$.\footnote{These fibres are defined by keeping internal parameterization variables constant, as shown in our derivation in Section \ref{sec:generalderivation}, however they could have an expression as a limit of a gradient flow, as in \cite{Sharma.2021} -- we leave such a construction for future work.} This is one of the main novel results of the paper -- a derivation of the ``softly softly'' constrained dynamics where the mobility varies on the same scale as the confining potential.

%\mhc{I'm still not 100\% sure that this average is written correctly -- we haven't defined what the fibres are, ie what the paths are we are integrating over, in the multidimensional case. Must keep the $\s$ variables constant during integration, and this is not specified in the integration the way it is written. To make this correct we might need to think about the gradient flow down $U^c$ which leads to $\mathcal M$, and use this gradient flow to define the fibres. But this seems too much for a physics paper, and the referees probably won't ask.}

 % \mhc{
 % , which is constructed as
 % \begin{equation}
 % \overline{M_P}(\x) = J^{-1}\begin{pmatrix} \overline{\tilde M} & 0 \\ 0 & 0 \end{pmatrix} (J^{-1})^T.
 % \end{equation}
 % Here $ \overline{\tilde M}$ is the averaged mobility in an intrinsic coordinate system, defined later in the section in Eqn.~\eqref{eq:Mtildevary}, and $J=\partial \y/\partial \x$ is a Jacobian of a variable transformation $\x\to\y$, which takes Cartesian coordinates $\x$ to a curvilinear coordinate system $\y = (\s,\c)$, and is discussed in more detail in Sec.~\ref{sec:generalderivation}.
 % }

Equation \eqref{eq:ODP0} has several nice properties that one would desire of constrained dynamics: 
(i) It preserves the constraints: $dc_i(X_t)/dt = 0$; (ii) It has stationary density (with respect to the usual Euclidean measure $d\x$)
\begin{equation}\label{eq:pidelta}
\pi(\x) = \frac{1}{Z}e^{-\frac{U^{\rm eff}(\x)}{k_BT}}\delta(\c(\x)),
\end{equation} 
where $Z$ is another normalization constant, and $\delta(\c(\x)) = \prod_{i=1}^m\delta(c_i(\x))$; and (iii) It satisfies detailed balance with respect to this stationary distribution. We verify these properties in Appendix~\ref{sec:proofs}.

%We say $\pi$ is ``locally'' stationary because it is unnormalized. 
%Written as a density with respect to the natural surface measure $\dsigmaM$ on $\mathcal M$, we have
%    \begin{equation}\label{eq:piM}
%\pi(\x) = e^{-\frac{V(\x)}{k_BT}}|\nabla c|^{-1}\dsigmaM,
%    \end{equation}

\paragraph{Intrinsic form of the equations.} 
%I reorganized the explanation a bit - please check.}

We can alternatively write the dynamics in terms of a collection of variables $\s=(s_1,s_2,\ldots,s_{d})$ which form a local parameterization of the manifold $\mathcal M$; such variables are sometimes referred to as generalized coordinates. For example, if $\mathcal M$ is a circle, then it can be parameterized by an angle $\theta$, as $s_1= a \,\theta$ where $a$ is a lengthscale, for instance the radius of the circle. %In the following we give the general expression for the dynamics, but here we recapitulate, for conciseness, only a simplified result, in the case where the metric tensor $g_{\s} = ((\nabla \s)(\nabla \s)^T)^{-1}$ on the parametrized manifold is the identity matrix. 
The dynamics in local variables is
\begin{equation}\label{eq:seff}
\dd{\s}{t} = -\tilde M\nabla_{\s}U^{\s, \rm eff} + k_BT \divM \tilde M + \sqrt{2k_BT}\tilde M^{1/2} \eta.
\end{equation}
Here $\nabla_{\s}$ means the usual collection of partial derivatives with respect to the $s_i$. 
The operator $\divM$ is the divergence operator with respect to the natural volume form on $\mathcal M$:  $\divM \v = |g_{\s}|^{-1/2}\sum_{i=1}^d \partial_{s_i}(|g_{\s}|^{1/2}\v^i)$, where we write $\v^i$ to mean the $i$th component of a vector field $\v$, and where $g_{\s} = ((\nabla \s)(\nabla \s)^T)^{-1}$ is  the natural metric tensor  on $\mathcal M$. (Appendix~\ref{sec:diffgeom} has a summary of relevant geometry formulas.)

The effective potential is 
\begin{equation}\label{eq:effpot}
U^{\s, \rm eff} = U + k_BT \log |C|- k_BT \log \kappa,
\end{equation}
where $\kappa$ is defined in Eq.~\eqref{eq:kappa0}, and $|C|$ is the pseudodeterminant of $C$, defined after Eq.~\eqref{eq:C}. 

The projected mobility converted to $\s$-variables is
 \begin{equation}\label{eq:Mtilde}
 \tilde M = (\nabla \s)M_P(\nabla \s)^T.
 \end{equation}
  If the mobility is rapidly-varying, then the mobility is $\overline{\tilde M} = \overline{(\nabla \s)M_P(\nabla \s)^T} = (\nabla \s)\overline{M_P}(\nabla \s)^T$, where the averaging operator and $\overline{M_P}$ are defined in \eqref{eq:MPvary}. 
 Alternatively, we can work with the projected friction tensor $\tilde \Gamma$ which is again the Moore-Penrose pseudoinverse of the projected mobility, $\tilde M=\tilde \Gamma^\dagger$, and which can be shown to be (Sec.~\ref{sec:GammaP})
 % In analytic calculations it can be easier to work with the projected friction tensor, which in $\s$-coordinates may be shown (see Section \ref{sec:projection}) to be $(\nabla_\s\x)^T\Gamma (\nabla_\s\x)$, hence is related to the mobility as 
 \begin{equation}\label{eq:GammaS}
     \tilde \Gamma=(\nabla_\s\x)^T\Gamma (\nabla_\s\x).
 \end{equation}
 Thus, the projected friction is obtained by simply evaluating the friction in each of the tangent vector directions, where the tangent vectors are the column space of $\nabla_\s\x$. 

% For soft-soft constraints where the mobility is rapidly-varying, the mobility is replaced by  an averaged version  $\overline{\tilde M}$, defined by
% \begin{equation}\label{eq:Mtildevary}
% \overline{\tilde M} = \frac{1}{\int e^{\frac{-\Uc(\x)}{k_BT}}d\c}\int \tilde M e^{\frac{-\Uc(\x)}{k_BT}}d\c.
% \end{equation}
% The notation above means we integrate over the $c_i$, assuming these are variables; for example $\int e^{-\half k_1c_1^2(\x)/k_BT}dc_1 = \sqrt{2\pi k_BT/ k_1}$.
% That is, the mobility should be replaced by the equilibrium average of $\tilde M$ with respect to $\Uc$, averaged over the fibres normal to $\mathcal M$. This is one of the main novel results of the paper -- a derivation of the ``softly softly'' constrained dynamics where the mobility varies on the same scale as the confining potential. 

In the intrinsic coordinates, there is an extra contribution of $k_BT \log |C|$ in the potential, in Eq.~\eqref{eq:effpot}, which is not present in the extrinsic form of the equations, in Eq.~\eqref{eq:Ueffsoft}. This extra contribution has been the source of much discussion in the literature \cite{Kampen.1981,hinch1994Brownian,fatkullin2010reduced}, and it is a well-known phenomenon in sampling and free energy calculation problems that arise in molecular dynamics\cite{sprik1998free,Ciccotti.2005,ciccotti2008projection}.
It is notable that this additional force \emph{depends} on the particular choice of the constraint functions -- multiplying these functions by another function can change $|C|$ without changing the manifold $\mathcal M$ that the dynamics is constrained to. In this sense, the effective dynamics are affected by the nature of the constraining forces, even when they have been averaged out of the equations. We discuss this point more in subsequent sections especially Sec.~\ref{sec:2dcurve}.

We remark that although the extra drift $k_BT \nabla_\s\log |C|$ does not appear explicitly in the extrinsic equations, it is still encoded indirectly in these equations as part of the divergence term. We show in Appendix~\ref{sec:divM} that 
\begin{equation}\label{eq:divM}
    k_B T \nabla \cdot M_P=- M_P \nabla \log |C| + k_B T (P \cdot \nabla) M_P .
\end{equation}
This comes after first showing the key identity $M_P\nabla\cdot P = -M_P\nabla \log |C|$.
Hence, the divergence term in Cartesian coordinates gives both a divergence on the manifold, and a drift on the manifold.
Such relationships may be useful to simplify calculations in specific cases; especially in the case of \textit{hard} constraints, where the term $\log |C|$ is involved in the extrinsic form of the equations and one may want to avoid calculating it (see Table~\ref{tab:formulas}). 

% Finally, for completeness we present the general form of the intrinsic dynamics, which is most completely described in terms of the backward equation for the dynamics \sophie{Miranda can we give this in terms of the FPE? not the backward? or just give the equilibrium distribution?}:
% \begin{equation}\label{eq:limgeneral}
% \partial_t a = k_BT \pi_{\s}^{-1}\divM (\pi_{\s} \tilde M D_{\s} a). 
% \end{equation}
% Here $D_{\s}a = (\partial_{s_1}a,\ldots,\partial_{s_d}a)$, and $\divM$ is the divergence on manifold $\mathcal M$ with respect to metric tensor $g_{\s}$. % = ((\nabla \s)(\nabla \s)^T)^{-1}$ 
% An explicit formula for the divergence in an arbitrary coordinate system is given in Section \ref{sec:diffgeom}. 
The stationary distribution, with respect to the natural surface measure $\dsigmaM = |g_{\s}|^{1/2}ds_1\cdots ds_d$ on $\mathcal M$, is $\pi_{\s}(\s) = e^{-U^{\s, \rm eff}/k_BT}$ or 
\begin{equation}
    \pi_{\s}(\s) = \frac{1}{Z}\kappa |C|^{-1}e^{-\frac{U(\s)}{k_BT}}.
    %= e^{-(\Ue - k_BT \log \kappa + k_BT\log |C|)/k_BT}
\end{equation}
This stationary distribution can be obtained from Eq.~\eqref{eq:pidelta} using the coarea formula $\delta(\c(\x)) = |C|^{-1}\dsigmaM$, which essentially comes from changing variables  with Jacobian $|C|^{-1}$. 

%where $\kappa$ is defined in Eq.~\eqref{eq:kappa0}, and $|C|$ is the pseudodeterminant of $C$, defined in Eq.~\eqref{eq:C}.
%The 

%From the backward equation we can write out the stochastic dynamics with a particular metric tensor $g_{\s}$. We will write it out here 

\section{Physical examples and consequences of constraint dynamics}
\label{sec:consequences}

In this section, we explore four physical situations where constrained situations arise. We show how to use the above equations as summarized in Table \ref{tab:formulas} to obtain the constrained equations of motion. More importantly, we discuss the subtleties that arise in the projected dynamics and highlight original physical effects.  

%\mhc{none of this has the effective drift term. should we move part of the parabola section here? and use it to show the extra drift? or should we say somewhere, that the drift has been considered by many authors so we don't emphasize it...}

%\mhc{in general we need to be a bit more precise about the difference between this section and the previous -- this one has a derivation and result, previous has just result, but how will a reader navigate this?}

%\mhc{another thought is we could include only the "physical" examples in the previous section, and move the toy stuff to this one (e.g. the 2d random walkers)}

%\mhc{but then, should we put a "physical" example where we have to include an extra drift term, in the previous section? what example would be itneresting enough (and not covered in previous references)?}

\subsection{Particle confined by gravity near a wall: an example of a ``soft soft constraint''.}

%\sophie{SOPHIE TIDIES THIS SECTIONS}

Many soft matter systems investigate particle suspensions in confined geometries, often as they sediment/or are sedimented above a wall or as they interact with other confining interfaces such as porous media~\cite{gompper20252025,levitz2025probing}. As they diffuse near these interfaces, particles exhibit hindered mobility due to increased hydrodynamic friction between the confining interface and the particle~\cite{brenner1961slow,faxen1921einwirkung}. This affects all components of the mobility matrix. Resolving hindered motion near interfaces is costly, since it requires to resolve local motion near the interface at scales smaller than a particle size, while the simulation domain may be much larger~\cite{fish2025libmobility}. It is thus natural to seek simplified coarse-grained dynamics near interfaces.

\begin{figure}[h!]
    \centering
    \includegraphics[width=0.99\linewidth]{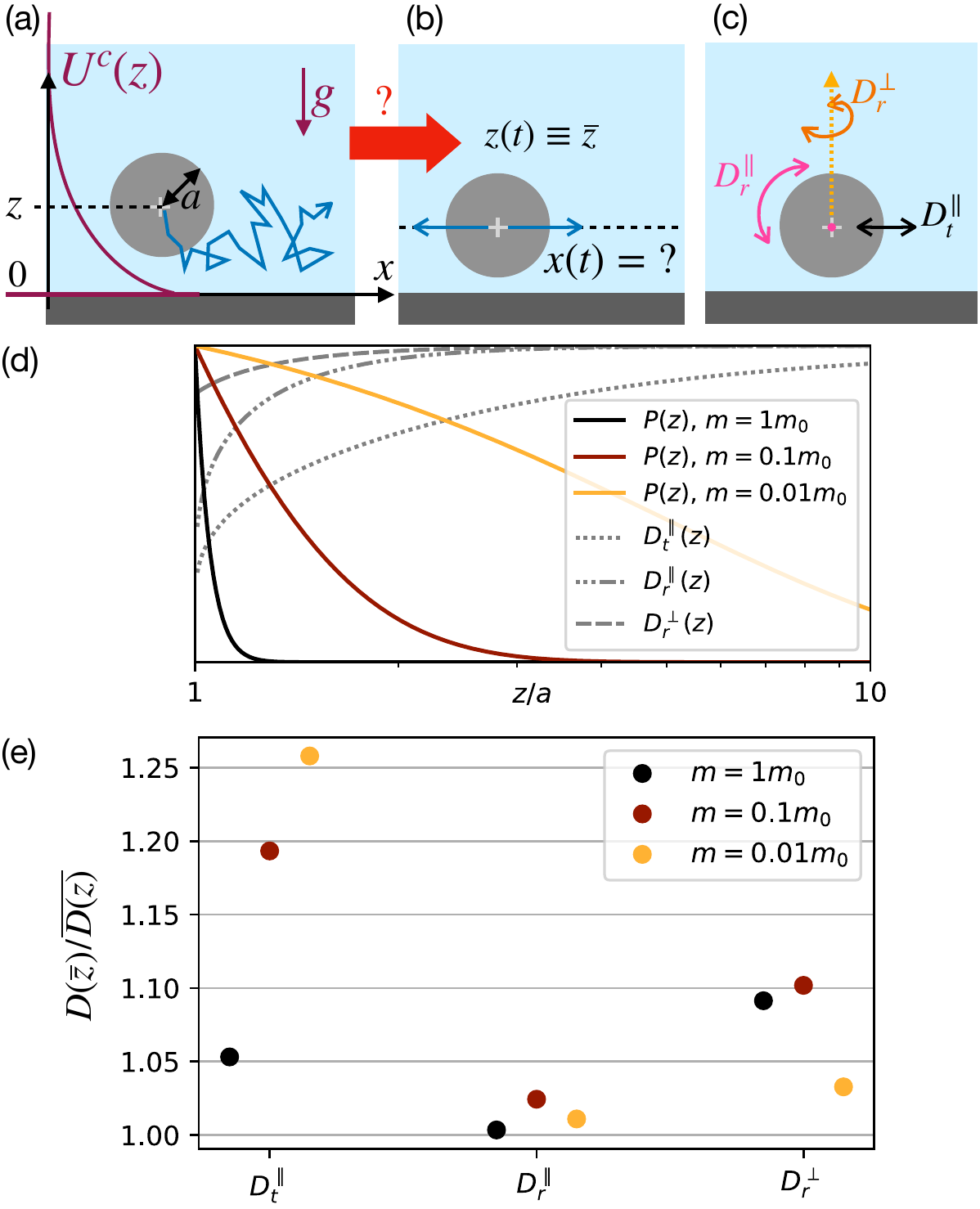}
    \caption{\textbf{Physical implications of constraining particle motion near a wall.} (a) Example of particle (gray) diffusing and confined by gravity near a bottom wall (dark gray). (b) Effective dynamics of the constrained motion would consider only motion along the wall. (c) Definition of  different diffusion coefficients near the wall. (d) Analytical predictions of the effective diffusion coefficients $\overline{D_i(z)}$ versus $D_i(\overline{z})$ where $\overline{ z}$ is the mean height for different values of particle-fluid mass imbalance $m$. (e) Associated equilibrium distributions $P(z)$ and formulas for $D_i(z)$ on the same horizontal axis. Note that the $D_i(z)$ are renormalized by their values at $z = \infty$. Numerical parameters: reference mass $m_0 g = 0.0592~\mathrm{pN}$, with $g$ gravity; particle radius $a = 1.395~\mathrm{\mu m}$; viscosity $\eta = 1.4\times 10^{-3}~\mathrm{Pa.s}$ and temperature $T = 300~\mathrm{K}$. }
    \label{fig:DofzBrenner}
\end{figure}

Here, we explain how to do so on an example of a particle sedimented above a wall (Fig.~\ref{fig:DofzBrenner}a) and what the effective dynamics would be if we were to consider only motion in the direction parallel to the wall (Fig.~\ref{fig:DofzBrenner}b). Near the wall, a particle can diffuse vertically in $z$, horizontally in $x$, and can also rotate, on an axis parallel ($r_\parallel$) or orthogonal ($r_\perp$) to the wall. This creates a $4\times 4$ diffusion matrix $D$. In Fig.~\ref{fig:DofzBrenner}c, we highlight the diagonal components of this matrix (except the $z$ component), written $D_i$, where $i$ refers to a degree of freedom, and where $i = t,\parallel$ corresponds to the horizontal translational degree of freedom $x$. We also give the dependence of these diagonal elements with $z$ in Fig.~\ref{fig:DofzBrenner}d. Gravity creates an effective confining potential near the wall as 
\begin{align*}
    \begin{cases}
        \Uc(z) = - m g (z-a) \,\, &\text{for} \,\, z \geq a, \\
        \Uc(z) = \infty  \,\, &\text{for} \,\, z \leq a,
    \end{cases}
\end{align*}
where $z$ is the particle height above the wall (measured from the particle center), $a$ is the particle radius, $m$ is the mass imbalance (between the mass of the particle and the mass of the displaced fluid), and $g$ is gravity.
Note, that this confining potential does not have the quadratic shape proposed in Eq.~\eqref{eq:Uc}, but the results are general and apply to any confining potential shape provided it confines motion of a given degree of freedom to a small lengthscale. Here, on average the particle is located at the so-called sedimentation height $\overline{z}$ where, for this section, we notice the average simplifies to $\overline{\cdot}  = \int \cdot  \exp(- \Uc(z)/k_B T) \mathrm{d}z / \int \exp(- \Uc(z)/k_B T) \mathrm{d}z $. Due to the confining gravity potential, a particle only explores moderately space around its mean height $\overline{z}$. Fig.~\ref{fig:DofzBrenner}d shows typical probability distribution curves $P(z) \propto \exp \left( - \Uc(z)/k_B T \right)$. 

We now seek an effective description of motion over long time scales and parallel to the wall.
The constraint in this example is simply $c(z) = z - \overline{z}$. The projection matrix $P$ is thus simply a block diagonal matrix with a $0$ in the diagonal $z$ component, and 1's on all other diagonal components. The diffusion matrix has no off-diagonal components between $z$ and another degree of freedom, and hence the projected diffusion matrix $D_P = k_B T M_P$ is simply the block matrix related to the ``other'' degrees of freedom. However, there is significant variation of the diffusion components on scales similar to that of the confining potential, as highlighted in Fig.~\ref{fig:DofzBrenner}d. We must thus modify all coefficients of the projected diffusion matrix by averaging them over the confining gravity potential according to Eq.~\eqref{eq:MPvary}. To estimate them we use analytical expressions for mobility above a wall using the tabulated expansions in Ref.~\onlinecite{perkins1992hydrodynamic},  which agree well with experimental data~\cite{sprinkle2020driven}. Fig.~\ref{fig:DofzBrenner}e reports the effective, averaged, (diagonal) diffusion coefficients $\overline{ D_i(z)} $, and compares them to a ``naive'' picture where one would simply have evaluated the diffusion coefficients at the mean height $D_i(\overline{z})$. 

Because the confining potential $\Uc(z)$ is indeed quite soft, $\overline{D_i(z)} $ and $D_i(\overline{z} )$  are markedly different, reaching up to $10-20\%$ difference depending on parameter values (Fig.~\ref{fig:DofzBrenner}e). The coefficient that changes the most is the translational diffusion coefficient parallel to the wall $D_{t}^{\parallel}$, because its variations with $z$ extend the furthest away from the wall (see dotted line in Fig.~\ref{fig:DofzBrenner}e). In comparison, $D_{r}^{\parallel}$ barely changes, also because its spatial dependence on $z$ is quite mild (dashed line in Fig.~\ref{fig:DofzBrenner}d). To test these effects further, we test a situation where the mass imbalance between the fluid and the particle is varied by several orders of magnitude (a situation which can be realized by using fluid/gaz droplets instead of colloids). According to the value of $m$, one may notice a more or less drastic difference between $D_i(\overline{z})$ and $\overline{ D_i(z)}$. For instance, for the heaviest particle explored (black), one notices more differences in $r_{\perp}$ than on ${t_\parallel}$ components. This is because $P(z)$ in that case varies the most on the same scales as $D_r^{\perp}(z)$, while  $D_t^{\parallel}(z)$ only barely changes on these scales. The situation is reversed for the lightest particle (yellow). 
Eq.~\eqref{eq:MPvary} thus provides a valuable recipe to coarse-grain, explore and simulate colloidal motion including hydrodynamics.

\subsection{Particle approaching a stuck or trapped particle: off-diagonal mobility terms.}
\label{sec:consequences:offdiag}

We now consider a situation where a particle approaches another ``stuck'' or trapped particle. This is a common situation for instance in colloidal sciences, where particles can adhere to surfaces while others are freely diffusing; or when a particle is trapped in an optical trap~\cite{thorneywork2020direct,franosch2011resonances,grier1997optical}. We note that the particles' diffusion originates from the fact that they are embedded in an underlying fluid. Thus when two particles are immersed in water they interact via hydrodynamic interactions, \textit{e.g.} when a particle is moved in the vicinity of another particle, it can induce motion ``at a distance'' of the other particle. 
%Basically, motion of particle 1 affects particle 2 due to long-range hydrodynamic interactions and reciprocally. 
We now investigate how to account for effective interparticle interactions, within a model where one of the particles is constrained. 

\begin{figure}
    \centering
    \includegraphics[width=0.99\linewidth]{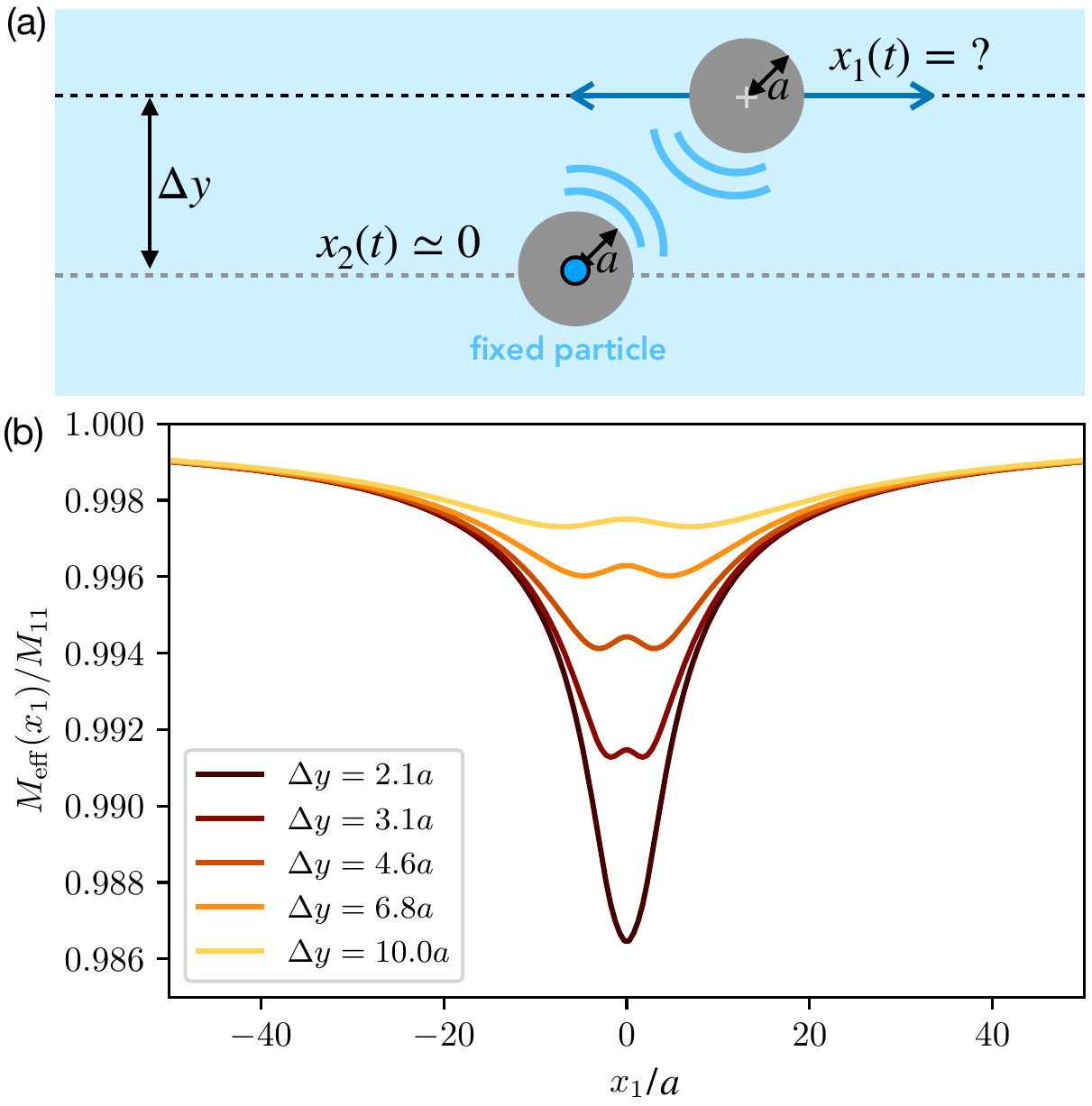}
    \caption{\textbf{Particle mobility near another trapped particle.} (a) Setup of two particles moving on separated lines and where particle 2 is trapped in a confining potential around $x_2(t) \simeq 0$. (b) Effective long time mobility of particle 1 as a function of its position along the line $x_1$, as given by Eq.~\eqref{eq:offdiagproj}. The colors indicate different separation distances $\Delta y$ between the lines. }
    \label{fig:2particles}
\end{figure}

To illustrate this, we consider the following simplified situation, where 2 random walkers, or particles, can diffuse on parallel lines, separated by a distance $\Delta y$, and their coordinates are described by $(x_1(t),x_2(t))$. Particle 2 is trapped by a stiff potential $U^c(x_2)$ to a point of the line, such that $x_2(t) \simeq 0$ (see Fig.~\ref{fig:2particles}a). The mobility tensor that describes this system contains non-zero off-diagonal components,
\[
M = \begin{pmatrix} M_{11} & M_{12}(x_1,x_2) \\  M_{12}(x_1,x_2) & M_{22} \end{pmatrix}
\]
where $M_{12}(x_1,x_2)$ depends on space and accounts for long range hydrodynamic interactions between the two particles. For simplicity, we will consider here the potential $U^c$ varies over much shorter lengthscales compared to $M_{12}$. 

%\mhc{the projected mobility is a nontrivial projection of the actual mobility, which is reduced from the value it takes in simply restricting it due to off-diagonal terms in the mobility tensor which induce correlations that reduce the mobility. We now show the simplest example that illustrates this mobility reduction, and then we will show an example of how this manifests in a physical system with physical measurements.}

We first see how we can use the tabulated equations to obtain the effective motion of particle 1. The
 ``constraint'' here is $c(x_1,x_2) = x_2$. We thus have $C = (\nabla c)^T = (0,1)$. The mobility formula Eq.~\eqref{eq:MP} is useful if you have access to the original mobility matrix and if calculating the inverse is hard. For small analytic calculations, however, the friction route is often easier. We use
$$ \Gamma = \begin{pmatrix}
     \Gamma_{11} & \Gamma_{12} \\ \Gamma_{12} & \Gamma_{22}
 \end{pmatrix}, \,\, P = \begin{pmatrix}
     1 & 0 \\ 0 & 0 
 \end{pmatrix}, \,\, \Gamma_P = P \Gamma P = \begin{pmatrix}
     \Gamma_{11} & 0 \\ 0 & 0
 \end{pmatrix}.$$
% and its inverse $\Gamma_P^\dagger = \begin{pmatrix}
 %    \Gamma_{11}^{-1} & 0 \\ 0 & 0
 %\end{pmatrix}$. 
 Since $\Gamma = M^{-1}$, naturally $\Gamma_{11} = M_{22}/(M_{11} M_{22} - M_{12}^2)$, and $M_{P,11} = \Gamma_{11}^{-1}$.
% The limiting dynamics involve the projected mobility tensor Eq.~\eqref{eq:MP}, $M_P =M-MC^T(CMC^T)^{-1}CM$. 
% We compute $C M C^T = M_{22}$ and  
% \begin{align*}
%     MC^TCM &= \begin{pmatrix}
%         M_{12}^2 & M_{12}M_{22}\\
%         M_{12}M_{22} & M_{22}^2
%     \end{pmatrix}
% \end{align*}
% and hence
% \[
% M_P = \begin{pmatrix}
%     M_{11}-\frac{(M_{12})^2}{M_{22}} & 0 \\ 0 & 0
% \end{pmatrix}.
% \]
Thus, the effective mobility in the $x_1$-direction is 
\[
M^{\rm eff} = M_{P,11} = M_{11}-\frac{(M_{12})^2}{M_{22}}. \label{eq:offdiagproj}
\]
In the constrained dynamics,  the diffusivity of $x_1$ is therefore \emph{reduced} by what it would be from considering $x_1$ on its own, without accounting for the coupling with $x_2$. This is true even though particle 2 doesn't move. The reduction is associated with the off-diagonal term in the mobility -- the larger this term, the larger the reduction. 
In Sec.~\ref{sec:2dnondiag} we will show in detail that this reduction arises because of the anti-correlation in the dynamics between the 2 degrees of freedom. Here, clearly the effective mobility of $x_1$ has to be reduced by the presence of $x_2$ which increases hydrodynamic friction. 

We now quantify the typical impact of the additional term in the mobility in Eq.~\eqref{eq:offdiagproj} in a physical example of two nearby colloids. We suppose, for the sake of simplicity, that we can neglect motion in other directions than $x$ (a case of hard constraints) and take mobilities only on the $x$ axis. We use again the analytical expressions for mobility of a particle near a similar particle in the tabulated expansions in Ref.~\onlinecite{perkins1992hydrodynamic}. Note, in these ``far field'' expansions of hydrodynamic interactions, the self mobility of a particle ($M_{11}$, $M_{22}$) is \textit{not} affected by the presence of another particle. In these ``far field'' expansions, the only effect of hydrodynamic interactions is that when particle $1$ is displaced, this may induce motion of particle $2$ at a distance. This yields a contribution that depends depends on the relative distance between the particles, only in $M_{12}(x_1,x_2)$.

In Fig.~\ref{fig:2particles}b we estimate what this effective mobility might be as a function of the horizontal distance between the 2 particles, at different separations between the lines $\Delta y$.  We find, naturally, that the effective mobility is most reduced when the particles are closer to one another. 
The effect is due to an effective increased friction associated with the resistance of particle 2 to being displaced at a distance by particle 1.
While we obtain an apparently small reduction of about $2\%$, one should bear in mind that we used here ``far field'' expressions for hydrodynamic mobilities. This relatively small mobility reduction could be  enhanced at smaller separations, using more accurate predictions with short-range lubrication corrections; and should eventually decay to 0 when particles nearly touch. In that case,  $M_{11}$ should also depend on space. Eventually, when multiple particles are stuck, a colloid diffusing near them would exhibit an effective mobility reduced by contributions from all the stuck particles; in essence approaching the effective mobility reduction observed near a wall, which is significant, even in the ``far field'' regime (see $D_{\parallel}(z)$ in Fig.~\ref{fig:DofzBrenner}e).  

Note that a similar effect may occur at the level of a single particle, when it exhibits 2 coupled degrees of freedom (or a non-zero off-diagonal mobility term) in the sense that if you apply a force on 1 of the degrees of freedom, that force propagates to both degrees of freedom. For instance, for a particle near a wall, there exists a single non zero off-diagonal mobility term: between the rotation on the axis parallel to the wall and translation parallel to the wall~\cite{perkins1992hydrodynamic}. If one traps a particle near a wall, with $x(t) \simeq cst, z(t) \simeq cst$, then the effective rotational diffusion is modified by this coupling. This is a small contribution for spherical particles, but it could be significantly larger for particles with anisotropic shapes.

% Figure \ref{fig:example1} (bottom) shows the mean-squared displacements $\Delta(\tau)$ for simulations using $k=1$ and different values of $M_{12}$. The mean-squared displacements are notably not diffusive at short times, but at longer times they increase linearly with $\tau$. The time to approach diffusive behaviour increases with $M_{12}$; by scaling time we know this time will be proportional to $1/k$. This illustrates that the effective diffusivity is only meaningful at sufficiently long times. 

%\bigskip \mhc{sophie -- add your table here?}
% \sophie{depending on paper split up or not no? If the focus of this paper is soft constraints, the table has also hard constraints and the gamma version but we can leave a short version of it?}

\subsection{Tethered particles: how their mobility changes, and the importance of defining constraints}
\label{sec:tethered}

Many biological or synthetic systems contain ligand-receptor contacts, often made of polymers, such as DNA pairs between synthetic colloids~\cite{mirkin1996dna,feng2013specificity,wang2015crystallization,rogers2016using,cui2022comprehensive,gehrels2022programming}, protein linkers which bind microscale white blood cells to blood vessel walls~\cite{alon2002rolling,ley2007getting,korn2008dynamic} or spike proteins that viruses use to adhere to mucus~\cite{mammen1998polyvalent,sakai2017influenza,sakai2018unique,muller2019mobility}. 
The mobility of tethered particles is altered from that of the particles separately. In this section we show two examples of tethered particles: (1) a simple example to illustrate that the effective mobility of tethered particles is a harmonic average of the original particles' mobilities, and (2) a more geometrically complex example involving a particle tethered to a surface with a floppy spring, a situation commonly encountered when modelling particles with ligand-receptor contacts. This example also illustrates the importance of choosing constraints to accurately model the physical nature of the constraining forces. %\mhc{added some more introduction here}

\subsubsection{Two particles tethered by a stiff spring.}
We investigate the case where two particles, numbered 1 and 2, with diffusion coefficients $D_1 = k_B T/\Gamma_1$ and $D_2 = k_B T/\Gamma_2$ diffuse in free space but while being \textit{tethered}, \textit{i.e.} attached by a spring with spring constant $k$ to one another.  The physical question we ask is ``what is the effective mobility of the tethered particles?'' The answer to that question is non trivial. Reasonable choices include that the effective diffusivity is given via $D = k_BT/\Gamma$ where $\Gamma = \Gamma_1 + \Gamma_2$ or $\Gamma^{-1} = \Gamma_1^{-1} + \Gamma_2^{-1}$.

\begin{figure}
    \centering
    \includegraphics[width=0.99\linewidth]{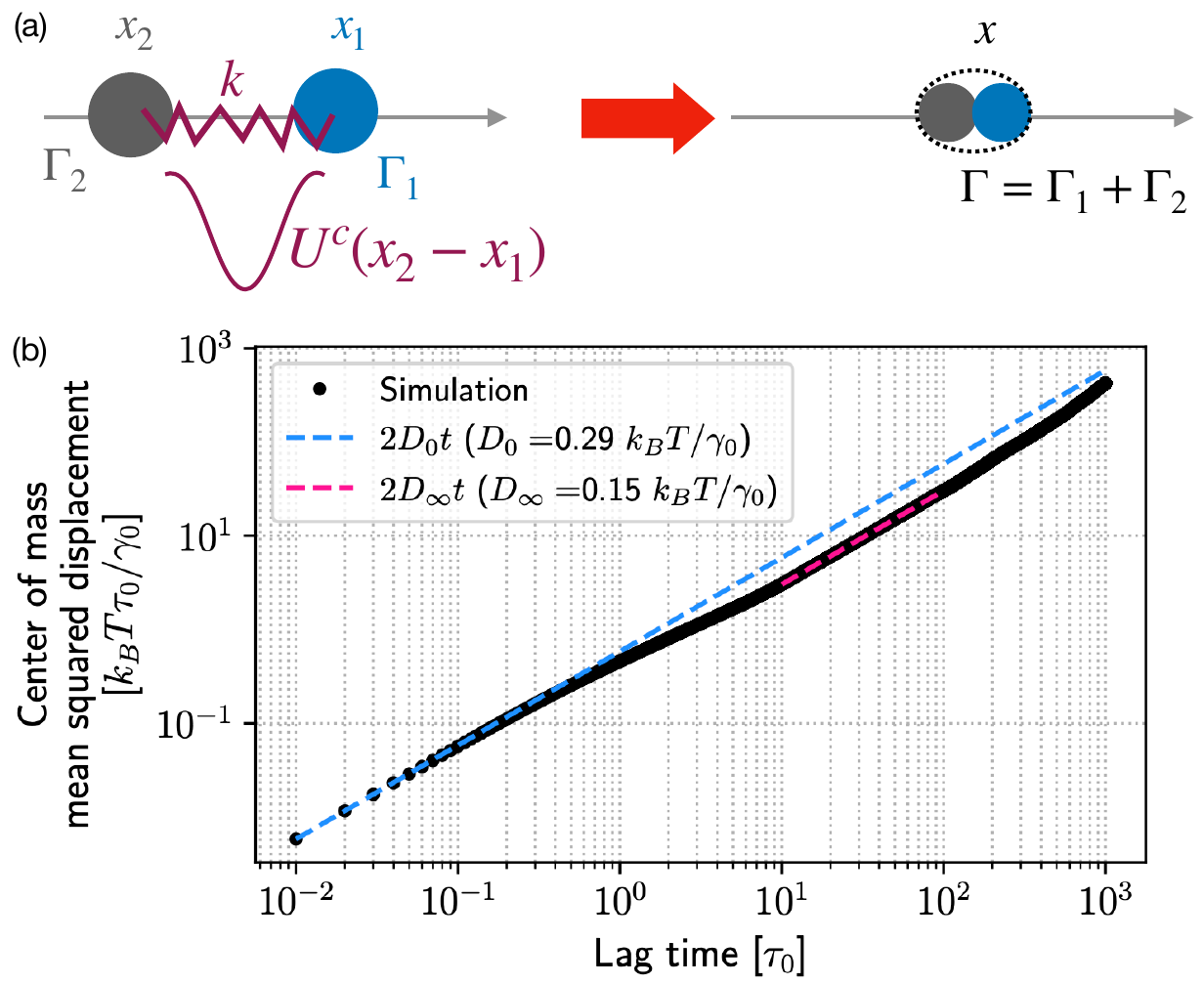}
    \caption{\textbf{Effective mobility of tethered particles.} (a) Setup of two particles moving on a line with a soft constraint between them. (b) Mean-squared displacement of the center of mass of the pair depicted in (a) with $k = 1 \gamma_0/\tau_0$, $\Gamma_1 = \gamma_0$ and $\Gamma_2 = 6 \gamma_0$ such that the effective bound friction coefficient predicted by the theory is $\Gamma = \Gamma_1 + \Gamma_2 = 7 \gamma_0$ and $D_{\infty} = k_B T/\Gamma \simeq 0.14 k_B T/\gamma_0$ close to the measured value. The lines show fits at short (done on the 2 first time points) and long timescales (done over the range where the pink line is plotted). These results correspond to an average of $100$ independent runs of $10^6$ time steps with time step $0.01 \tau_0$. The mean-squared displacement is calculated as in Eq.~\eqref{eq:msd}. In this simulation, $k_BT$, $\tau_0$, $\gamma_0$ are used as units.}
    \label{fig:tethers}
\end{figure}

We investigate this situation in 1D to illustrate the subtleties. Let $x_1$ and $x_2$ be the position of the particles on a line. In the absence of a constraint interaction, they evolve as
\begin{equation}
    \begin{cases}
    &\frac{dx_1}{dt} =  \frac{f_1}{\Gamma_1} + \sqrt{\frac{k_BT}{\Gamma_1}} \eta_1 \\
    &\frac{dx_2}{dt} =    \frac{f_2}{\Gamma_2} + \sqrt{\frac{k_BT}{\Gamma_2}} \eta_2
    \end{cases}
\end{equation}
where $f_1$ and $f_2$ are external forces, which could depend on space. 

To add a soft constraint between the particles, we can ask that the two particles remain at a fixed distance, so we consider the distance constraint $c(x_1,x_2) = x_1 - x_2$ and the constraint matrix $C = (\nabla c)^T = (1 , -1)$, and $CC^T = 2$. The projected friction is
\begin{equation*}
    \Gamma_P = P \Gamma P , \, \text{with} \,\, \Gamma = \begin{pmatrix}
    \Gamma_1 & 0 \\ 0 & \Gamma_2 
\end{pmatrix}\, \text{and} \,\, P = \frac{1}{2} \begin{pmatrix}
    1 & 1 \\ 1 & 1 
\end{pmatrix}.
\end{equation*}
We easily find that $\Gamma_P = \frac{\Gamma_1 + \Gamma_2}{2} P $ whose Moore-Penrose pseudo-inverse is  $\Gamma_P^{\dagger} = \frac{2}{\Gamma_1 + \Gamma_2} P$. One can work out a square root from the  Cholesky decomposition,
\begin{equation*}
    (\Gamma_P^{\dagger})^{1/2} = \sqrt{\frac{1}{\Gamma_1 + \Gamma_2}} \begin{pmatrix}
    1 & 0 \\ 1 & 0
\end{pmatrix}
\end{equation*}
and so we obtain the following equations for the constrained dynamics:
\begin{equation}
    \frac{dx_1}{dt} = \frac{dx_2}{dt} = \frac{1}{\Gamma_1 + \Gamma_2} \left( f_1 + f_2 \right) + \sqrt{2k_B T} \sqrt{\frac{1}{\Gamma_1 + \Gamma_2}} \eta_1 .
\end{equation}

The effective diffusion coefficient of tethered particles is thus proportional to the inverse of the sum of the friction coefficients. That is, the effective friction is a regular average of the two friction coefficients, while the effective diffusion coefficient is a harmonic average of the two diffusion coefficients. 
This makes sense physically, since if one of the particles, say particle number 1, is not very mobile, so $\Gamma_1 \gg \Gamma_2$, then we expect that the diffusion coefficient of the tethered particles should also be slow. So indeed it should be proportional to $(\Gamma_1 + \Gamma_2)^{-1}$. 

In Fig.~\ref{fig:tethers}b we confirm this theoretical prediction on an example simulation of particles connected by a spring. At long enough timescales, the center of mass of the tethered particles indeed diffuses with a diffusion coefficient close to that predicted by theory.

\subsubsection{Particle bound to a surface by a ligand-receptor contact: how you choose your constraints matters.}\label{sec:ligandreceptor}

%\mhc{this could be part of the previous section, on mobility changes? }

\begin{figure}[h!]
    \centering
    \includegraphics[width=0.6\linewidth]{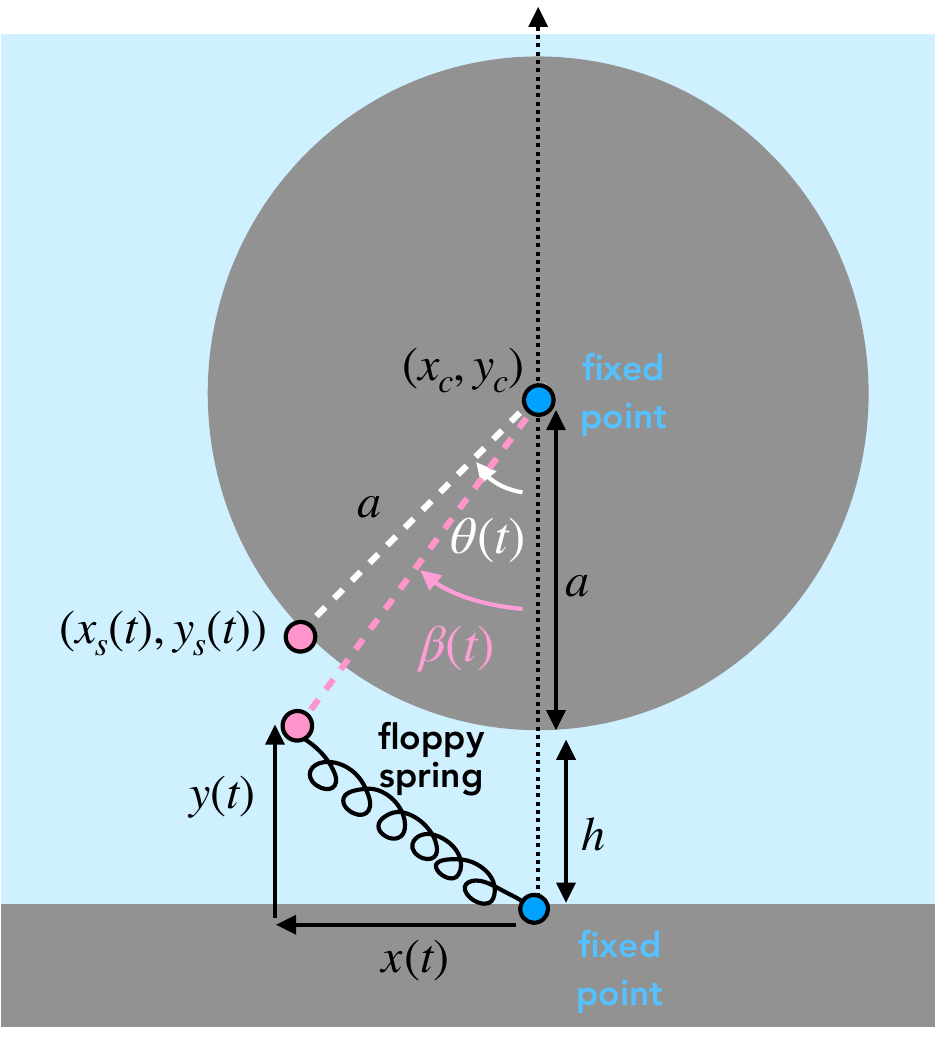}
    \caption{\textbf{Rotation of a particle tethered by a floppy spring to a wall.} Setup of notations used in the main text.}
    \label{fig:rolling}
\end{figure}

We consider a more realistic example with a large colloid whose only degree of freedom is rotation on an axis parallel to a wall, as $\theta(t)$ (see Fig.~\ref{fig:rolling}). We also consider a long floppy spring, with spring constant $K$, attached by one end right below the center of the colloid, and whose free tip can jiggle in $(x(t),y(t))$. The equations of motion are %\mhc{I removed negative signs in front of the forces}
\begin{align*}
        &\frac{dx}{dt} =  \frac{D}{k_B T} F_x + \sqrt{2D} \eta_{x} \\
        &\frac{dy}{dt} =  \frac{D}{k_B T} F_y + \sqrt{2D} \eta_{y}  \\
        &\frac{d\theta}{dt} = \sqrt{2D_\theta} \eta_{\theta} 
\end{align*}
where the forces are $F_x  = -K (x - \frac{\ell_0}{\sqrt{x^2 + y^2}})$ and $F_y = - K (y - \frac{\ell_0}{\sqrt{x^2 + y^2}})$, with $\ell_0$ the rest length of the floppy spring, and all other notation follows our usual conventions. 
The equation for $\theta$ should be interpreted as having periodic boundary conditions on $[-\pi,\pi]$. 
We consider here the tip of the floppy spring (in pink) is constrained by another stiff spring to a point $(x_s(t),y_s(t))$ on the surface of the colloid (the other pink dot). We note that the floppy spring, which jiggles slowly, should not be mistaken with the stiff spring defining the constraint, which could jiggle quite rapidly (the latter one is akin to the stiff spring between the two particles in the previous paragraph). There are several ways to impose such a constraint.

\paragraph{Distance constraints.}
One way is to impose distance constraints such that $x(t) = x_s(t)$ and $y(t) = y_s(t)$, giving the constraints
\begin{align*}
    &c_1(x,y,a\theta) = x - a \sin \theta  \\
    &c_2(x,y,a\theta) = (y - h) - a( 1 - \cos \theta)
\end{align*}
where $a$ is the colloid's radius and $h$ the minimum separation between the colloid and the wall. 
The constraint matrix is thus 
\begin{align*}
    C = \begin{pmatrix}
        1 & 0 & - \cos \theta \\
        0 & 1 & - \sin \theta
    \end{pmatrix}
\end{align*}
so that 
\begin{align*}
    CC^T = \begin{pmatrix}
        1 +  \cos \theta &  \sin \theta \cos \theta \\
         \sin \theta \cos \theta & 1 +  \sin^2 \theta
    \end{pmatrix}
\end{align*}
with pseudo-determinant $|C| = \sqrt{2}$.

One can evaluate the projected mobility  as 
\begin{equation*}
    M_P = \frac{D^{\rm eff} }{k_B T}  \begin{pmatrix}
        \cos^2 \theta & \cos \theta \sin \theta & \cos \theta \\
        \cos \theta \sin \theta & \sin^2 \theta & \sin \theta \\
        \cos \theta & \sin \theta & 1
    \end{pmatrix}
\end{equation*}
where $D^{\rm eff} = \frac{a^2D_{\theta}D}{a^2D_{\theta} + D}$ is the harmonic average of the diffusivities $D$ and $a^2D_{\theta}$. This 
 has a Cholesky decomposition  
\begin{equation*}
    \sigma_P = \sqrt{\frac{D^{\rm eff}}{k_B T}  }\begin{pmatrix}
        \cos \theta & 0 & 0 \\
        \sin \theta & 0 & 0 \\
        1 & 0 & 0 
    \end{pmatrix}
\end{equation*}
such that $\sigma_P \sigma_P^T = M_P$.

The constrained dynamics for $\theta$ are then %evolution of the system through a single coordinate, $\theta$, which evolves as
\begin{align}
    a\frac{d\theta}{dt} = &- \frac{D^{\rm eff}}{k_B T} \left( \cos \theta F_x  + \sin \theta F_y \right) + \sqrt{2 k_B T D^{\rm eff} } \eta_{\theta}.
    \label{eq:angletethered}
\end{align}
These constrained equations are obtained in the original Cartesian coordinate system using Eq.~\eqref{eq:ODP0}, however we can also consider $\theta$ to be an intrinsic variable and  write $F_x, F_y$ in terms of $\theta$ only, using the constraints. %\mhc{added -- please check} 

In this example, we notice that the effective torque on the particle is now projected from the floppy spring force. The effective mobility is a combined mobility as $\frac{1}{k_B T} \frac{a^2D_{\theta}D}{a^2D_{\theta} + D}$. We notice again that here the friction coefficients $k_B T/D$ and $k_B T/ a^2 D_{\theta}$ were summed to get an effective friction, much like in the previous example of the two particles, except with the slight modification that the angular diffusion coefficient has to be multiplied by $a^2$.

\paragraph{Polar constraints.}
Another way to impose the constraint is using polar coordinates. We can impose that the distance between $(x(t),y(t))$ and the colloid's center is exactly $a$ and that the angle $\beta(t) = \theta(t)$, such that the constraints (written with dimensions of lengths) are now
\begin{align*}
    &c_1(x,y,a\theta) = \frac{1}{a}\left(x^2 + (a+h -y)^2 - a^2 \right) \\
    &c_2(x,y,a\theta) = a\theta - a\beta = a\theta - a\arctan(x/(a+h-y)).
\end{align*}
In that case, one can show that the projected dynamics are exactly the same as in Eq.~\eqref{eq:angletethered} (see Appendix~\ref{app:tethered}). This will not be the case in general (we show momentarily another form of the constraints where the dynamics are not the same), but can happen in specific examples. 

An important practical derivation skill to note in these examples is that one should not substitute any relationship between variables, such as given by the constraints, until the end of the derivation of the constrained equations of motion. For instance, in the above example, in the constraint space $x = a \sin \theta$ and $a + h - y = a \cos \theta$ and so it is tempting to substitute these relations, for instance when calculating $\nabla \cdot M_P$. However, doing so gives the incorrect mobilities -- one has to note that the derivatives of $M_P$ have to be taken relative to the unconstrained coordinates and so one should not do the replacement too early on. 

% where $a$ is the colloid's radius and $h$ the minimum separation between the colloid and the wall. 
% The constraint matrix is thus 
% \begin{align*}
%     C = \begin{pmatrix}
%         2 x & - 2(a+h-y) & 0 \\
%         - \frac{1}{a+h-y} \frac{1}{1 + x^2/(a+h-y)^2}  & - \frac{x}{(a+h-y)^2} \frac{1}{1 + x^2/(a+h-y)^2}& 1
%     \end{pmatrix}
% \end{align*}
% and then 
% \begin{align*}
%     CC^T = \begin{pmatrix}
%         4x^2 + 4(a+h -y)^2 & 0 \\
%         0 & 1 +  \frac{1}{x^2 + (a+h-y)^2}
%     \end{pmatrix}
% \end{align*}
% and the pseudo-determinant is $|C| = 2 \sqrt{1 + x^2 + (a + h-y)^2}$. It does indeed depend on space. 

% The projected mobility is
% \begin{equation*}
%     M_P = \frac{D}{k_B T} \frac{D_{\theta}}{(x^2+ (a+h-y)^2)D_{\theta} + D}  \begin{pmatrix}
%         (a+h-y)^2 & x (a+h-y) & (a+h-y) \\
%         x (a+h-y) & x^2  & x \\
%         (a+h-y) & x & 1
%     \end{pmatrix}
% \end{equation*}
% such that a Cholesky decomposition is 
% \begin{equation*}
%     \sigma_P = \sqrt{\frac{D}{k_B T} \frac{D_{\theta}}{(x^2 + (a+h-y)^2)D_{\theta} + D}}  \begin{pmatrix}
%         (a+h-y) & 0 & 0 \\
%         x & 0 & 0 \\
%         1 & 0 & 0 
%     \end{pmatrix}
% \end{equation*}
% such that $\sigma_P \sigma_P^T = M_P$.
% Now we want the projected equations. 

% \vspace{2mm}
% \sophie{Here is a paradox that is making my life tricky...} 
%  \sophie{still point that out somewhere.}

\paragraph{Modified polar constraints.}
A third option for constraints in this example is to impose $\sin \theta = \sin \beta$ such that the constraints are now 
\begin{align*}
    &c_1(x,y,a\theta) = (x^2 + (a+h -y)^2 - a^2)/a  \\
    &c_2(x,y,a\theta) = a\sin \theta - a x/\sqrt{x^2+(a+h-y)^2}.
\end{align*}
In this case one obtains a pseudo-determinant on the manifold which is $|C| = 8 \cos^2(\theta)$. Thus, we expect from Eq.~\eqref{eq:seff} that the intrinsic variable $\theta$ would include an additional drift term $-k_BT\tilde M\partial_\theta \log|C|$, and the stationary distribution is correspondingly multiplied by $(8 \cos^2(\theta))^{-1}$. We report the details of the calculation in Appendix~\ref{app:tethered}, where we use the extrinsic projected equations \eqref{eq:ODP0} to show that the effective dynamics is the same as Eq.~\eqref{eq:angletethered}, up to an additional drift term $V$ given by 
\begin{equation}
V =  \frac{ D^{\rm eff}}{a} \tan \theta.
\label{eq:drift}
\end{equation}
This drift tends to push the tethered particle away from the center where $\theta = 0$ and towards $\theta = \pm \pi/2$.  Whether or not this is reasonable depends on the physical nature of the forces imposing the microscopic constraints -- is it more reasonable for the constraining potential to behave as $U^c\sim \theta^2$ or $U^c\sim \sin^2\theta$, near the constraint surface? Although the constraint surface is the same, the small deviations in the soft potential away from the constraint surface, are sufficient to change the macroscopic behaviour. 

Overall, this tethered example highlights the importance of knowing where the constraints come from physically, since the form of the constraints can have an impact on the effective dynamics. 

% \sophie{discuss with Miranda why, and maybe conclude on it is important to know where your constraints come from. }
%\sophie{Think about how to get the equations of motion after that either in extrinsic or intrinsic form}
%\mhc{should write down extrinsic (full set of variables), and intrinsic (just theta on its own), equations, to show get something different from previous case.}

\subsection{Particles with distance constraints -- additional drift}\label{sec:square}

\begin{figure}[h!]
    \centering
    \includegraphics[width=0.42\linewidth]{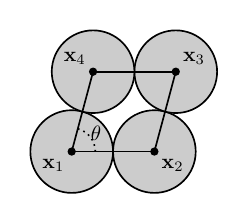}
    \includegraphics[width=0.54\linewidth]{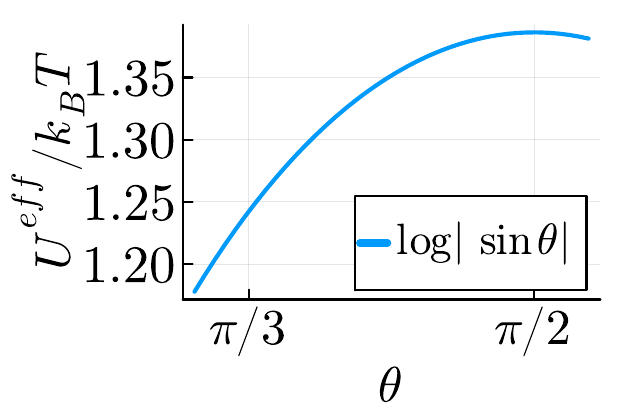}
    \caption{\textbf{Four discs interacting with stiff-spring constraints.} (Left) Schematic and (right) the corresponding effective potential  along the internal angle $\theta$. }
    \label{fig:discs}
\end{figure}

Next we give one physical example to illustrate how an additional drift term naturally enters the softly constrained dynamics, using intrisic variables as in  Eq.~\eqref{eq:seff}. Consider 4 discs with short-ranged interactions which keep them in close contact, as in Fig. \ref{fig:discs}. This example is reminiscent of similar structures formed via DNA-bonds between DNA-coated colloids~\cite{melio2024soft}. We treat the interactions as distance constraints: 
\begin{align*}
    c_1(\x) &= |\x_1-\x_2| - d\\
    c_2(\x) &= |\x_2-\x_3| - d\\
    c_3(\x) &= |\x_3-\x_4| - d\\
    c_4(\x) &= |\x_4-\x_1| - d
\end{align*}
where $\x=(\x_1,\x_2,\x_3,\x_4)$ and $\x_j = (x_j,y_j)$ is the center of the $j$th disc, and $d$ is the diameter of the discs. 
There are 8 degrees of freedom to start with, and 4 constraints, so the constraint manifold $\mathcal M$ is 8-4=4-dimensional, and can be parameterized by the center of mass $\x_c= (\x_1+\x_2+\x_3+\x_4)/4$, an overall rotation $\phi$, and an angle representing the internal deformation $\theta$. Those 4 coordinates form intrinsic coordinates for the constraint manifold. 

To evaluate the effective drift  in these intrinsic coordinates, we start by computing $|C|$, aided by the following
computation: 
\[
\nabla |\x_i-\x_j| = 
\begin{pmatrix} \cdots \overbrace{\frac{\x_i-\x_j}{|\x_i-\x_j|}}^{\x_i-\text{coords}} \cdots 
\overbrace{-\frac{\x_i-\x_j}{|\x_i-\x_j|}}^{\x_j-\text{coords}} \cdots 
\end{pmatrix}.
\]
Here $\cdots$ represents 0s. 
After some algebra we find
%
% We may compute \mhc{move to appendix}
% \[
% C = \begin{pmatrix}
%     \frac{x_1-x_2}{|\x_1-\x_2|} & \frac{y_1-y_2}{|\x_1-\x_2|} & 
%     -\frac{x_1-x_2}{|\x_1-\x_2|} & -\frac{y_1-y_2}{|\x_1-\x_2|} &
%     0 & 0 & 0 & 0 \\
%     0 & 0 & 
%     \frac{x_2-x_3}{|\x_2-\x_3|} & \frac{y_2-y_3}{|\x_2-\x_3|} & 
%     -\frac{x_2-x_3}{|\x_2-\x_3|} & -\frac{y_2-y_3}{|\x_2-\x_3|} & 0 & 0 \\
%     0 & 0 & 0 & 0 & 
%     \frac{x_3-x_4}{|\x_3-\x_4|} & \frac{y_3-y_4}{|\x_3-\x_4|} & 
%     -\frac{x_3-x_4}{|\x_3-\x_4|} & -\frac{y_3-y_4}{|\x_3-\x_4|}\\
%     -\frac{x_4-x_1}{|\x_4-\x_1|} & -\frac{y_4-y_1}{|\x_4-\x_1|} & 
%     0 & 0 & 0 & 0 & 
%     \frac{x_4-x_1}{|\x_4-\x_1|} & \frac{y_4-y_1}{|\x_4-\x_1|}
% \end{pmatrix}
% \]
%
\[
CC^T = \begin{pmatrix}
    2 & -\cos \theta & 0 & \cos\theta\\
    -\cos\theta & 2 & \cos\theta & 0 \\
    0 & \cos\theta & 2 & -\cos\theta\\
    \cos\theta & 0 & -\cos\theta & 2
\end{pmatrix}.
\]
This matrix has determinant $|CC^T| = 16-16\cos^2\theta$ hence $|C| = 4|\sin\theta|$. Thus, the non-constant contribution to the effective potential is $k_BT\log |\sin\theta|$ (Fig.~\ref{fig:discs}). Note that this effective potential was computed by a different method in  Ref.~\cite{waszkiewicz2025trimer}.

This contribution to the effective potential corresponds physically to the vibrational entropy -- it measures the relative amount of space that particles have to jiggle around their bonds infinitesimally. Vibrational entropy has been shown to play an important role in determining which configurations of colloidal particles are observed experimentally when the colloids interact with a short-ranged attractive potential \cite{meng2010free,perry2015two, HolmesCerfon:2013jw}, particularly when the entropy becomes infinite (when $|C|\to 0$, corresponding to the disc centers gathering on a line) \cite{kallus2017,mannattil2022Thermal}. This additional term in the free energy has been measured experimentally for colloids in the square configuration above, where it is  associated with an eigenvalue of the correlation matrix that is 3-4 orders of magnitude smaller than the other eigenvalues \cite{melio2024soft}.

%\mhc{now integrate over rotations, show get constant factor (=$\sqrt{2}$), so don't need to consider -- in this case only}

To obtain the full set of dynamical equations in the local variables $\s=(\x_c,\phi,\theta)$, we calculate the effective friction tensor with Eq.~\eqref{eq:GammaS}.  To this end let us explicitly construct the parameterization, as
\[
\x(\x_c,\phi,\theta) = R(\phi)\x_0(\theta) + \x_t(\x_c).
\]
Here $\x_0$ is a baseline configuration, obtained by rotating the square so that $\x_1,\x_3$ lie on the horizontal axis and $\x_2,\x_4$ lie on the vertical axis:
\[
\x_0(\theta) = d(-\cos\frac{\theta}{2}, 0,0,-\sin \frac{\theta}{2}, \cos \frac{\theta}{2},0,0,\sin \frac{\theta}{2})^T.
\]
Recall $d$ is the diameter of the discs. 
Note that $\x_0$ has been constructed so its center of mass is at the origin. This baseline configuration is then rotated by an angle $\phi$, via a rotation matrix $R(\phi)\in \R^{8\times 8}$ which has the 2x2 rotation matrices $\begin{pmatrix}\cos\phi & -\sin \phi \\ \sin\phi & \cos\phi \end{pmatrix}$ on its diagonal. Finally, it is translated by vector
\[
\x_t(x_c,y_c) = (x_c,y_c,x_c,y_c,x_c,y_c,x_c,y_c)^T.
\]

From this parameterization, we may directly calculate $\nabla_\s\x$
\begin{multline*}
    \nabla_\s\x = \\
    \begin{pmatrix}
        1 & 0 & 1 & 0 &1 & 0 &1 & 0 &\\
        0 &1 & 0 &1 & 0 &1 & 0 & 1\\
        &&& \multicolumn{2}{c}{R'(\phi)\x_0(\theta)}\\
        \frac{d}{2}(\sin\frac{\theta}{2}& 0&0&-\cos \frac{\theta}{2}& -\sin \frac{\theta}{2}&0&0&\cos \frac{\theta}{2})
    \end{pmatrix},
\end{multline*}
and then substitute into Eq.~\eqref{eq:GammaS} to obtain the projected friction tensor. In the case when $\Gamma=\gamma I$, \textit{i.e.} each coordinate has the same friction coefficient and there is no coupling between coordinates, the projected friction is 
\[
(\nabla_\s\x)^T\Gamma (\nabla_\s\x) = 
\gamma (\nabla_\s\x)^T(\nabla_\s\x) 
= \gamma\begin{pmatrix}
    4 & 0 & 0 & 0\\ 
    0& 4 & 0 & 0\\
    0& 0& 2d^2 &0 \\
    0&0&0&\frac{1}{2}d^2
\end{pmatrix}.
\]
In this case, the internal deformation angle evolves as
\[
\dd{\theta}{t} = -\frac{2k_BT}{ \gamma d^2 }\frac{1}{|\tan \theta|} + \sqrt{\frac{4k_BT}{\gamma d^2}}\,\eta^\theta(t).
\]
%
%Note that the translational diffusivity is consistent with our results in Sec.~\ref{sec:tethered}, where the factor $4$ comes from having 4 particles.
%
The drift pushes the angle $\theta$ toward smaller values, where the vibrational entropy is larger. In fact, as $\theta\to 0$ (which is not possible for discs due to steric interactions), the vibrational entropy diverges, and our results are no longer applicable because the constraint matrix $C$ fails to have full rank. 

The other variables simply diffuse, with the expected translational ($k_BT/(4\gamma)$) and rotational ($k_BT/(2\gamma d^2)$) diffusivities:
\[
\dd{\x_c}{t} = \sqrt{\frac{k_BT}{2\gamma}}\eta^{\x_c}(t), \quad
\dd{\phi}{t} = \sqrt{\frac{k_BT}{\gamma d^2}}\eta^{\phi}(t).
\]

% We have
% \begin{align*}
%     \x_c &= \frac{\x_1+\x_2+\x_3+\x_4}{4}\\
%     \phi &= \cos^{-1}\left(\frac{x_2-x_1}{|\x_2-\x_1|}\right)\\
%     \theta &= \cos^{-1}\left(\frac{(\x_2-\x_1)\cdot(\x_4-\x_1)}{|\x_2-\x_1||\x_4-\x_1|}\right)\\
% \end{align*}
% Thus
% \[
% \nabla_\x \x_c = \frac{1}{4}\begin{pmatrix}
%     1 & 0 & 1 & 0 & 1 & 0 & 1 & 0 \\
%     0 & 1 & 0 & 1 & 0 & 1 & 0 & 1
% \end{pmatrix}
% \]
% and
% \[
% \nabla_\x \phi = -\frac{1}{\sin\phi(\x)|\x_2-\x_1|}\left((-1,0,1,0,0,0,0,0) - \sin\phi(\x)\nabla_\x |\x_2-\x_1| \right)
% \]

% Figure \ref{fig:example2} shows the diffusion coefficient estimated as for the previous example is in good agreement  with the theoretical prediction. Notably, in this case the effective diffusion coefficient \emph{does} vary with $k$, because different potentials lead to different averages, even when their modes are all at the origin. 

%%%%%%%%%%%%%%%%%%%%%%%%%%%%%%%%%%%%%%%%%%%%%%%
%%%%%   Section: toy models               %%%%%
%%%%%%%%%%%%%%%%%%%%%%%%%%%%%%%%%%%%%%%%%%%%%%%
\section{Derivation of softly constrained dynamics on model examples.}
\label{sec:toymodels}

In this section our purpose is to show, on the one hand, how to use singular perturbation theory on pedagogical examples to obtain the effective constrained dynamics -- a technique we will use in full generality in Sec.~\ref{sec:derivation}. Our examples also serve to make this general geometrical derivation more concrete. On the other hand, we also wish to illustrate further how and why when implementing constraints one should be cautious about the \textit{underlying physics} the constraints refer to, since these can modify the effective dynamics that result from this constraint. 

To start, we consider 2D Brownian motion constrained to a flat line, using a stiff spring of typical scale $k$. The spring is chosen as a canonical physical force that would constrain motion to a specific location in space -- it could represent a physical tether made by \textit{e.g.} a polymer with sticky ends~\cite{rogers2011direct, cui2022comprehensive} or the minimum of an external potential occurring due to gravity or optical tweezers. Then we show how the obtained equations change if the spring is non-uniform, if the geometry is not a line, or if the mobility depends on space. 

An important aspect of these derivations is that they demonstrate the validity range of the effective constrained dynamics of Eq.~\eqref{eq:ODP0} and Eq.~\eqref{eq:seff}. These approximate dynamics are valid on timescales that are long enough such that particles or degrees of freedom have had the time to equilibrate in the confining potential, in the directions orthogonal to the manifold $\mathcal{M}$. Hence, we expect that no matter what the shape or magnitude of the confining potential is, we will obtain equations that accurately reproduce dynamics in the directions along the manifold $\mathcal{M}$ over long enough time scales. 

\paragraph*{Mathematical framework for the derivation.}

In the following, we will make heavy use of two partial differential equations (PDEs) associated with the dynamics Eq.~\eqref{eq:OD}. 
 One PDE is the Fokker-Planck equation, which can be written as 
\begin{equation}\label{eq:FP}
    \partial_t p = k_BT\nabla \cdot \left( e^{-\frac{\Utot}{k_BT}} M\nabla(e^{\frac{\Utot}{k_BT}}p)\right)\equiv \mathcal L^* p.
\end{equation}
%\sophie{we have to decide uniformly for $\pp{p}{t}$ or for $\partial_t p$.}
Here $p(\x,t)$ is the probability density to find the process at position $\x$ at time $t$, given initial probability density $p(\x,0) = p_0(\x)$. 
%\sophie{I know we need appropriate boundary condition but is this really useful for the setup?} This equation must be solved with an initial condition $p(\x,0) = p_0(\x)$, and with appropriate boundary conditions. For an unbounded domain, one typically requires the flux $\j = -k_BT e^{-\frac{U}{k_BT}} M\nabla(e^{\frac{U}{k_BT}}p)$ to vanish sufficiently quickly (e.g. exponentially quickly) as $|\x|\to \infty$. \mhc{insert refs for math stmts, eg pavliotis}
Another PDE is the Kolmogorov backward equation, which is the adjoint of Eq.~\eqref{eq:FP}, and takes the form 
\begin{equation}\label{eq:back}
\partial_t f = k_BTe^{\frac{\Utot}{k_BT}}\nabla\cdot (e^{-\frac{\Utot}{k_BT}}M\nabla f) \equiv \mathcal L f.
\end{equation}
This equation describes the evolution of a statistic $f(\x,t) = \E[h(\x(t)) | \x(0) = \x]$, given initial condition $f(\x,0) = h(\x)$. %It must be solved with boundary conditions, which on an unbounded domain typically require $|f|$ to be bounded as $|x|\to \infty$. 
We will often refer to the generator,  $\mathcal L$, which is the operator on the right-hand side of Eq.~\eqref{eq:back}.
Its adjoint, $\mathcal L^*$, is the operator on the right-hand side of Eq.~\eqref{eq:FP}.

\begin{figure}[h!]
    \centering
    \includegraphics[width=0.99\linewidth]{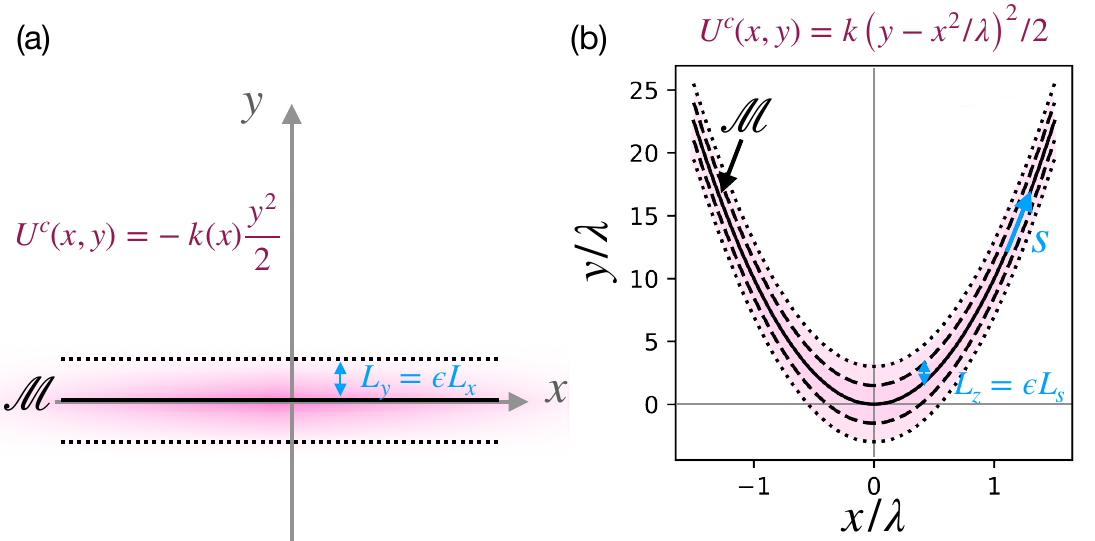}
    \caption{(a) Non-uniform constraint to a flat line; (b) Constraint to a curved line. The full line represents the manifold where $y = x^2/\lambda$; the dashed lines satisfy $y = x^2/\lambda \pm 1.5 \lambda$ and the dotted lines $y = x^2/\lambda \pm 3 \lambda$. }
    \label{fig:setup}
\end{figure}

%\subsection{Canonical example of a 2D random walker constrained to a flat line}

%\mhc{remove this section?}
%There is thus an effective potential $\mathcal{U} = \frac{k}{2} y^2 $ where $k$ is a spring constant. 

% \begin{equation}
%     \partial_t a = D \partial_{xx} a.
% \end{equation}
% Going back to real units, this corresponds with the following constrained Langevin equation
% \begin{equation}
% \frac{dx}{dt} = \sqrt{2D} \eta_x(t)
% \end{equation}
% which is simply, and naturally, the equation describing a single particle diffusing in 1D. 

% While this example appears trivial, it takes its importance as soon as the constrained domain or constraining function becomes complex. 

\subsection{2D walker with a non-uniform constraint to a flat line}
\label{sec:2dflatline}

We start by an example of a 2D Brownian walker diffusing freely in $(x,y)$ space, constrained -- softly -- to a flat line,  $y = 0$ -- see Fig.~\ref{fig:setup}a. 
The constraint is prescribed via a spatially dependent potential, $U(x,y) = k(x) y^2/2 $ where $k(x)$ is a spring constant which varies with position. This could correspond to a physical scenario where a particle binds to a non-uniform sticky polymer distribution on a surface or diffuses within a specific optical trap. We assume $k(x)$ is of typical scale $k$ and is a smooth enough function. For instance, one could imagine $k(x) = k (1 + \exp(-x^2/2\lambda^2))/2$ where $\lambda$ is a spatial scale, such that the constraint is stiffer towards the center of the domain. 
%\mhc{technically that model is not of ``typical scale $k$'' when $x$ is far away from the origin}

The overdamped Langevin equations satisfied by this random walker are 
\begin{align*}
    \frac{dx}{dt} &= - \frac{D}{k_B T} k'(x) \frac{y^2}{2} + \sqrt{2D} \eta_x(t) \\
    \frac{dy}{dt}  &= - \frac{D}{k_B T} k(x) y  + \sqrt{2D} \eta_y(t) 
\end{align*}
where $\eta_{i}(t)$ are independent Gaussian random noises, $D$ is the diffusion coefficient, and $k_BT$ the unit of energy. The generator for this system is
\begin{equation}\label{eq:backward}
\mathcal{L} = - \frac{D}{k_BT} k(x) y \partial_y - \frac{D}{k_BT} k'(x) \frac{y^2}{2} \partial_x + D \partial_{yy} + D \partial_{xx}.
\end{equation}

We are interested in the limiting equations when the spring constant becomes large. Formally, let us define a scale $L_y = \sqrt{k_B T/k}$ associated with motion in the $y$ direction, where $k$, we recall, is the typical magnitude of the spring constant. We are then interested in motion along the scale $L_x$ in the $x$ direction, over long enough time scales where the motion of $x$ is diffusive, \textit{i.e} over timescales $\tau = L_x^2/D$. We then take the nondimensionalization, $\tilde{k}(x) = k(x)/k$, $\tilde{x} = x/L_x$, $\tilde{y}  = y/L_y$,  and $\tilde{t} = t/\tau$. Assuming the spring is ``stiff enough'' corresponds to assuming that variations in $y$ are equilibrated, and much smaller than in $x$, or that $L_y \ll L_x$. We thus define the small number $\epsilon = L_y/L_x$. This is equivalent to looking at times that are ``long enough''. With these notations, and dropping the $\tilde{\cdot}$ for simplicity, the generator in Eq.~\eqref{eq:backward} becomes 
%\begin{equation}
%    \partial_t f = - \frac{1}{\epsilon^2} \kappa(x)  y \partial_y  - \kappa'(x) \frac{y^2}{2} \partial_x + \frac{1}{\epsilon^2}  \partial_{yy} +  \partial_{xx}. 
%\end{equation}
%If we choose $L_y = \epsilon L_x$, then we obtain
\begin{equation*}
    \partial_t f = \frac{1}{\epsilon^2} \mathcal{L}_0 f + \mathcal{L}_2f .  
\end{equation*}
where $\mathcal{L}_0  = k(x) y \partial_y + \partial_{yy}$ and $\mathcal{L}_2 = - k'(x) \frac{y^2}{2}\partial_x +  \partial_{xx}$.
We search for a solution perturbatively as $f = f_0 + \epsilon f_1 + \epsilon^2 f_2 + \ldots $, such that at lowest order $O(1/\epsilon^2)$,
\begin{equation*}
    \mathcal{L}_0 f_0 = 0.
\end{equation*}
The solution is $f_0 = a(x,t)$. The associated equilibrium distribution at this order is $\pi_0 =  \exp(- k(x) y^2 / 2)/Z$ where $Z$ is a normalization prefactor (on $y$). There are no terms at order $O(1/\epsilon)$. 

At order $O(1)$, to obtain a solution for $f_2$, we must satisfy the so-called Fredholm alternative~\cite{pavliotis2008multiscale}, as 
\begin{equation*}
    \left\langle \partial_t f_0 - \mathcal{L}_2 f_0, \pi_0 \right\rangle = 0
\end{equation*}
where $\left\langle f(x,y,t) , p(x,y,t) \right\rangle = \int f(x,y,t) p(x,y,t) \mathrm{d}y$ is the inner  product between functions $f,p$, in $y$-space. 
We arrive at the following equation characterizing the effective dynamics -- going back to real units -- 
\begin{equation*}
    \partial_t a = D \partial_{xx} a - D \frac{k'(x)}{2k(x)} \partial_x a.
\end{equation*}
The latter corresponds to the effective Langevin equation
\begin{equation}
\frac{dx}{dt} = - D \frac{k'(x)}{2k(x)} +\sqrt{2D} \eta_x(t).
\label{eq:constrainedFlatLine}
\end{equation}

There are a few physical interpretations of Eq.~\eqref{eq:constrainedFlatLine}. First, when the spring constant is uniform in $x$, $k(x) \equiv k$, then the first term on the right hand side vanishes, and we simply recover 1D Brownian diffusion on the $x$ coordinate, as we would expect. 

Second, when the spring constant is varying, there is an additional drift in the constrained equations, that corresponds to that from an effective potential as $U^{\rm eff}(x) = k_B T \log(k(x))/2$. This is consistent with Eq.~\eqref{eq:Ueffsoft} giving $U^{\rm eff}(x) = - k_B T \log \kappa$ where $\kappa$ is given by Eq.~\eqref{eq:kappa0}, which here is $\kappa(x) = \sqrt{k_B T/ k(x)}$. Suppose $k(x)$ decreases with increasing $|x|$, then $k'(x) <0$ and the force will be directed outwards around $x = 0$, such that the particle will be effectively repelled by the center. 

Finally, the stationary distribution, according to Eq.~\eqref{eq:pidelta} is  $\pi(x) = \exp(-U^{\rm eff}(x)/k_BT) /Z \simeq 1/\sqrt{k(x)}$. This expression also reflects that the particle spends more time far from the center. This is consistent with the fact that away from the center, the constraining potential is ``wider'', so there is more entropy or more microscopic configurations to explore. In essence, this is exactly the parallel of the Fick-Jacobs entropic trapping, where a particle spends a long time exploring wide spaces before it can escape through constrictions~\cite{zwanzig1992diffusion,jacobs_diffusion_1967,reguera_kinetic_2001, kalinay_corrections_2006,rubi2019entropic}.

\subsection{2D walker constrained to a curved line}\label{sec:2dcurve}

%\mhc{one role of this section is to introduce the extra drift velocity, which arises due to geometrical properties of the constraint. this is important!!! we haven't emphasized it anywhere, but Fig 7 shows how it arises entropically, because the pink region is thicker near the origina than further away. So, there has to be an effective force, which pushes stuff towards the origin, if we're going to diffuse with constant diffusivity in $s$. This needs to be stated clearly somewhere.}

We now constrain the 2D random walker motion to the curved line $\mathcal M = \{(x,y):y = x^2 /\lambda\}$ where $\lambda$ is a lengthscale,  such that $\lambda \rightarrow\infty$ corresponds to the flat line previously introduced -- see Fig.~\ref{fig:setup}b. We suppose that the potential energy is $ U(x,y) = k c(x,y)^2/2$ where $c(x,y) = y - x^2/\lambda$ is the constraint, and we assume $k=cst$ in this section.
%To make notations easier, in the following we non-dimensionalize all lengthscales by $\lambda$, meaning $y \rightarrow y/\lambda$, $x \rightarrow x/\lambda$ and quantities that include lengthscales such as $k$ are also modified as $k \rightarrow \lambda^2 k$ such that now  $U(x,y) = k c(x,y)^2/2$ with $c(x,y) = y - x^2$. \sophie{do this to simplify down the road?}

The equations satisfied by this random walker are 
\begin{align*}
    \frac{dx}{dt} &=  2 \frac{D}{k_BT} k \frac{x}{\lambda} (y - x^2/\lambda) + \sqrt{2D} \eta_x(t)\\
    \frac{dy}{dt} &%= - D \frac{\partial_y \mathcal{U}}{k_B T}  + \sqrt{2D} \eta_y(t) \\
     = - \frac{D}{k_BT} k (y - x^2/\lambda) + \sqrt{2D} \eta_y(t).
\end{align*}
%where $\eta_{i}(t)$ are Gaussian random noise, $D$ is the diffusion coefficient, and $k_BT$ the unit of energy. 
The generator for this system is 
\begin{equation*}
    \mathcal{L} = - \frac{D}{k_BT} k (y - x^2/\lambda) (\partial_y - 2 (x/\lambda) \partial_x) + D \partial_{yy} + D \partial_{xx}. 
\end{equation*}
Again, we are interested in the limiting equations when the spring constant becomes large. At this stage, one must notice that the previous nondimensionalization reveals the appearance of 3 lengthscales, $L_x$, $L_y$ and $\lambda$ and that there is no intuitive ratio of lengthscales that can be approximated with a small number, because of the constrained line's curvature. 
%We then take the same nondimensionalization, giving
%\begin{equation}
%    \partial_t f = - \frac{L_x^2}{L_y^2} ( y - \frac{L_x}{L_y} \frac{L_x}{\lambda} x^2) (\partial_y - 2 \frac{L_y}{L_x} \frac{L_x}{\lambda} x \partial_x) +  \frac{L_x^2}{L_y^2}   \partial_{yy} +  \partial_{xx}. 
%\end{equation}
%At this stage, it is clear we have a problem of scales because if we choose $L_y = \epsilon L_x$ we will have to assume some connection between $\lambda$ and $L_x$, knowing that we are trying to enforce the constraint, which is not naturally enforced in that way. This example basically highlights that we can not simply
To explore the case where the particle is ``not too far away'' from the constrained line we must change variables to the intrinsic coordinates that best describe the constrained space. 

%Given the constraint we are writing actually the constraint will only be close to $0$ if $L_y \sim L_x^2/\lambda$. To enforce this via a single scale, we must change variables. 

% \mhc{I think this might be too complicated to include -- see section IIIA which does essentially the same calculation but with mobility -- unless you can figure out a way to do this explicitly, without changing variables first. 

% This is one issue with IIIA -- it is implicit, but we want to eventually get an explicit equation. One option is to show they explicit one has the same form when you change variables. But it would be nicer to get this in a more direct way. 

% Eric has a derivation from lagrange multipliers, but it's not clear why one has to add the particular form he adds. Other forms for the lagrange multiplier seem to give nonsense.}

\paragraph{Derivation in the intrinsic coordinate system on the curved line}

A natural coordinate in the constrained space is $s$, the arc-length parameterization along the curve, 
$$s = \int_0^x \sqrt{1 + 4\tilde{x}^2/\lambda^2} \mathrm{d}\tilde{x}.$$ 
As an arc-length parameterization, this has the nice property that on the curve $\mathcal M$, $|(\partial_sx,\partial_sy)| = 1$, where $x(s)$ is defined via the inverse of the relationship above, and $y(s) = x^2(s)/\lambda$. 
The other coordinate we will take to be the value of the constraint itself, $z = (y-x^2/\lambda)$, since this variable is associated with fluctuations on a small lengthscale. 
Locally, we have a mapping $(x,y) \to (s,z)$ which is invertible. 
Again because we chose an arc-length parameterization,  on the curve we have $|(\partial_x s, \partial_y s)|=1$.
Using formulas derived in Appendix~\ref{app:changeparabolic}, we can obtain the generator in the intrinsic coordinate system
\begin{align*}
    \mathcal{L} &= -\frac{D k}{k_B T} z \left( (1+w^2)\frac{\partial}{\partial z}  - w \sqrt{1+w^2} \pp{}{s} \right) \\
    &\quad + D (1 + w^2) \pp{^2}{z^2} + D (1 + w^2)\pp{^2}{s^2} - 2 D w \sqrt{1 +w^2} \pp{^2}{s \partial z}  \\
    &\quad- \frac{2}{\lambda} D\pp{}{z} + \frac{2}{\lambda} D \frac{w}{\sqrt{1+w^2}} \pp{}{s} ,
\end{align*}
where we wrote for simplicity $w = 2x/\lambda$. Note that since $\partial_z{x} = 0$, then $w=w(s)$ is only a function of $s$ in the following. 
To make progress, we assume again a separation of lengthscales $L_z = \epsilon L_s$, where $L_z = \sqrt{k_B T/k}$ is associated with motion along the spring and $\tau = L_s^2/D$ is the timescale over which we expect motion to be diffusive.  We then obtain the expansion
\begin{align*}
    \mathcal{L} &= \frac{1}{\epsilon^2}\mathcal{L}_0 + \frac{1}{\epsilon} \mathcal{L}_1 + \mathcal{L}_2 \\
    &= \frac{1}{\epsilon^2} (1+w^2) \left[ - z  \frac{\partial}{\partial z} + \pp{^2}{z^2} \right] \\
    &+ \frac{1}{\epsilon} \left[  z w \sqrt{1+w^2} \pp{}{s}  - \frac{2L_s}{\lambda} \pp{}{z} - 2 w \sqrt{1 +w^2} \pp{^2}{s \partial z}  \right] \\
    & + (1 + w^2)\pp{^2}{s^2} + \frac{2L_s}{\lambda} \frac{w}{\sqrt{1+w^2}} \pp{}{s} .
\end{align*}
We assume no specific separation of lengthscales between $\lambda$ and $L_s$ as long as $\epsilon$ is small enough.  Solving for $\partial_t f = \mathcal{L}f$ perturbatively on $f = f_0 + \epsilon f_1 + \epsilon^2 f_2$, we obtain at lowest order in $O(1/\epsilon^2)$, $f_0 = a(s,t)$ and the associated equilibrium distribution $\pi_0 =  \exp(-z^2/2)/\sqrt{2\pi}$. 
%\mhc{missing a $k$ -- unless this went into the nondimensionalization?} \sophie{it did} 
At order $O(1/\epsilon)$ we need to solve $$\mathcal{L}_0 f_1 = - \mathcal{L}_1 f_0 = - z w \sqrt{1+w^2} \partial_s a(s,t), $$ whose solution is $f_1 = \frac{w}{\sqrt{1+w^2}} z \partial_s a(s,t)$. 
At order $O(1)$ one must satisfy Fredholm's alternative, which here is $\langle \partial_t f_0 - \mathcal{L}_1 f_1 - \mathcal{L}_2f_0 , \pi_0 \rangle = 0$. 
From this expression, one should be careful in calculating $\mathcal{L}_1f_1$ (see Appendix~\ref{sec:l1f1}), and one obtains
\begin{align*}
   \pp{a}{t}  = &(1+w^2) \pp{^2a}{s^2}  + \frac{2L_s}{\lambda} \frac{w}{\sqrt{1+w^2}} \pp{a}{s}  + \\
   & - w^2  \pp{^2a}{s^2}   - \frac{2L_s}{\lambda} \frac{w}{\sqrt{1+w^2}} \pp{a}{s} - \frac{2 L_s}{\lambda} \frac{w}{\left( 1+w^2\right)^{3/2}} \pp{a}{s}.
\end{align*}
Simplifying out terms, and going back to real dimensions, we obtain the effective dynamics in the intrinsic coordinate system
\begin{align}
   \pp{a}{t}   =& D \pp{^2a}{s^2}  - \frac{2 D}{\lambda} \frac{w}{\left( 1+w^2\right)^{3/2}} \pp{a}{s} . %- \frac{4L_s}{\lambda} \frac{w}{\sqrt{1+w^2}} \pp{a}{s}.
   \label{eq:1Dcurved}
\end{align}
This is the generator for the following stochastic differential equation:
\[
\dd{s}{t} =  - \frac{2 D}{\lambda} \frac{w(s)}{\left( 1+w^2(s)\right)^{3/2}} + \sqrt{2D}\eta(t).
\]
That is, 
we obtain a process which diffuses along the parabola  with diffusion coefficient $D$, and (recalling that the mobility is $D/k_BT$) which is subject to a drift force $- \frac{2k_BT}{\lambda} \frac{w(s)}{\left( 1+w^2(s)\right)^{3/2}}$.

\paragraph{Consistency with the intrinsic formulas.}
Let's check that this is consistent with the expression Eq.~\eqref{eq:seff} for the dynamics in intrinsic variables exposed in Sec.~\ref{sec:setup}.
First we calculate the diffusion coefficient. Given our initial friction tensor $\Gamma = \frac{k_BT}{D}I$, where $I$ is the $2\times 2$ identity matrix, the projected friction coefficient in Eq.~\eqref{eq:GammaS} is 
\[
\begin{pmatrix}\partial_s x & \partial_s y \end{pmatrix}\Gamma
\begin{pmatrix}\partial_s x \\ \partial_s y \end{pmatrix}
= \frac{k_BT}{D}|(\partial_s x, \partial_s y)|^2
= \frac{k_BT}{D},
\]
since $|(\partial_s x, \partial_s y)|=1$ because $s$ is an arc-length parameterization. Hence, the diffusion coefficient is still $D$. This is a useful fact: for isotropic diffusions, arc-length parameterizations (or metric-preserving parameterizations more generally) leave the diffusion coefficient unchanged. 

Next we calculate the effective drift. We have $|\nabla c| = \sqrt{1+4x^2/\lambda^2}$ so the effective potential in Eq.~\eqref{eq:effpot} is $U^{\rm eff}(s) %= k_BT\log \sqrt{1+\frac{4x^2}{\lambda^2}}
=\frac{1}{2}k_BT\log (1+w^2)$. Thus, the effective drift is 
\[
-\partial_s U^{\rm eff} = -k_BT\frac{w\partial_sw}{1+w^2} 
= -\frac{2k_BT}{\lambda}\frac{w}{(1+w^2)^{3/2}},
\]
where we used the fact that $\partial_s w = 2 \partial_s x/\lambda$ and $\partial_s x = (\partial_x s)^{-1} = 1/\sqrt{1+w^2}$. This is exactly equal to the drift term of Eq.~\eqref{eq:1Dcurved}.
The corresponding stationary distribution is $\pi(s) \propto e^{-U^{\rm eff(s)}} = (1+4x^2(s)/\lambda^2)^{-1/2}$.

\medskip

%\mhc{sophie -- I added this, please check if it makes sense.}
This extra drift term, which was discussed in Section \ref{sec:square} as being physically related to vibrational entropy when constraints are distance constraints, can be seen in this low-dimensional example to be related to a geometrical property of the constraint functions. Fig.~\ref{fig:setup} shows contour lines of the constraint function $c(x,y)$ offset by some amount $\pm\epsilon$ (for different $\epsilon$s), and shows that near the bottom of the parabola, the space between the offset contour lines is thicker, whereas further up the parabola, the space narrows. The distance between the offset contour lines at some position $s$ on the parabola is proportional to $1/|C|$ for small offsets, and hence so is the local volume near a small portion of the curve, $[s,s+\Delta s]$. This volume plays the role of an entropy, so the free energy density of being at position $s$ on the curve must include this entropic contribution, which is $-k_BT\log(1/|C|)$.

This example helps to explain why the particular form of the constraints matters for the effective potential. For example, if we were to consider the constraint function 
\[
\tilde c(x,y) = f(x,y)c(x,y)
\]
for some function $f(x,y)>0$, then the curve $\mathcal M$ where $\tilde c(x,y)=0$ is the same as the curve where  $c(x,y)=0$, however the effective potential on this curve becomes $k_BT\log |\nabla \tilde c| = k_BT\log f + k_BT\log |\nabla c|$. Hence, this effective potential can be changed arbitrarily by a suitable choice of function $f$ (which, one can see is the same thing as changing $\kappa(\x)$). Essentially, the effective potential inherits from the physical phenomena from which the constraints derive. 

%\mhc{if time: add $k_BT\log f$ to eq (23), see what extrinsic equations are, etc.}

%  The projected diffusion coefficient is simply the original diffusion coefficient, since the original diffusion tensor was diagonal and we chose an arc-length parameterization. %, where $|\nabla s| = 1$.  \sophie{not sure what the meaning of $\nabla s$ is though}. \mhc{$(\partial_x s, \partial_y s)$}
% According to Eq.~\eqref{eq:effpot}, the effective potential in the intrinsic coordinate system is $U^{\rm eff}(s) = k_BT \log|C|$. The constraint $c(x,y) = y - x^2/\lambda$ corresponds to a constraint matrix $C = (\nabla c)^T = (-2x/\lambda, 1)$. Since $C C^T = 1 + 4x^2/\lambda^2$, we have $|C| = \sqrt{1+4x^2/\lambda^2}$. Then, the effective drift term should be \sophie{is this first - sign ok?}
% \[ - D \frac{\partial_s|C|}{|C|} \partial_s a = 
% - D \frac{\partial_s(\sqrt{1+w^2})}{\sqrt{1+w^2}}\partial_s a
% \quad = - D\frac{w \partial_s w}{(1+w^2)^2} \partial_s a
% \]
% and using the fact that $\partial_s w = 2 \partial_s x/\lambda$ and $\partial_x s = 1/\sqrt{1+w^2}$, we recover the drift term of Eq.~\ref{eq:1Dcurved}. 

\paragraph{Equations of motion in the extrinsic coordinate system.}
From the backward equation for the effective dynamics on the  curve in the parameterized coordinates, Eq.~\eqref{eq:1Dcurved}, we can use the change of variable formulas from $s, z$ to $x,y$, detailed at the end of Appendix~\ref{app:changeparabolic}-1, to recover the backward equation for the effective dynamics in the original coordinate system (recall $w=2x/\lambda$): 
% \begin{align*}
%     \pp{a}{t} =& \frac{1}{1+w^2} \left( \pp{^2a}{x^2} + w^2 \pp{^2a}{y^2} + 2 w \pp{^2a}{x \partial y}\right) \\
%     & + \frac{2}{\lambda} \frac{1}{(1+w^2)^2}\pp{a}{y} -\frac{2}{\lambda} \frac{w}{(1+w^2)^2}\pp{a}{x} \\
%     & - \frac{2}{\lambda} \frac{w^2}{(1+w^2)^2}\pp{a}{y} -\frac{2}{\lambda} \frac{w}{(1+w^2)^2}\pp{a}{x} \\
% \end{align*}
% and simplifying
\begin{align}
    \pp{a}{t} =& \frac{D}{1+w^2} \left( \pp{^2a}{x^2} + w^2 \pp{^2a}{y^2} + 2 w \pp{^2a}{x \partial y}\right) \nonumber\\
    & + \frac{2 D}{\lambda} \frac{1}{(1+w^2)^2}\left( (1-w^2)\pp{a}{y} - 2w \pp{a}{x}\right).
    \label{eq:finalcurved}
\end{align}
This corresponds to the following system of stochastic differential equations:
\begin{align*}
    \dd{x}{t} &= \frac{D}{\lambda}\left( \frac{-8 x/\lambda }{(1+ 4x^2/\lambda^2)^2}\right) + \sqrt{\frac{2D}{1+ 4x^2/\lambda^2}} \eta_x(t)\\
    \dd{y}{t} &= \frac{2D}{\lambda}\left( \frac{1 - 4 x^2/\lambda^2}{(1+ 4x^2/\lambda^2)^2}\right) + \sqrt{\frac{2D(4x^2/\lambda^2)}{1+ 4x^2/\lambda^2}} \eta_x(t).
\end{align*}
Naturally, in the extrinsic coordinate system, there is diffusion in both $x$ and $y$. 
This system of equations can also be recovered from Eq.~\eqref{eq:ODP0} and the associated formulas for $M_P$.

The equation for $x$ is decoupled from the equation for $y$, and can be written as 
\begin{equation}
    \frac{dx}{dt} =  \partial_x \tilde D(x) + \sqrt{2\tilde D(x)}\eta(t)
    \label{eq:curvedline}
\end{equation}
where the effective diffusivity in $x$ is $\tilde D(x) = D/(1 + 4x^2/\lambda^2)$. 
This equation makes it clear that the stationary distribution in $x$ is 
\[
\pi(x) \propto 1,
\]
namely the stationary distribution is constant in $x$. This makes sense, since the volume between the level sets $y(x) = \pm \delta$ and above a small element $dx$ is $2\delta$, which is constant in $x$.  This result is consistent with the stationary distribution in $s$ being 
\[
\pi(s)\propto |C|^{-1} = (1+4x^2(s)/\lambda^2)^{-1/2},
\]
since the Jacobian of the transformation is $\partial s / \partial x = (1+4x^2/\lambda^2)^{1/2}$.

\subsection{2D walker with non-diagonal mobility constrained to a line}\label{sec:2dnondiag}

We now consider a 2D walker with non-diagonal mobility confined to a line; and demonstrate analytically the form of its effective mobility. 
%\sophie{change $A$ to $\sigma$}
%\mhc{mention this is the same example considered in section **} 
% We now consider a more general mobility tensor and think about how the mobility may be affected in the constrained dynamics. We consider a 2D random walker $(x(t),y(t))$ moving in a potential $U(x,y) = \half ky^2$, with a constant but now non-diagonal mobility tensor \mhc{don't need to say this because we said it before, just reference eqn/section/example, then put equations.}
% \[
% M = \begin{pmatrix} M_{11} & M_{12} \\  M_{12} & M_{22} \end{pmatrix}.
% \]
%  For simplicity we scale energy such that $k_BT=1$. 
 The equations of motion are
\begin{align}
\dd{x}{t} &= -M_{12}ky + \sqrt{2k_B T}\sigma_{11} \eta_x(t)   \nonumber \\
\dd{y}{t} &= -M_{22}ky + \sqrt{2k_B T}\sigma_{21} \eta_x(t) + \sqrt{2k_B T} \sigma_{22}\eta_y(t), \label{eq:2dnondiag}
\end{align}
where $k$ is the spring constant characterizing confinement to the line, and $\sigma$ is the Cholesky decomposition of $M$, \textit{i.e.} a lower triangular matrix such that $\sigma \sigma^T = M$. (We could use any square root of $M$ for $\sigma$, however the Cholesky decomposition simplifies certain calculations later.)

When $k$ is large, the random walker is strongly confined near the $x$-axis and we wish to compute its effective mobility $M^{\rm eff}$ along the $x$-axis, or equivalently its effective diffusion coefficient $D^{\rm eff} = k_BT M^{\rm eff}$. In this section we will derive the expression for $D^{\rm eff}$, and then verify numerically that it is the correct expression for the effective dynamics.

\subsubsection{Derivation of constrained dynamics via the mean-squared displacements}

We may derive the limiting dynamics by singular perturbation theory as we have done before; details are included in Appendix~\ref{sec:OUextra} for pedagogical purposes. However, there is more physics to be learned using another derivation, via the mean-squared displacements. %To start with, we state the result using the formulas given in Table~\ref{tab:formulas} and then do the derivation. 

% \mhc{$k_BT$ seems to be missing in derivation below. just say we set it to 1?}
 
To this aim, let's compute the effective diffusivity $D^{\rm eff}=k_BT M^{\rm eff}$ directly from the stochastic dynamics Eq.~\eqref{eq:2dnondiag}, using the fact that the effective diffusivity is 
\[
D^{\rm eff} = \lim_{t\to \infty} \half\dd{}{t}\langle x^2(t) \rangle.
\]
Notice that we can write Eq.~\eqref{eq:2dnondiag} as 
\begin{align}
    \dd{x}{t} &= v_y(t) + v_1(t)  \nonumber\\
    \dd{y}{t} &= v_2(t) + \frac{A_{21}}{A_{11}}v_1(t), \label{eq:xy2}
\end{align}
where 
\begin{align*}
    v_y(t) &= -M_{12}ky(t)\\
    v_1(t) &= \sqrt{2k_B T}\sigma_{11}\eta_x(t)\\
    v_2(t) &= -M_{22}ky +  \sqrt{2k_B T} \sigma_{22}\eta_y(t).
\end{align*}

%\sophie{@Miranda, read, I had a go at simplifying}\mhc{good}
The standard expression for the effective diffusivity of a process moving in a random velocity field $v_y + v_1$ is 
\begin{equation}\label{eq:x2b}
D^{\rm eff}
= \int_0^\infty \Big(C_{y,y}(\tau) + C_{1,1}(\tau) + C_{y,1}(\tau) +  C_{1,y}(\tau)\Big) d\tau,
\end{equation}
where $C_{i,j}(\tau) = \langle v_i(t) v_j(t+ \tau) \rangle$ is the velocity autocorrelation function, which is independent of $t$ for large enough $t$.

We first express the correlation functions. Simply, $C_{11}(\tau) = 2 M_{11}\delta{\tau}$. The correlation functions involving $y$ can be calculated using the analytical solution for $y$ (assuming $x(0){=}y(0){=}0$):
\[
y(t) = \sqrt{2k_B T}\int_0 ^t e^{-M_{22}k(t-t')}\left( \sigma_{21}\eta_x(t') + \sigma_{22}\eta_y(t')  \right)\mathrm{d}t'.
\]
This gives $C_{y,1}(\tau) = 0$ since $y(t)$ can not be correlated with the noise $\eta_x(t+\tau)$ sampled in the future; 
\[
C_{1,y}(\tau) = -2k_B TM_{12}^2ke^{-M_{22}k\tau}, \]
where we used $\sigma_{11}\sigma_{21} = M_{12}$, and for large $t$, 
\[
C_{y,y}(\tau) \sim M_{12}^2k_B T ke^{-M_{22}k\tau}. 
\]

The effective diffusivity Eq.~\eqref{eq:x2b} consists of 4 terms: the first is 
\[
\int_0^\infty C_{1,1}(\tau)d\tau = k_B T M_{11},
\]
which is the effective diffusivity that would be induced by the forcing from the white noise $v_1$ on its own, if there were no correlation with $y$. 
 The second is
\[
\int_0^\infty C_{y,y}(\tau)d\tau = k_B T\frac{M_{12}^2}{M_{22}},
\]
which represents the effective diffusivity that would be induced from the additional forcing from $v_y$ on its own.
The forces $v_1,v_y$ are anti-correlated, so the effective diffusivity is altered by the terms coming from their cross-correlation: 
\[
\int_0^\infty (C_{y,1}(\tau) + C_{1,y}(\tau))d\tau = -2k_B T\frac{M_{12}^2}{M_{22}}.
\]
Combining these terms, gives a total effective diffusivity of 
\[
\frac{D^{\rm eff}}{k_B T} = M_{11} - \frac{M_{12}^2}{M_{22}}.
\label{eq:projMobOff}
\]
Thus, the non-diagonal term in the mobility tensor causes the stochastic process $y(t)$ to alter the velocity of $x(t)$, and this has two effects: one, it increases the mobility of $x$ by an amount $M_{12}^2/M_{22}$, by adding an additional noisy forcing, and two, it decreases the mobility by twice that amount, because this force is anticorrelated with the white noise forcing. Overall, the anticorrelation wins, and causes the effective mobility to decrease. Notably, when the off-diagonal coefficient $M_{12}$ is large enough, this diffusivity can be significantly \emph{reduced} compared with the original diffusivity $M_{11}$ along the $x$-axis.

These subtle correlations are more easily understood within the example provided in Sec.~\ref{sec:consequences:offdiag}. Let $x(t)$ and $y(t)$ now correspond to the positions of 2 different particles where $y(t)$ is trapped in a constraining potential. The particles' motion feeds back onto one another due to hydrodynamic interactions, which ``act at a distance'': when moving particle $x$, this displaces all the fluid surrounding $x$ and hence induces motion of particle $y$. When a random jiggle of $x$ occurs, this translates into a jiggle in $y$, which self-consistently feedsback onto $x$: this corresponds to the increase in mobility $M_{12}^2/M_{22}$. However, the random jiggle of $y$ also increases the spring recall force in $y$, which translates at a distance in a recall force on $x$ giving this $-2M_{12}^2/M_{22}$ decrease of the mobility. Effectively, this yields a decrease of the mobility. While the origin of the factor $2$ is not obvious, clearly the total change in mobility has to be negative: the effective mobility of $x$ should be hindered by the presence of the other particle $y$, which acts like a surface which enhances hydrodynamic friction. 

%\sophie{factor 2 still a mistery to me.}
%Why if $v_y$ anti-correlated with $v_1$? Looking at Eq.~\eqref{eq:xy2}, we see that when $v_1$, the noise from the degree of freedom 1, is large, this causes $y$ to increase, which causes $v_y$ to become more negative. Hence, $v_1$ and $v_y$ are anticorrelated. 
%\mhc{this is just explaining the structure of the equations. It doesn't really give insight -- is there a reason why we should expect a factor of 2, between the forcing, and the anticorrelation? Is there a thermodynamic reason why?} \sophie{isn't it simply that it's 1 with y and y with 1? like also when $v_y$ increases, this causes $v_1$ to decrease? Not really but it's not obvious... }

%\subsubsection{Numerical validation}
\paragraph*{Numerical validation.}

\begin{figure}[h!]
\centering
\includegraphics[width=0.8\linewidth]{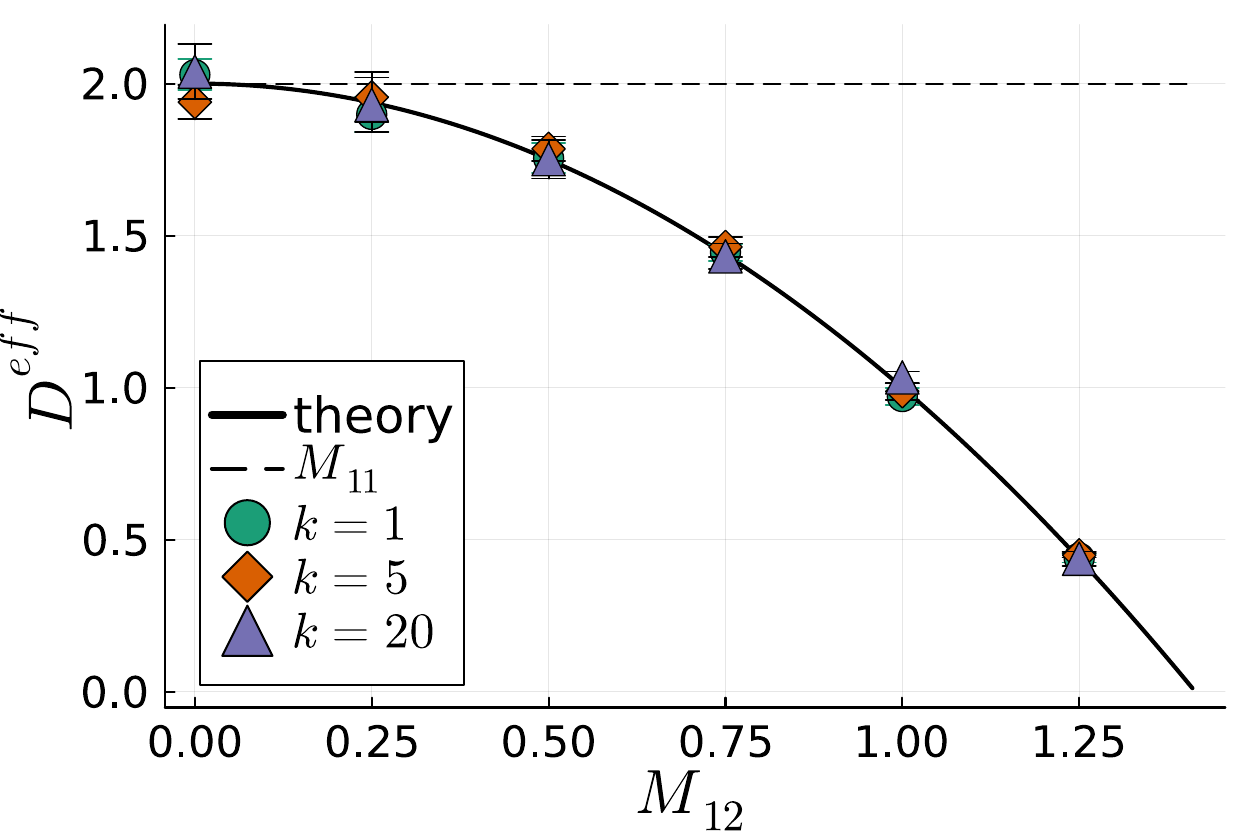}
\includegraphics[width=0.8\linewidth]{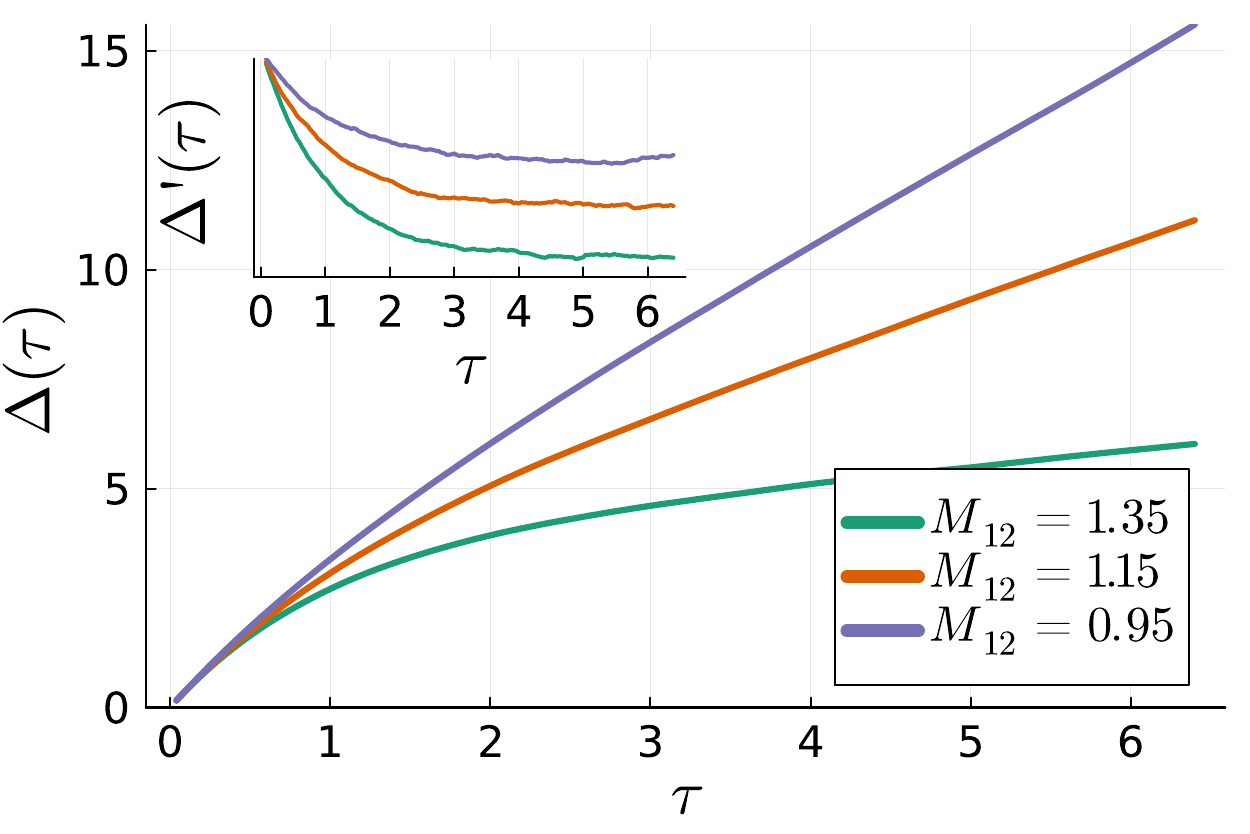}
\caption{Effective mobility decreases with off-diagonal contribution $M_{12}$. Top: Numerically estimated diffusivity (markers) for different spring constants $k$, as a function of $M_{12}$, and the theoretical prediction $D^{\rm eff}/ k_B T= M_{11}-\frac{(M_{12})^2}{M_{22}}$ (black line). Since this model is a toy model, we choose variable values arbitrary, here: $k_B T = 1$, $M_{11}=2,M_{22}=1$, a numerical timestep of $\Delta t = 1/k/50$, and saved at intervals of $\delta t=2/k$. Simulations were run for a total time of $t=10^5/k$ and were repeated 10 times at each set of parameters. Error bars are 2 standard deviations.
Bottom: mean-squared displacements $\Delta(\tau)$ with spring constant $k=1$ and several values of $M_{12}$ as shown in the legend. Inset shows the slopes $\Delta'(\tau)$, estimated by finite difference. }\label{fig:example1}
\end{figure}

We verified this reduction numerically, by simulating the full equations of motion Eq.~\eqref{eq:OD} (explicitly given in Eq.~\eqref{eq:2dnondiag}), and estimating the effective diffusivity from the mean slope of the mean squared displacement (Appendix~\ref{app:numerics}).
The result agrees with our prediction, as expected (Fig. \ref{fig:example1}(top)), and further shows that the effective diffusivity is independent of $k$, as our later derivation will show for this flat constraint, provided we average the increments over a long enough time window. 
Figure~\ref{fig:example1} (bottom) shows that effective diffusive behaviour only arises over a sufficiently long timescale, with the timescale depending on both $M_{12}$ (shown) and $1/k$ (not shown, but which comes from nondimensionalizing the equations). 
This illustrates that the effective diffusivity is only meaningful at sufficiently long times.

% the mean-squared displacements $\Delta(\tau)$ for simulations using $k=1$ and different values of $M_{12}$. The mean-squared displacements are notably not diffusive at short times, but at longer times they increase linearly with $\tau$. The time to approach diffusive behaviour increases with $M_{12}$; by scaling time we know this time will be proportional to $1/k$.  

% Figure \ref{fig:example1}(top) compares the theoretical projected diffusivity to the estimated diffusivity, for varying $M_{12}$, with different spring constants $k$. The numerical results agree perfectly with the theoretical prediction. Notably, the agreement is independent of $k$, as predicted by our theory, provided we average the increments over a long enough time window. 

\subsubsection{Project-then-average: Numerical validation for ``soft soft'' constraints with an off-diagonal, spatially-varying mobility.}

\begin{figure}[h!]
\centering
\includegraphics[width=0.95\linewidth]{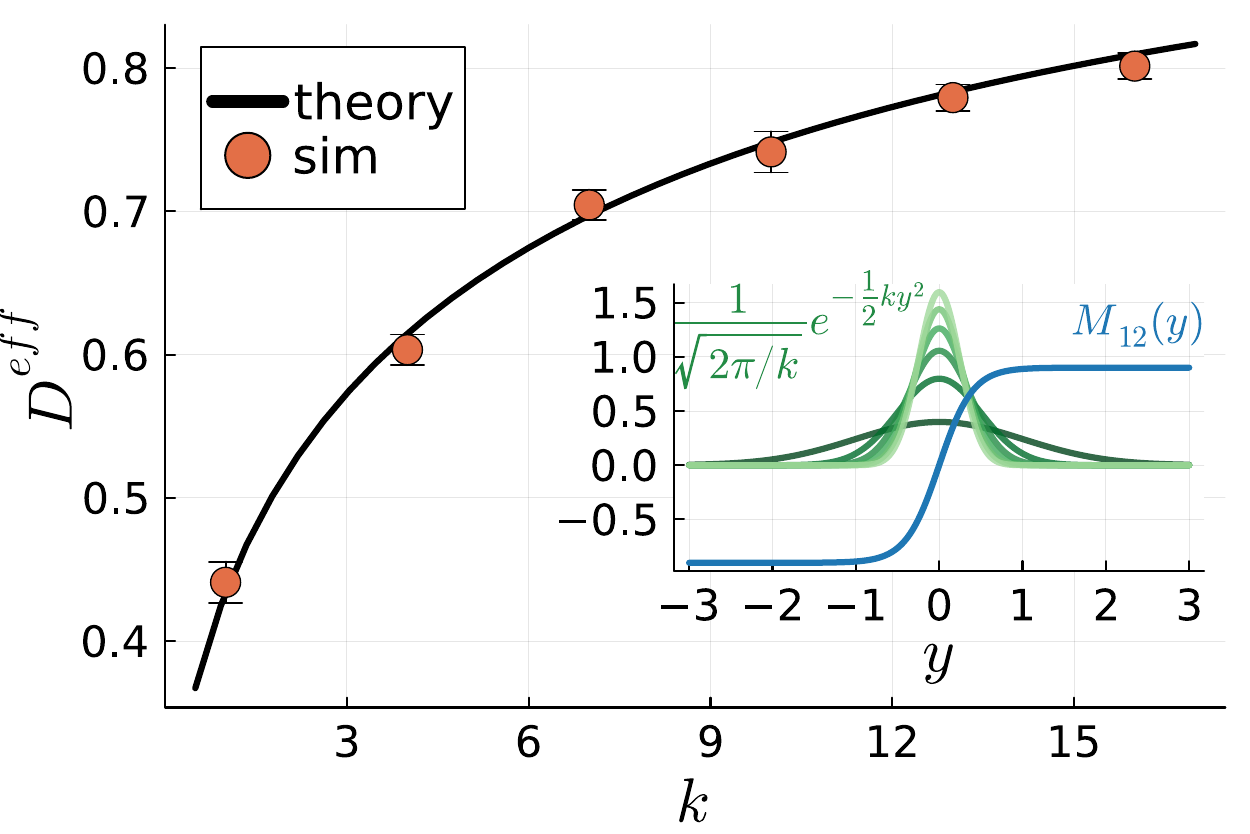}
\caption{Soft-soft constraints: the theoretically predicted effective diffusivity $D^{\rm eff}$ (solid black line), compared with 
numerically estimated diffusivity (markers) for different spring constants $k=1,4,7,10,13,16$. As $k$ increases, $D^{\rm eff}\to 1$, the value it would take in the merely softly-constrained case.  
Simulations used a numerical timestep of $\Delta t = 1/k/250$, ran over a time of $t=10^4$, saved data at intervals of $\delta=2/k$, and were repeated 10 times at each set of parameters. Error bars are 2 standard deviations.
Inset shows $M_{12}(y)$ and the probability density functions for the different values of $k$.}\label{fig:example2}
\end{figure}

As a final, albeit crucial numerical validation, we explore the case of ``soft soft'' constraints where we allow the off-diagonal term to vary with $y$. That is, we consider a mobility matrix
\[
M(x,y) = \begin{pmatrix} M_{11} & M_{12}(y) \\ M_{12}(y) & M_{22} \end{pmatrix}.
\]
We use a harmonic potential
$U(x,y) = \half ky^2$. 
% The equations of motion with $k_BT=1$ are thus :
% \begin{align}
% dX_t &= -M_{12}kY_t dt + \partial_y M_{12}dt + \sqrt{2}A_{11}dW^1_t + \sqrt{2}A_{12}dW^2_t, \nonumber\\
% dY_t &= -M_{22}kY_t dt + \sqrt{2}A_{21}dW^1_t + \sqrt{2}A_{22}dW^2_t.  \label{eq:dynamicsvary}
% \end{align}
Following our calculations in the previous section, and our result Eq.~\eqref{eq:MPvary} for varying mobility, we expect the effective diffusivity along the $x$-axis to be given by
\begin{equation}\label{eq:Mvary_example}
\frac{D^{\rm eff}}{k_BT} = M_{11} - \frac{1}{M_{22}}\int M_{12}^2(y)\frac{1}{\sqrt{2\pi/k}}e^{-\half ky^2}dy.
\end{equation}
The effective mobility is thus first \emph{projected}, and subsequently \emph{averaged}; conclusively answering Aleksandar Donev's interrogation. Overall, the effective mobility \emph{depends on the particular choice of constraining potential}, highlighting the importance of choosing contraining potentials that reflect physical reality.  

This point is illustrated in Fig. \ref{fig:example2}, which shows that increasing $k$ in this example leads to increasing $D^{\rm eff}$, even though the potentials all have the same mean and mode. As $k\to\infty$, $D^{\rm eff}$ approaches the value it would take in the softly-constrained case, obtained by evaluating $M(y)$ at $y=0$.  This figure also compares the predictions against numerical simulations of the dynamics, 
\begin{align}
dX_t &= -M_{12}kY_t dt + \partial_y M_{12}dt + \sqrt{2}A_{11}dW^1_t + \sqrt{2}A_{12}dW^2_t, \nonumber\\
dY_t &= -M_{22}kY_t dt + \sqrt{2}A_{21}dW^1_t + \sqrt{2}A_{22}dW^2_t,  \nonumber %\label{eq:dynamicsvary}
\end{align}
writing  $A(y) = M^{1/2}(y)$, using the following choices for parameters: $M_{11}=M_{22}=1$, and $M_{12}$ a sigmoid function with  $A=0.9, \alpha=5$:
$
M_{12}(y) = 2A\left( \frac{1}{1+e^{-\alpha y}} - \frac{1}{2}\right).
$
The comparison is favourable, verifying that our asymptotic calculations are correct, and one should  first project and then average.

%%%%%%%%%%%%%%%%%%%%%%%%%%%%%%%%%%%%%%%%%%%%%%%
%%%%%   Section: Derivation               %%%%%
%%%%%%%%%%%%%%%%%%%%%%%%%%%%%%%%%%%%%%%%%%%%%%%
\section{Derivation of the constrained overdamped equations from soft constraints} \label{sec:derivation}

In this section we derive the constrained overdamped equations, from the stiff limit of a soft potential.
The derivation is first performed in instrinsic coordinates, and we subsequently show the result is equivalent to the formulation in extrinsic coordinates. %We also interpret the obtained equations in terms of \sophie{?}.

%We start with the overdamped equation Eq.~\eqref{eq:OD} with a potential of the form Eq.~\eqref{eq:Uc}. We then make a change of variables in these equations to describe the manifold and the orthogonal directions. At this point it is convenient to first consider two-dimensional case, and then the general case. 

\begin{figure}[h!]
\begin{tikzpicture}[x=0.75pt,y=0.75pt,yscale=-1,xscale=1]
%uncomment if require: \path (0,300); %set diagram left start at 0, and has height of 300

%Shape: Axis 2D [id:dp9545680714385423] 
\draw [line width=1.1]  (99.95,161.18) -- (153.72,161.18)(99.95,111.87) -- (99.95,161.18) -- cycle (146.72,156.18) -- (153.72,161.18) -- (146.72,166.18) (94.95,118.87) -- (99.95,111.87) -- (104.95,118.87)  ;
%Shape: Arc [id:dp48037741052816263] 
\draw  [draw opacity=0][line width=1.5]  (98.99,70.65) .. controls (130.22,59.69) and (170.75,73.7) .. (197.58,107.32) .. controls (211.1,124.26) and (218.99,143.48) .. (221.24,161.97) -- (139.84,153.4) -- cycle ; \draw  [line width=1.5]  (98.99,70.65) .. controls (130.22,59.69) and (170.75,73.7) .. (197.58,107.32) .. controls (211.1,124.26) and (218.99,143.48) .. (221.24,161.97) ;  
%Shape: Axis 2D [id:dp8632517955970849] 
\draw [line width=1.1]  (184.09,92.66) -- (222.77,130.02)(218.34,57.19) -- (184.09,92.66) -- cycle (221.21,121.56) -- (222.77,130.02) -- (214.26,128.75) (209.88,58.75) -- (218.34,57.19) -- (217.08,65.7)  ;
\draw   ;

% Text Node
\draw (156,158) node [anchor=north west][inner sep=0.75pt]    {$x$};
% Text Node
\draw (96.67,97.4) node [anchor=north west][inner sep=0.75pt]    {$y$};
% Text Node
\draw (228,130) node [anchor=north west][inner sep=0.75pt]  [rotate=-44]  {$s$};
% Text Node
\draw (222.48,45.91) node [anchor=north west][inner sep=0.75pt]  [rotate=-44]  {$z$};
% Text Node
\draw (201.33,165.07) node [anchor=north west][inner sep=0.75pt]    {$c(x,y) =0$};

\end{tikzpicture}
\caption{A sketch of the change of variables used in the derivation of the softly-constrained dynamics in Section \ref{sec:generalderivation}.
%\mhc{add: level sets of s/z}
}\label{fig:varchange}
\end{figure}
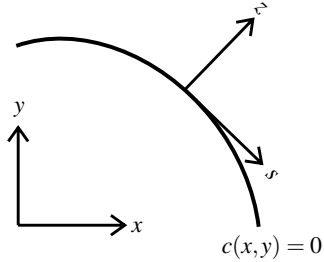

\subsection{Derivation of softly constrained dynamics in instrinsic coordinates}\label{sec:generalderivation}

To consider the limit of the overdamped dynamics Eq.~\eqref{eq:OD} as the potential $U^c$ in Eq.~\eqref{eq:Uc} becomes increasingly confining, we first 
change variables in the  Kolmogorov backward Eq.~\eqref{eq:back}, to a collection of $m$ variables lying along the manifold, and $n-m$ variables orthogonal to the manifold. A similar change of variables has been considered many times in the context of constrained dynamics\cite{froese2001realizing, morse2003Theory,HolmesCerfon:2013jw}; we describe it here for completeness. 

% Recall the Kolmogorov backward equation is
% \begin{equation}
% \partial_t f = k_BTe^{\frac{\Utot}{k_BT}}\nabla\cdot (e^{-\frac{\Utot}{k_BT}}M\nabla f) \equiv \mathcal L f.
% \end{equation}
We start with the tangential variables: let $\s=(s_1,s_2,\ldots,s_{d})$ be a collection of variables that form a local parameterization of $\mathcal M$.  That is, every point $\x\in \mathcal M$ maps to one set of coordinates $\s$.\footnote{It could be the case that they do not form a global parameterization of $\mathcal M$. While it is possible to handle this case using an atlas of such local parameterizations, combined with a smoothing function to move between parameterizations, this is highly technical and not a situation we will consider here. The equations we write are valid locally, wherever the parameterization is valid.} 
In the example in Section \ref{sec:2dcurve}, the single variable $s(x)$ was an arc-length parameterization of the parabolic curve, however in general we consider an arbitrary parameterization. 
We may assume this parameterization extends smoothly to a neighbourhood of $\mathcal M$, by the tubular neighbourhood theorem. 

On $\mathcal M$, each $\x\in \mathcal M$ maps to one point $\s$, and this also holds true in a neighbourhood of $\mathcal M$, so we may think of $\s$ as a function of $\x$. From this function $\s(\x)$ and its Jacobian $J_{\s}~=~\nabla \s$ (the $i$th row of $J_{\s}$ is $(\nabla s_i)^T$ and the gradient is taken with respect to extrinsic coordinates) we may determine several geometric quantities that will be useful in the following: 
\begin{itemize}[nosep]
\item The tangent vectors to $\mathcal M$ are the set of vectors $\{\nabla s_i\}_{i=1}^d$;
\item The metric tensor on $\mathcal M$ is $g_{\s}=(J_{\s}J_{\s}^T)^{-1}$ with inverse $g^{-1}_{\s} = J_{\s}J_{\s}^T=\nabla \s (\nabla \s)^T$;
\item The natural surface measure is $\dsigmaM = |g_{\s}|^{1/2}d\s$. 
\end{itemize}
Appendix~\ref{sec:diffgeom} has more information about these geometric quantities. 

We must also extend the parameterization to a neighbourhood of $\mathcal M$, which we do by defining $z_i=c_i(\x)$ for $i=1,\ldots,m$. For the example in Section \ref{sec:2dcurve}, we had $z_1 = y-x^2/\lambda$. 
The $z_i$ parameterize the normal directions to $\mathcal M$. See Figure \ref{fig:varchange} for a sketch of the variable change.

 Write $\y=(s_1,\ldots,s_{d},z_1,\ldots,z_m)$ for the whole collection of variables related to the manifold $\mathcal{M}$. The transformation $\x\to \y$ exists and is differentiable in a neighbourhood of $\mathcal M$. We write $J = \nabla \y = \frac{\partial \y}{\partial \x}$ for the Jacobian of the full transformation, and note that the transformation induces a metric tensor $g=(J^{-1})^TJ^{-1}$ in the new parameterization. An important property of $g$ is that it is block diagonal for $\x\in\mathcal M$: 
\begin{equation}\label{eq:gdiag}
g|_{\mathcal M} = \begin{pmatrix}
    (\nabla \s(\nabla \s)^T)^{-1} & 0\\
    0 & (\nabla \c(\nabla \c)^T)^{-1}
\end{pmatrix}.
\end{equation}
 This is because  the vectors $\{\nabla c_i\}_{i=1}^m$ are orthogonal to any tangent vector to $\mathcal M$. %\footnote{To compute this: compute $g^{-1} = JJ^T = \begin{pmatrix} \nabla \s(\nabla \s)^T & 0 \\ 0 & \nabla \c(\nabla \c)^T \end{pmatrix}$ on $\mathcal M$, because of the orthogonality property, then invert.}

 We change variables in the Kolmogorov backward equation \eqref{eq:back} to obtain a right-hand side characterized by the application of the generator $\mathcal L$ given by:
\begin{multline}\label{eq:back_s_multivar}
    \mathcal L f = 
    k_BT\frac{e^{\frac{\Utot}{k_BT}}}{|g|^{\frac{1}{2}}}
    \Bigg( 
    \sum_{i=1}^{d}\partial_{s_i}\Big(|g|^{\frac{1}{2}}e^{-\frac{\Utot}{k_BT}}(M^{\y}D_{\y}f)^{s_i}\Big)\\
    +
    \sum_{i=1}^m\partial_{z_i}\Big(|g|^{\frac{1}{2}}e^{-\frac{\Utot}{k_BT}}(M^{\y}D_{\y}f)^{z_i}\Big)
    \Bigg).
\end{multline}
We have defined the mobility matrix acting on the new local coordinates as 
\begin{equation}\label{eq:JMJT}
M^{\y} = JMJ^{T},
\end{equation}
and the vector of partial derivatives as 
$D_\y f= (\partial_{s_1}f,\ldots,\partial_{s_d}f, \partial_{z_1}f,\ldots,\partial_{z_m}f)^T$. We write $\v^{\alpha}$ for the $\alpha$th component of vector $\v$.
Note that the potential in the new coordinates could depend on both $\s$ and $\z$, as $\Utot(\s,\z)$.

As in Section \ref{sec:2dcurve}, we assume a separation of lengthscales $L_z=\epsilon L_s$ between a typical scale $L_z$ in the confined $\z$-variables and a scale $L_s$ characterizing a lengthscale of interest in the $\s$-variables. In the case where the confining potential is given by the harmonic form in Eq.~\eqref{eq:Uc}, then $L_z =\sqrt{k_BT/k}$ , where $k=\max{k_i}$ is the largest spring constant in the confining potential.
We assume the manifold's curvature is less than $L_s^{-1}$ so it varies on lengthscales larger than $L_s$.  
Other quantities (denoted as $\alpha$), such as the components of the mobility tensor or metric tensor, have variations characterized by a lengthscale $L_\alpha = \min_i|\alpha/\partial_{z_i} \alpha|$ (or $L_\alpha = \min_i\sqrt{|\alpha/\partial_{z_iz_i} \alpha|}$ when $\partial_{z_i} \alpha|_\mathcal M\approx 0$); we say such quantities are \emph{slowly-varying} if $L_\alpha \gg L_z$, and \emph{rapidly-varying} if $L_\alpha \sim L_z$. 

Then we change variables again, to $Z_i=z_i/\epsilon$, to zoom in to the quickly changing normal directions near the manifold. We  assume the metric tensor varies slowly near the manifold so we may replace it with its evaluation at $Z_i=0$: $g_0 \equiv g|_{\mathcal M}$ defined in Eq.~\eqref{eq:gdiag}. 
Further, we assume the external potential also varies slowly near the manifold, so it is evaluated at $Z_i=0$; we continue to write $\Utot$ for brevity, but we mean $\Uc(\s,\Z)+\Ue(\s;0)$, where $\Z=(Z_1,\ldots,Z_m)$; and we abbreviate $\utot = \Utot/k_B T$. 
We do \emph{not} make any assumptions about the mobility tensor at this point, as we wish to allow for situations where it varies rapidly in $\z$, \textit{i.e.} it varies on the same scale as the potential. 
The rescaled equation has 
\begin{equation}
    \partial_t f = (\frac{1}{\epsilon^2}\mathcal{L}_0 + \frac{1}{\epsilon} \mathcal{L}_1 + \mathcal{L}_2)f
    \label{eq:backscale} 
\end{equation} where now the rescaled operators are
\begin{align*}
    \mathcal L_0 f&= k_BT\frac{e^{\utot}} {g_0}\sum_{i=1}^m\partial_{Z_i}\left(e^{-\utot} g_0(M^{\y}_{\z\z}D_{\Z} f)^{z_i}\right)\\
    \mathcal L_1f &= k_BT\frac{e^{\utot}}{g_0}\Bigg( \sum_{i=1}^d\partial_{s_i}\left(e^{-\utot} g_0(M^{\y}_{\s\z}D_\Z f)^{s_i}\right) \\
    &\qquad \qquad\quad
    + \sum_{i=1}^m\partial_{Z_i}\left(e^{-\utot} g_0(M^{\y}_{\s\z}D_\s f)^{z_i}\right)
    \Bigg)\\
    \mathcal L_2 f &= k_BT\frac{e^{\utot}}{g_0}\sum_{i=1}^d\partial_{s_i}\left(e^{-\utot} g_0(M^{\y}_{\s\s}D_\s f)^{s_i}\right) .
\end{align*}
Here $D_{\s}f = (\partial_{s_1}f,\ldots,\partial_{s_d} f)^T$, $D_\Z f= (\partial_{Z_1}f,\ldots,\partial_{Z_m}f)^T$, and we have decomposed the mobility tensor into blocks as 
\[
M^{\y} = \begin{pmatrix}
    M^{\y}_{\s\s} & M^{\y}_{\s\z}\\
    M^{\y}_{\s\z} & M^{\y}_{\z\z}
\end{pmatrix}.
\]

We make the antatz $f = f_0 + \epsilon f_1 + \epsilon^2 f_2 + \ldots$ and substitute into Eq.~\eqref{eq:backscale}. The $O(\epsilon^{-1})$ equation is $\mathcal L_0f_0=0$ which has (bounded) solution  $f_0(\s,\Z) = a(\s,t)$ for some function $a$.

The $O(\epsilon^{-1})$ equation is $\mathcal L_0f_1 = -\mathcal L_1f_0$, which becomes
\[
\sum_{i=1}^m\partial_{Z_i}\left(e^{-\utot} g_0(M^{\y}_{\z\z}D_{\Z} f_1)^{z_i}\right) 
= 
-\sum_{i=1}^m\partial_{Z_i}\left(e^{-\utot} g_0(M^{\y}_{\s\z}D_\s a)^{z_i}\right).
\]
 If we can find a solution such that
\[
M^{\y}_{\z\z}D_{\Z}f_1 = -M^{\y}_{\s\z}D_\s a,
\]
then this will be a desired solution. 

For \emph{slowly-varying mobility}, where the mobility tensor varies slowly with $\Z$, then we may replace the mobility with its value at $\Z=0$, and a solution is
\[
f_1(\s,\Z,t) = -(M^{\y}_{\z\z})^{-1}M^{\y}_{\s\z}\Z D_{\s} a.
\]

For \emph{rapidly-varying mobility}, there is a solution for $f_1$ provided a function of the mobility is curl-free: $\partial_j ((M^{\y}_{\z\z})^{-1}M^{\y}_{\s\z})_{i\alpha} - \partial_i ((M^{\y}_{\z\z})^{-1}M^{\y}_{\s\z})_{j\alpha} = 0$ for  every $i,j,\alpha$. In this case the solution is
\[
f_1(\s,\Z,t) =-\left(\int^\Z(M^{\y}_{\z\z})^{-1}M^{\y}_{\s\z}\,d\phi\right)D_\s a,
\]
where the integral is along any path $\phi$ ending at $\Z$ (and starting at the same point). 
The curl-free condition always holds when $s$ is one-dimensional; we were not able to show it holds in general so in general it must be checked (though our final result will be independent of this condition).

Continuing, at $O(1)$ we must solve 
\begin{equation}\label{eq:O1}
\mathcal L_0 f_2 = \partial_t f_0 - \mathcal L_1 f_1 - \mathcal L_2 f_0. 
\end{equation}
By the Fredholm Alternative, this equation only has a solution when a solvability condition is satisfied: we require $\int \left(\partial_t f_0 - \mathcal L_1 f_1 - \mathcal L_2 f_0 \right)\cdot \pi_0(\Z)d\Z=0$, where $\pi_0$ is the stationary solution for the fast dynamics, \textit{i.e.} the solution to $\mathcal L^*_0\pi_0=0$ where $\mathcal L^*_0$ is the adjoint of $\mathcal L_0$ in $\Z$ (with $\s$ thought of as a parameter).
Hence, the solvability condition means we must integrate the right hand side term against $\pi_0$, the function in the nullspace of $\mathcal L_0^*$. We find that 
\begin{equation}\label{eq:kappa}
\pi_0(Z;\s) = \kappa^{-1}(\s)e^{\frac{-\Uc(\s,\Z)}{k_BT}}, \quad \text{where} \;\; \kappa(\s) = \int e^{\frac{-\Uc(\s,\Z)}{k_BT}}d\Z.
\end{equation}
Thus, since $\int \partial_ta\, \pi_0(Z)dZ = \partial_ta$, the dynamics are given by 
\[
\partial_t a = \int \pi_0(\Z)\left(\mathcal L_2 f_0 + \mathcal L_1 f_1 \right)d\Z.
\]
The first term is 
\[
    \int \pi_0(\Z)\mathcal L_2 f_0 d\Z
    = \frac{k_BTe^{\frac{\Ue}{k_BT}}}{g_0\kappa}\sum_{i=1}^d\partial_{s_i}\left( g_0\kappa e^{\frac{-\Ue}{k_BT}} (\overline{M^{\y}_{\s\s}}D_\s)^{s_i} a\right),
\]
where the external potential is evaluated at $Z=0$, i.e. $\Ue=\Ue(\s,0)=\Ue(\s)$,  and we have defined the equilibrium averaging operator 
\begin{equation}\label{eq:eqavg}
    \overline{h} \equiv \frac{\int e^{-\frac{U^c(s,Z)}{k_BT}}h(\Z)d\Z }{ \int e^{-\frac{U^c(s,\Z)}{k_BT}}dZ}
    = \kappa^{-1}\int e^{-\frac{U^c(s,\Z)}{k_BT}}h(\Z)d\Z.
\end{equation}

In the second term, the integral over terms of the form $\partial_{Z_i}(\cdot)$ is zero, by the divergence theorem applied to the $\Z$-variables. We have that $D_\Z f_1 = -M^{\y}_{\s\z}(M^{\y}_{\z\z})^{-1}M^{\y}_{\s\z}D_{\s} a$. Hence, we end up with 
\begin{multline*}
\int \pi_0(\Z)\mathcal L_1f_1d\Z 
= \\
-\frac{k_BTe^{\frac{\Ue}{k_BT}}}{g_0\kappa}\sum_{i=1}^d\partial_{s_i}\left( g_0\kappa e^{-\frac{\Ue}{k_BT}}
\Big(\overline{M^{\y}_{\s\z}(M^{\y}_{\z\z})^{-1}M^{\y}_{\s\z}}D_{\s}a\Big)^{s_i}
\right).
\end{multline*}
Putting this together gives equation
\[
\partial_t a = \frac{k_BTe^{\frac{\Ue}{k_BT}}}{g_0\kappa}\sum_{i=1}^d\partial_{s_i}\left( g_0\kappa e^{-\frac{\Ue}{k_BT}}
(\overline{\tilde M} D_\s a
)^{s_i}\right) 
\]
with modified mobility tensor
\begin{equation}\label{eq:Mtilde_general}
\tilde M = M^{\y}_{\s\s} - M^{\y}_{\s\z}(M^{\y}_{\z\z})^{-1}M^{\y}_{\s\z}.
\end{equation}
Recognizing that
\[
g_0 
= |\nabla\s|^{-1}|\nabla\c|^{-1}, \qquad 
    g_\s^{-1} = (\nabla \s)(\nabla \s)^T,
    %\tilde M(s) &= M^{\y}_{ss}-\frac{(M^{\y}_{sz})^2}{M^{\y}_{zz}},\label{eq:Mtilde}
\]
where $g_{\s}$ is the metric tensor on $\mathcal M$,
and defining
\begin{equation}
    \pi_{\s}(\s) = \kappa |\nabla \c|^{-1}e^{-\frac{\Ue(\s)}{k_BT}},
\end{equation}
% \begin{align}
% g_0 %&=|g|^{1/2}|_{z=0}
% &= |\nabla\s|^{-1}|\nabla\c|^{-1}\nonumber\\
%     \pi_{\s} &= \kappa|\nabla_{\x}\c|^{-1},\nonumber\\
%     g_s^{-1} &= (\nabla \s)(\nabla \s)^T,
% \end{align}
%where $g_{\s}$ is the metric tensor on $\mathcal M$, and $\pi_{\s}$ is the stationary distribution on $\mathcal M$, 
the limiting equation can be written as
\begin{equation}\label{eq:limgeneral}
\partial_t a = k_BT \pi_{\s}^{-1}\divM (\pi_{\s} \overline{\tilde M} D_{\s} a). 
\end{equation}
This is the backward equation corresponding to dynamics Eq.~\eqref{eq:seff}, and is the general form for the backward equation of a reversible, overdamped diffusion on $\mathcal M$, with stationary distribution proportional to $\pi_{\s}$, and mobility tensor $\overline{\tilde M}$.  Note that when the mobility varies slowly, \textit{i.e.} we have soft constraints but not soft-soft constraints, then $\overline{\tilde M}=\tilde M$. %\mhc{added sentence here}

%%%%%%%%%%%%%%%%%%%%%%%%%%%%%%%%%%%%%%%%%%%%%%%
%%%%%   Section: Interpretation  (Linked comments)         %%%%%
%%%%%%%%%%%%%%%%%%%%%%%%%%%%%%%%%%%%%%%%%%%%%%%
%\section{Interpretation of the constrained equations}

% \todo{  discussion of the projection, mathematical properties of it (relates to friction)
%     \begin{itemize}
%            \item transforming the general derivation result back to extrinsic coordinates. (or earlier). relates to projection -- could introduce at the same time. 
%             \item Showing that an equivalent form of the constraint equations may also be obtained by projecting the friction coefficient \sophie{pretty sure this works need to check}
%        \end{itemize}
% }

% \todo{discuss:
% \begin{itemize}
%     \item General physical intuition on why these formulas are ok?
%       \item Validating that the result has the ``right properties'' (appendix for details)
%     \end{itemize}
% }

\subsection{Link to the extrinsic form of the constrained equations}\label{sec:extrinsic}

%\sophie{sophie read until here.}

%\mhc{updated this section}

%\mhc{add more discussion earlier on, about what we are really doing -- just changing variables and showing that equation for z is 0, ie z isn't changing}

%\mhc{not sure how $\kappa(s)$ links to delta-functions -- should check link between $\delta(q_i)$ and form of potential}\\

We now show that the 
 constrained overdamped dynamics Eq.~\eqref{eq:limgeneral} derived in the previous section, which were written in intrinsic coordinates,  are equivalent to the extrinsic form Eq.~\eqref{eq:ODP0} of the equations. We first consider the softly-constrained equations, involving $M_P$, and then consider the softly-softly constrained equations, involving $\overline{M_P}$. 

\paragraph{Softly-constrained equations.} 
We start with the extrinsic equations Eq.~\eqref{eq:ODP0}, and change variables to show they are equivalent to Eq.~\eqref{eq:limgeneral}. Eq.~\eqref{eq:ODP0} has a corresponding backward equation which has general form Eq.~\eqref{eq:back} but with $\Utot\to U^{\rm eff}$ and $M\to M_P$. We then make a change of variables $\x\to\y = (s_1, .. s_d, z_1, ...z_m)$, using the same change of variables as in Section \ref{sec:generalderivation}, obtaining an equation similar to \eqref{eq:back_s_multivar}. The generator becomes
%Details are the the Appendix, Section \ref{sec:proofs}. 
\begin{multline}\label{eq:genS}
    \mathcal L f = 
    k_BT\frac{e^{\frac{U^{\rm eff}}{k_BT}}}{|g|^{\frac{1}{2}}}
    \Bigg( 
    \sum_{i=1}^{d}\partial_{s_i}\Big(|g|^{\frac{1}{2}}e^{-\frac{U^{\rm eff}}{k_BT}}(M^{\y}_PD_{\y}f)^{s_i}\Big)\\
    +
    \sum_{i=1}^m\partial_{z_i}\Big(|g|^{\frac{1}{2}}e^{-\frac{U^{\rm eff}}{k_BT}}(M^{\y}_PD_{\y}f)^{z_i}\Big)
    \Bigg).
\end{multline}
Here
$
M^{\y}_P = JM_PJ^{T},
$ is the variable-transformed mobility matrix, 
and $J = \frac{\partial \y}{\partial \x}$ is the Jacobian of the transformation, and $g = (J^{-1})^TJ^{-1}$.

To recover Eq.~\eqref{eq:limgeneral} from the expression of the generator, we first show that the terms in the second line of Eq.~\eqref{eq:genS} vanish. For this we will use one of the properties of the projected mobility $M_P$, namely that
%In the extrinsic equations Eq.~\eqref{eq:ODP0}, we assumed a specific form of the mobility, but for the variable change that follows, we only need to assume:
\begin{itemize}
    \item[(A)] $M_P\nabla c_i(\x) = 0$ (the 0-vector) for $i=1,\ldots,m$, for  $\x$ in a neighbourhood of $ \mathcal M$. 
\end{itemize}
This is true for the mobilities Eq.~\eqref{eq:MP} or Eq.~\eqref{eq:MPvary} proposed, but the assumption holds for many other mobility tensors, including any mobility tensor of the form $M_P = PMP^T$, where $P(\x)$ is a projection matrix onto level sets of the span of the collection of vectors $\{M\nabla c_i\}_{i=1}^m$, for $\x$ in a neighbourhood of $\mathcal M$. 

In the new coordinate system, the condition $M_P\nabla c_i = 0$ becomes $M^{\y}_PD_{\y}z_i = 0$, which implies, since $D_{\y}z_i = (0,\ldots,0,1,0,\ldots)^T$ (a 1 in the component corresponding to $z_i$), that the new mobility matrix has the block form
\begin{equation}\label{eq:Mzblock}
M^{\y}_P = \begin{pmatrix}
    M^{\s}_P & 0 \\ 0 & 0
\end{pmatrix}
\end{equation}
in a neighbourhood of $\mathcal M$. 
Here $M^{\s}$ represents the nonzero components, which are in the rows and columns corresponding to the $\s$ variables. For $\x$ in a neighbourhood of $\mathcal M$, we thus have that only the terms in the first line of Eq.~\eqref{eq:genS} contribute to $\mathcal Lf(\x)$; the terms in the second line vanish, by the assumed form of $M^{\y}_P$. 

Further, the terms in the first line of Eq.~\eqref{eq:genS} can be written in divergence form, using the block diagonal form of the metric tensor Eq.~\eqref{eq:gdiag}. This shows that on $\mathcal M$, 
$|g| = |g_{\s}||C|^{-2}$, where $g_{\s}=((\nabla \s)(\nabla \s)^T)^{-1}$ is the metric tensor on $\mathcal M$. 
% Thus, for $\x\in \mathcal M$, the generator can be written as 
% \begin{equation}\label{eq:Lu_scoords}
% \mathcal Lf = k_BT|\nabla \c|e^{\frac{U}{k_BT}}\divM\left( |\nabla \c|^{-1}e^{-\frac{U}{k_BT}}M^{\s}_PD_{\s}f  \right).
% \end{equation}
% It can also be written using intrinsic gradients as 
% \[
% \mathcal Lf = k_BT\frac{e^{\frac{U}{k_BT}}}{|\nabla \c|^{-1}}\divM\left( |\nabla \c|^{-1}e^{-\frac{U}{k_BT}}M^{\s}_Pg_{\s}\gradM f  \right).
% \]
% From this, we can compute the adjoint of $\mathcal L$ on $\mathcal M$: 
% \begin{equation}\label{eq:Lstar_s}
% \mathcal L^* p = 
% k_BT\divM\left( 
% |\nabla \c|^{-1}e^{-\frac{U}{k_BT}}
% M^{\s}_Pg_{\s}\gradM(|\nabla \c|e^{\frac{U}{k_BT}}p)
% \right).
% \end{equation}

We find that on $\mathcal M$, under assumption (A), the generator  has the form 
\begin{equation}\label{eq:Lu_s}
\mathcal Lf = k_BT|C|e^{\frac{U^{\rm eff}}{k_BT}}\divM\left( |C|^{-1}e^{-\frac{U^{\rm eff}}{k_BT}}M^{\s}_PD_{\s}f  \right),
\end{equation}
where $M^{\s}_P \equiv (JM_PJ^T)_{\s\s}$. So far, what we have done is to show that under assumption (A), the variables $\z$ in the orthogonal directions to the manifold don't change under the dynamics. 

Comparing Eqns.~\eqref{eq:Lu_s} and~\eqref{eq:limgeneral}, we see these are the same equations provided $M^{\s}_P = \tilde M$ on $\mathcal M$, where $\tilde M$ is defined in Eq.~\eqref{eq:Mtilde_general}.
% \begin{enumerate}[(i),nosep]
% \item $V = \Ue - k_BT\log \kappa + cst$, where $cst$ denotes an arbitrary constant;
% \item $M^{\s}_P = \tilde M$. 
% \end{enumerate}\medskip
%Item (i) is satisfied by choosing $V$ as in Eq.~\eqref{eq:V}. Often $\kappa(s)$ is constant, so we may simply choose $V=\Ue$, the external potential. 
We may show that $M^{\s}_P = \tilde M$ by direct calculation. 
We have that
\[
M^{\s}_P = JP_MMJ^T= (JMJ^T - JMC^T(CMC^T)^{-1}(JMC^T)^T)_{\s\s}
\]
where $M$ is the mobility tensor for the original, unconstrained dynamics, Eq.~\eqref{eq:OD}.
Also recall that the block form of the mobility tensor in $(\s,\z)$ coordinates is 
\[
JMJ^T = 
\begin{pmatrix} (\nabla \s)^T\\ (\nabla \c)^T \end{pmatrix} 
M
\begin{pmatrix} \nabla \s& \nabla \c \end{pmatrix}
=
\begin{pmatrix}
    M^{\y}_{\s\s} & M^{\y}_{\s\z} \\ M^{\y}_{\s\z} & M^{\y}_{\z\z}
\end{pmatrix}.
\]
Extracting the required components of this identity:
\begin{itemize}
    \item $(\nabla \c)^T M\nabla\c = M^{\y}_{\z\z}$
    \item $JM\nabla\c = \begin{pmatrix} M^{\y}_{\s\z} \\ M^{\y}_{\z\z}\end{pmatrix}$
\end{itemize}
Hence 
    \begin{multline*}
        JMC^T(CMC^T)^{-1}(JMC^T)^T)=
        \begin{pmatrix}
    M^{\y}_{\s\z}(M^{\y}_{\z\z})^{-1}M^{\y}_{\s\z} & M^{\y}_{\s\z} \\ M^{\y}_{\s\z} & M^{\y}_{\z\z}\end{pmatrix}.
    \end{multline*}
Putting this together we obtain 
\[
JP_MMJ^T = \begin{pmatrix}
        M^{\y}_{\s\s}-M^{\y}_{\s\z}(M^{\y}_{\z\z})^{-1}M^{\y}_{\s\z} & 0 \\ 0 & 0 
    \end{pmatrix}
\]
and hence $JP_MMJ^T_{\s\s} = \tilde M$ on $\mathcal M$, as claimed. 

%\mhc{Q: need it to be true on $\mathcal M$? Or near it as well????}

\paragraph{Softly-softly constrained dynamics.}
%\mhc{this is new}

Now consider the softly-softly constrained dynamics. After the change of variables in the previous section we obtain \eqref{eq:genS} with $M_P^{\y} =J\overline{M_P}J^T$. This tensor satisfies property (A) (since $\overline{M_P}$ is invariant in $\z$, by construction), and so it has block form 
\eqref{eq:Mzblock} with $M_P^{\s} = \nabla \s \overline{M_P}(\nabla \s)^T$.

Now, we have that, for  $\x\in\mathcal M$,
 \[
\nabla \s \overline{M_P}(\nabla \s)^T \Big|_{\x}= \overline{\nabla \s|_{\x} M_P (\nabla \s|_{\x})^T},
\]
since for the purpose of the averaging operator, $\nabla \s|_{\x\in\mathcal M}$ is constant. 
We showed in the previous section that $\nabla \s M_P (\nabla \s)^T\big|_{\x} = \tilde M\big|_{\x}$  for $\x\in\mathcal M$, hence, $M_P^{\s} = \overline{\tilde M}$, which is what we need to show to show equivalence of the two equations, so we are done.

\medskip

We now briefly explain how to go in the other direction: how to construct the extrinsic mobility \eqref{eq:MPvary}, using only knowledge of the mobility $\tilde M$  or $\overline{\tilde M}$ on $\mathcal M$. 

We start with some mobility, say $\tilde M$, on $\mathcal M$. This  is a tensor defined only on $\mathcal M$, and acts only on the $\s$-variables. We extend this to a tensor acting on the $\z$-variables as well, with the desired block form \eqref{eq:Mzblock}, by defining an extended tensor acting on the $\y$-variables
 \[
 \tilde M^{\rm ext} = \begin{pmatrix}\tilde M & 0 \\0 & 0\end{pmatrix}.
 \]
 This may be trivially extended to be defined on a neighbourhood of $\mathcal M$ (not just on $\mathcal M$) by assuming it is constant in $\z$. Then, we transform it to Cartesian variables using $J^{-1} = \partial\x/\partial\y =\begin{pmatrix}
     \nabla_{\s}\x & \nabla_{\z}\x
 \end{pmatrix} $, to obtain
 \[
 M_P^{\rm ext} \equiv J^{-1}\tilde M^{\rm ext} (J^{-1})^T.
 \]
 This gives a projected mobility in the original variables. When 
$\tilde M = \nabla \s M_P(\nabla \s)^T$, then straightforward calculations show that $M_P^{\rm ext} = M_P$.

\subsection{Interpreting the projected mobility / projected friction tensor}\label{sec:projection}

We end this section with a brief discussion of the geometrical properties of the projected mobility Eq.~\eqref{eq:MP}, and its link to the corresponding projected friction Eq.~\eqref{eq:GammaP}. We restrict this section to the softly-constrained case, to focus on the projection, and not the additional averaging required for the softly-softly constrained dynamics.

\subsubsection{Interpretation of $M_P$ as a weighted $L_2$ projection}

%\mhc{make it clear that projection depends on $\x$ -- its not just to the tangent space, rather to the level sets of $c$}

Although the projection operator $P_M$ used to obtain the projected mobility $M_P$ in Eq.~\eqref{eq:MP} is not the most familiar projection in the Euclidean space, it can still be interpreted as a ``special'' projection. 
We claim that $P_M$ is an orthogonal projection in a weighted inner product space. To show this, let's first consider a general weighted inner product space, by letting $N$ be any symmetric, positive definite $n\times n$ matrix, and defining an inner product 
\[
\langle v,w\rangle_N = v^TNw.
\]
The norm induced by this inner product is $|v|_N = v^TNv$. 

Now let $P_N^A$ be a projection matrix which projects onto the column space of matrix $A$, denoted $\col(A)$, and which is orthogonal with respect to the inner product $\langle \; , \;\rangle_N$. That is, $P_N^A$ is defined by 
$
P_N^Av  = \argmin_{y\in \col(A)} |v-y|_N.
$
for each $v\in \R^n$. 
Standard calculations show that 
\[
P_N^A = A(A^TNA)^{-1}A^TN.
\]

Now, comparing with expression Eq.~\eqref{eq:MP} for $P_M$ (and choosing $A=MC^T, N=M^{-1}$), and similarly considering its transpose $P_M^T$ (and choosing $A=C^T,N=M$), we see that
\begin{align*}
    I-P_M &= MC^T(CMC^T)^{-1}C= P_{M^{-1}}^{MC^T},\\
    I-P_M^T &= C^T(CMC^T)^{-1}CM = P_M^{C^T}.
\end{align*}

Thus, $P_M = I-P_{M^{-1}}^{MC^T}$ is a projection onto the subspace defined by $M(\x)$ multiplied by the tangent space to $\mathcal M$ at $\x$, and which is orthogonal in an inner product weighted by $M^{-1}(\x)$. 
Similarly, $P_M^T = I-P_M^{C^T}$ is a projection onto the tangent space to $\mathcal M$ at $\x$, and which is orthogonal with respect to the inner product induced by $M(\x)$. 
%\sophie{As a physicist I want to know why this is cool, or maybe why this makes sense. Its nice to see that it indeed corresponds to a projection (so that's one thing) but why does it have to be that projection? }

 %The physical reason why $P_M$ is not simply $P$ is that the mobility is not the ``natural'' physical object one should project. Rather, the natural physical object to project is the friction matrix $\Gamma$, or even more to the point, the forces on the particles/degrees of freedom, which are linearly related to $\Gamma$.

%\mhc{I removed the sentence above, because it's covered in the next section}

\subsubsection{Interpretation of $\Gamma_P$, and link to $M_P$}\label{sec:GammaP}

The projection $P(\x)$ used to construct the projected friction $\Gamma_P$ in Eq.~\eqref{eq:GammaP} is clearly a standard orthogonal projection; \textit{i.e.} it projects to the tangent space to $\mathcal M$ at each $\x$ in a way that finds the vector with minimum norm. Using our notation from the previous section, 
\[
I-P = C^T(CC^T)^{-1}C = P^{C^T}_I.
\]

%\mhc{I modified the sentences below -- please check}
There is a physical reason why this projection is natural for the friction matrix $\Gamma$. This is because  the natural equations to project come from Newton's second law, which means we must project the forces acting on a system. Since forces are linearly related to velocities via the friction matrix, one should project the friction matrix. 

In overdamped dynamics, we instead express the particle velocities as being linearly related to the forces via the mobility matrix. But this is not the original physical law; the original law is on forces and thus one should project forces.

%we are used, rather, to express particle velocities as linearly related to forces via the mobility matrix.But, of course, this is not the original physical law; the original law is on forces and thus one should project forces. %\sophie{There's something very deep here in that when you write $F = \gamma v$, there's a right way and a ``wrong'' way to measure $\gamma$ in a MD simulation. Basically I think the correct way is you impose a force say $F$ on a particle, then measure the MEAN velocity $\langle v \rangle$ and compute $\gamma = F/\langle v \rangle$. However, ``stokes view'' of dragging a particle at constant speed, and measuring the mean force, does not provide the friction coefficient, namely $\langle F \rangle /v \neq \gamma$ (I don't know of an example where that is indeed the case but I remember having a long discussion with David Dean where he was convinced this should be wrong is some cases. I think you can imagine stupid cases where this is wrong like long anisotropic particles or something). I think this is related to our stuff but I can't see how.} \sophie{Put it in comment and write to DSD}

\medskip 

We show in the Appendix~\ref{app:equivalenceMobFriction}, that $M_P = \Gamma_P^\dagger$ is the Moore-Penrose pseudo-inverse of $\Gamma_P$. There is thus a nice correspondence where the projected mobility $M_P$ can be interpreted as the inverse of a friction matrix, the projected friction. Because this inverse has to be taken \textit{after} the projection, the effective projection operator $P_M$ for the mobility is therefore not the simple $P$. 

\medskip

The projected friction tensor has a simple expression in local (intrinsic) variables:
\[
(\nabla_\s\x)^T\Gamma (\nabla_\s\x) ,
\]
which gives Eq.~\eqref{eq:GammaS}, and which we show here starting from the expression for the projected mobility $\tilde M = (\nabla \s)M_P(\nabla \s)^T$. 
 Suppose we are given a local parameterization $\s$ of the manifold $\mathcal M$. Then, by changing variables, and by our result above that $M_P = \Gamma_P^\dagger$, we know that the mobility in intrinsic variables is 
\[
%\tilde M = 
\nabla_\x \s M_P(\nabla_\x \s)^T
= \nabla_\x \s \Gamma_P^\dagger(\nabla_\x \s)^T.
\]
We can also show that 
\[
\nabla_\x \s \Gamma_P^\dagger(\nabla_\x \s)^T = ((\nabla_\s\x)^T\Gamma_P \nabla_\s\x)^\dagger,
\]
which follows from the identity $\nabla_\x \s\nabla_\s\x = I$ and the properties of the Moore-Penrose pseudoinverse.
Now, we may develop an alternative expression for $\Gamma_P$. Noting that this is an orthogonal projection onto the tangent space to $\mathcal M$, and noting that the tangent space is spanned by the columns of $\nabla_\s\x$, we may alternatively construct the projection operator $P$ as 
\[
P = (\nabla_\s\x)((\nabla_\s\x)^T(\nabla_\s\x))^{-1}(\nabla_\s\x)^T.
\]
Substituting this expression to construct $\Gamma_P=P\Gamma P$  shows that 
\[
(\nabla_\s\x)^T\Gamma_P \nabla_\s\x = (\nabla_\s\x)^T\Gamma (\nabla_\s\x),
\]
which gives expression \eqref{eq:GammaS} for $\Gamma_P$. 
This is a simple expression for the projected friction, which shows the friction in any direction, is given simply by projecting $\Gamma$ onto that direction.

\section{Deriving the constrained dynamics using Lagrange multipliers}
\label{sec:lagrange}

A common method for imposing constraints in mechanical physical systems, is to use Lagrange multipliers. 
Previously, Lagrange multipliers have been considered in the context of the inertial Langevin equations \cite{hinch1994Brownian,morse2003Theory,holmes2016stochastic,lelievre2012langevin}, which are simpler than the overdamped Langevin equations because the noise acts on accelerations and not on velocities. From the constrained inertial overdamped Langevin equations one can derive the correct form of the overdamped Langevin equations. They have also been considered in the general context of equations with Poisson brackets and irreversible brackets \cite{ottinger2015preservation}, which could presumably be eventually applied to the overdamped Langevin equation. We have not seen them used directly in the context of the overdamped Langevin equations before, except in mathematical theories that are not connected to physical systems \cite{ciccotti2008projection}.

In this section we show an alternative derivation of the constrained dynamics, using Lagrange multipliers. 
The goal is to show how the projection $P_M$ arises naturally with \emph{deterministic} constraints, when imposed using Lagrange multipliers, but that applying \emph{stochastic} constraints is subtle and nontrivial -- the Lagrange multipliers must be applied in a specific way, which preserves their reversible structure; some seemingly reasonable ansatzes fail. 
This section thus serves as a cautionary tale, that applying Lagrange multipliers in a way that is standard for mechanical systems, does not always work for stochastic dynamics.

\subsection{Deterministic dynamics}

We first consider \emph{deterministic} dynamics, and show 
 how the form of the projected mobility arises naturally 
by imposing constraints using Lagrange multipliers.
 For simplicity we'll consider only one constraint, $c(\x)$. Suppose we start with the deterministic overdamped equation $\dd{\x}{t} = -M\nabla U$, and then project it to $\{\x:c(\x)=0\}$ by assuming the constraint forces are perpendicular to the constraint surface. This could be motivated by the principle of the constraint forces doing no ``virtual work'' \cite{Landau.1976}. 
Thus, we ask for a $\lambda$ such that
\[
\dd{\x}{t} = M(-\nabla U + \lambda \nabla c), \quad \text{and}\quad
\dd{c(\x)}{t} = 0.
\]
The constraint equation implies $\nabla c\cdot \dd{\x}{t} = 0$ or
\[
\nabla c \cdot M(-\nabla U + \lambda \nabla c) = 0.
\]
Solving for $\lambda$ gives
\[
\lambda = \frac{\nabla c^TM\nabla U}{\nabla c^TM\nabla c}.
\]
Substituting back into the overdamped ODE, and using the fact that $\nabla c^TM\nabla U$ is a scalar so it may be moved around, we get
\begin{align*}
\dd{\x}{t} %&= -M\nabla V - M\frac{\nabla c^TM\nabla V}{\nabla c^TM\nabla c}\nabla c \\
&= -\left(M - \frac{M\nabla c\nabla c^TM}{\nabla c^TM\nabla c}\right)\nabla U
= -\tilde M \nabla U,
\end{align*}
where the projected mobility is
\[
\tilde M = \PM M \quad\text{with}\quad
\PM  = I - \frac{M\nabla c\nabla c^T}{\nabla c^TM\nabla c}.
\]
This is the projection in Eq.~\eqref{eq:MP},  for a scalar constraint. 
%
%(Repeating the calculating with multiple constraints is straightforward and also gives Eq.~\eqref{eq:MP} as the projection.)

% \medskip
% \mhc{aleks writes down an explicit form of $\tilde M^{1/2}$, in terms of $M^{1/2}$ -- not sure if we should include this. Formula is (but I think there's a typo)
% \[
% \tilde M^{1/2} = M^{1/2} - \frac{M\nabla c\nabla c^T M^{1/2}}{\nabla c^T M \nabla c}
% \]}

\subsection{Derivation of the stochastic dynamics using Lagrange multipliers}

Next we will consider the overdamped dynamics, which we write as 
\[
\frac{d\x}{dt} = -M\nabla U + k_BT\nabla\cdot M + \sqrt{2k_BT} M^{1/2}\eta(t) .
\]
Again we focus on a single scalar constraint $c(\x)$. 
We will modify this process by adding a drift and diffusion term to ensure that $dc(\x_t)/dt~=~0$. 
We first show two \emph{incorrect}, but seemingly natural ways to do this, and then we will show the correct way -- part of our goal of this section is to show that  imposing constraints  in stochastic systems using Lagrange multipliers is subtle and nontrivial. 

\subsubsection{Incorrect method \#1 -- It\^{o} product.}

A seemingly natural way to impose the constraints, and one that has been followed in the mathematical literature\cite{ciccotti2008projection,lelievre2012langevin}, is to add a process of the form $y(t) = A + \sqrt{2k_BT}S \eta(t)$, where the drift vector $A$ and diffusion matrix $S$ will be found to impose the constraint: 
\begin{multline}\label{eq:lagr1}
\frac{d\x_t}{dt} = -M\nabla U + k_BT\nabla\cdot M + \sqrt{2k_BT} M^{1/2}\eta(t) \\+A + \sqrt{2k_BT}S \eta(t). 
\end{multline}
The constraint evolves as 
\begin{align*}
\frac{dc(\x_t)}{dt} &= \nabla c \cdot (-k_BTM\nabla U + k_BT \nabla \cdot M + A) \\
& \quad + \sqrt{2k_BT}\nabla c \cdot(M^{1/2}+S)\eta\\
&\quad + k_BT \nabla \nabla c:(M^{1/2}+S)(M^{1/2}+S)^T. 
\end{align*}
Preserving the constraint ($dc(\x_t)/dt = 0)$ thus requires 2 conditions
\begin{align*}
 \nabla c \cdot (-k_BTM\nabla U + k_BT \nabla \cdot M + A)\qquad \\+ k_BT \nabla \nabla c:(M^{1/2}+S)(M^{1/2}+S)^T &=0\\
 \nabla c \cdot(M^{1/2}+S) &= 0.
\end{align*}
Now we assume that both $A$ and $S$ are parallel to the same direction $N\nabla c$ for some symmetric positive definite matrix $N$. That is, we assume 
\begin{equation}\label{eq:AS}
    A=aN\nabla c,\qquad S = N\nabla c s^T,
\end{equation} 
for some scalar $a$ and vector $s$. The quantities $a,s$ are the equivalent of the Lagrange multiplier $\lambda$ in the previous section.  A straightforward calculation shows that 
\begin{align}
a &= -\frac{\nabla c \cdot (-M\nabla U + k_BT \nabla \cdot M)}{\nabla c \cdot N\nabla c} - k_BT\frac{\nabla\nabla c : (P_NMP_N^T)}{\nabla c \cdot N\nabla c},  \nonumber\\
s &= -\frac{M^{1/2}\nabla c}{\nabla c \cdot N\nabla c}.  \label{eq:as}
\end{align}
where 
\begin{equation}
P_N = I - \frac{N\nabla c\nabla c^T}{\nabla c^TN\nabla c}.
\end{equation}
Substituting back into \eqref{eq:lagr1} gives
\begin{align*}
\dd{\x_t}{t} =& -P_NM\nabla U + k_BT P_N \nabla \cdot M \\
&- k_BT\frac{\nabla\nabla c : (P_NMP_N^T)}{\nabla c \cdot N\nabla c}+ \sqrt{2k_BT P_NM^{1/2}}.
\end{align*}
This can be rewritten using the fact that
\begin{multline*}
P_NMP_N = M-\frac{(N\nabla c)^T(M\nabla c)}{\nabla c\cdot N\nabla c} 
-\frac{(M\nabla c)^T(N\nabla c)}{\nabla c\cdot N\nabla c}
\\+ \frac{(\nabla c \cdot M \nabla c)(N\nabla c)^T(N\nabla c)}{|\nabla c\cdot N\nabla c|^2}.
\end{multline*}
Thus, choosing $N=M$, so that $P_N=P_M$, Eq.~\eqref{eq:lagr1} can be written as 
\[
\dd{\x_t}{t} = -M_P\nabla U + k_BT \nabla \cdot M_P + \sqrt{2k_BT}M_P^{1/2}\eta - k_BT R,
\]
where 
\[
R = M\nabla \cdot P_M + \frac{\nabla \nabla c: P_M}{\nabla c \cdot M\nabla c}M\nabla c.
\]
This differs from Eq.~\eqref{eq:ODP0} by the additional drift $R$. There is no reason for this drift term to vanish, hence, there is no reason that this should be the correct dynamics.

\subsubsection{Incorrect method \#2 -- Stratonovich product.}

We can revisit our ansatz for the additional process imposing the constraint, and instead suppose $y(t) = A + \sqrt{2k_BT}S\circ \eta(t)$ where $\circ$ denotes the Stratonovich product. We rewrite the overdamped dynamics using
\[
\nabla \cdot M + \sqrt{2}M^{1/2} \eta = M^{1/2}\nabla \cdot M^{1/2} + \sqrt{2}M^{1/2}\circ \eta. 
\]
Thus, in Stratonovich form, the process evolves as 
\begin{multline}\label{eq:lagr2}
\frac{d\x_t}{dt} = -M\nabla U + M^{1/2}\nabla \cdot M^{1/2} + \sqrt{2k_BT} M^{1/2}\circ \eta(t) \\+A + \sqrt{2k_BT}S\circ \eta(t). 
\end{multline}
Again one can compute  $dc(\x_t)/dt$, and, assuming Eq.~\eqref{eq:AS}, one can solve for the $a,s$ which preserve the value of $c$: 
% Since the regular chain rule may be used with the Stratonovich product, we have that 
% \begin{multline*}
% \dd{c(\x_t)}{t} = -\nabla c \cdot M\nabla U + k_BT \nabla c \cdot M^{1/2}\nabla \cdot M^{1/2}  - \nabla c \cdot A \\
%  + \sqrt{2k_BT}(M^{1/2}+S)\circ \eta.
% \end{multline*}
% Assuming Eq.~\eqref{eq:AS}, we find that
\begin{align*}
a &= -\frac{\nabla c \cdot (-M\nabla U + k_BT \nabla c \cdot M^{1/2}\nabla \cdot M^{1/2}) }{\nabla c \cdot N\nabla c}\\
s&= -\frac{M^{1/2}\nabla c}{\nabla c \cdot N\nabla c}.
\end{align*}
% Thus, Eq.~\eqref{eq:lagr2} becomes
% \[
% \frac{d\x_t}{dt} =-P_NM\nabla U + k_BTP_NM^{1/2}\nabla \cdot M^{1/2} + \sqrt{2k_BT} P_NM^{1/2}\circ \eta.
% \]
Substituting into Eq.~\eqref{eq:lagr2}, 
converting to It\^{o}'s form, and setting $N=M$, gives %this equation is
% \[
% \frac{d\x_t}{dt} = -P_NM\nabla U + k_BT \nabla P_NMP_N^T   k_BTP_NM^{1/2}\nabla \cdot (M^{1/2}Q_N^T)+ \sqrt{2k_BT}P_NM^{1/2}\eta
% \]
% where we define $Q_N = I-P_N$. When $N=M$, this equation becomes
\begin{multline*}
\frac{d\x_t}{dt} =  -M_P\nabla U + k_BT \nabla \cdot M_P  + \sqrt{2k_BT}M_P^{1/2}\eta \\+  k_BTP_MM^{1/2}\nabla \cdot (M^{1/2}(I-P_N)^T).
\end{multline*}
%where we define $Q_N = I-P_N$.
This differs from the correct dynamics Eq.~\eqref{eq:ODP0} by the last term, which again doesn't need to vanish. %can be written as 
% \begin{align*}
%     P_MM^{1/2}\nabla \cdot (M^{1/2}(I-P_N)^T) &= M_PM^{-1/2}\nabla \cdot (M^{1/2}(I-P_N)^T)\\
%     &= M_PM^{-1/2}\nabla \cdot\left(\frac{(M^{1/2}\nabla c)(M\nabla c)^T}{\nabla c \cdot M\nabla c} \right).
% \end{align*}
%This doesn't need to vanish. \mhc{is numerator correct? why get different powers of $M$?}

\subsubsection{Correct method}

Finally, we consider an ansatz for the constraining process $y(t)$ which preserves the structure of the overdamped equations. That is, we search for a drift $A$ and diffusion matrix $S$ such that  
\begin{multline}\label{eq:lagr3}
\frac{d\x_t}{dt} = -M\nabla U + A + k_BT\nabla\cdot (M^{1/2}+S)(M^{1/2}+S)^T \\+ \sqrt{2k_BT} (M^{1/2}+S)\eta(t). 
\end{multline}
Compared to Eq.~\eqref{eq:lagr1}, we have included the noise $S$ in the divergence part of the drift as well. 
Proceeding before by assuming form \eqref{eq:AS} and solving for $a,s$ gives the same expressions \eqref{eq:as}, but with $\nabla\cdot M$ replaced by $\nabla\cdot (P_NMP_N)$ in the expression for $a$. 
%so that
\begin{align*}
\dd{\x_t}{t}= & -P_{N} M \nabla U +k_BT  P_{N} \nabla \cdot \left(P_{N} M P_{N}^{T}\right)  \\
& \quad -k_BT  \frac{\nabla \nabla  c : P_{N} M P_{N}^{T}}{\nabla  c  \cdot N \nabla  c } N \nabla  c  +\sqrt{2 k_BT } P_{N} M^{1 / 2} \eta 
\end{align*}
Using the following identity\footnote{ Here are the intermediate steps to show this:
\begin{align*}
& \left(P_{N} \nabla \cdot \left(P_{N} M P_{N}^{T}\right)\right)_{i} \\
& \quad=\left(P_{N}\right)_{i, j} \partial_{k}\left(P_{N} M P_{N}^{T}\right)_{j, k} \\
& \quad=\partial_{k}\left(\left(P_{N}\right)_{i, j}\left(P_{N} M P_{N}^{T}\right)_{j, k}\right)-\left(P_{N} M P_{N}^{T}\right)_{j, k} \partial_{k}\left(P_{N}\right)_{i, j} \\
& \quad=\partial_{k}\left(P_{N} M P_{N}^{T}\right)_{i, k}+\left(P_{N} M P_{N}^{T}\right)_{j, k} \partial_{k}\left(\frac{(N \nabla  c )_{i} \partial_{j}  c }{\nabla  c  \cdot N \nabla  c }\right)  \\
& \quad=\partial_{k}\left(P_{N} M P_{N}^{T}\right)_{i, k}+\left(\frac{\left(P_{N} M P_{N}^{T}\right)_{j, k} \partial_{j} \partial_{k}  c }{\nabla  c  \cdot N \nabla  c }\right)(N \nabla  c )_{i}
\end{align*}
where in the last step we used $P_NMP_N^T\nabla c =0$
(meaning that the only nonzero term arises when $\partial_k$ acts on $\partial_j c$).} 
(written using Einstein notation for summation over repeated indices): 
\begin{multline*}
\left(P_{N} \nabla \cdot \left(P_{N} M P_{N}^{T}\right)\right)_{i} = \partial_{k}\left(P_{N} M P_{N}^{T}\right)_{i, k}
\\+\left(\frac{\left(P_{N} M P_{N}^{T}\right)_{j, k} \partial_{j} \partial_{k}  c }{\nabla  c  \cdot N \nabla  c }\right)(N \nabla  c )_{i},
\end{multline*}
we can write this as 
\[
\dd{\x_t}{t} =  -P_{N} M \nabla U +k_BT  \nabla \cdot \left(P_{N} M P_{N}^{T}\right)+\sqrt{2 k_BT } P_{N} M^{1 / 2} \eta.
\]
Notice that when $N=M$, this equation equals Eq.~\eqref{eq:ODP0}.
Thus, we obtain the correct equation with ansatz \eqref{eq:lagr3} for the Lagrange multipliers. A posteriori this is expected, because it is the only ansatz which preserves the structure of the Brownian dynamics equations. 
%\mhc{moved derivation to a footnote and changed wording above slightly}

%One can ask what happens when $N\neq M$. In this case, the adjoint of the generator for the dynamics is 
%\[
%\mathcal L^*\rho = \nabla\cdot \left( k_BT P_NM\nabla U \rho + P_NMP_N^T\nabla \rho \right)
%\]
%so that
%\[
%\mathcal L^* e^{- U/k_BT} = k_BT \nabla \cdot \left( P_N(P_NM-MP_N^T)\nabla U e^{-U/k_BT}\right).
%\]
%In order to have zero current in steady-state, one must have $P_NM=MP_N^T$, which appears to be true only for $N=M$. 

%%%%%%%%%%%%%%%%%%%%%%%%%%%%%%%%%%%%%%%%%%%%%%%
%%%%%   Section: Conclusion               %%%%%
%%%%%%%%%%%%%%%%%%%%%%%%%%%%%%%%%%%%%%%%%%%%%%%
\section{Conclusion/Discussion}

In this paper we have derived the equations of motion for softly constrained overdamped Langevin equations. Our derivation, based on singular perturbation theory, demonstrates that the constrained equations are valid for long enough timescales, \textit{e.g.} longer than the time for particles or objects to relax over the stiffly constrained degrees of freedom. In addition, our derivation allowed us to find the form of the effective mobility in the case of ``softly-softly'' constrained dynamics, where the mobility varies on lengthscales similar to that of the constrained degrees of freedom. 

We provide practical formulas in Table~\ref{tab:formulas}, that we hope allow the reader to use softly constrained dynamics in applications. The pedagogical examples we have provided highlight a few subtleties in the applications: (i) how the effective mobility and equations of motion can depend on a particle's local environment through hydrodynamic interactions, (ii) how to use these equations either at the level of friction or mobility and make sense of the derived equations in particular of the emergence of local drift. 

The softly constrained equations may be used to simulate effectively complex examples, as they allow one to use longer timesteps and avoid resolving fast motion of the stiffly constrained degrees of freedom. In practice, using the equations on the extrinsic coordinates in Eq.~\eqref{eq:ODP0} might be easier computationally, as it does not require computing a determinant as is the case for intrinsic coordinates, in Eq.~\eqref{eq:seff}. Intrinsic coordinates might be better for modeling as they are more effective at describing the constrained space. 

%\mhc{added:}

The softly-softly constrained equations are appropriate when mobility varies on the same scale as the confining potential, as happens for example with particles interacting with lubrication hydrodynamics. In that case, %one must average the mobility tensor as well as project it. 
our calculations show that the correct order of operations is to first project the mobility tensor, %using projection $P_M$, 
and then average it with respect to the equilibrium distribution induced by the confining potential. %, over fibres normal to the manifold. 
The order of operations brings up a puzzling question: why is it appropriate to restrict \emph{friction}, but to average \emph{mobility}? That is, the averaging operation is linear in mobility but nonlinear in friction, unlike the projection operation which is nonlinear in mobility. The averaging process thus appears to average over velocities, rather than over forces. %While we believe the mathematical derivation is correct, one may wonder whether this is physically correct.
For now, we have not been able to identify the physical intuition that leads to this result.
It could be that more intuition lies in the coarse-graining process of the solvent dynamics, which is used to derive the original mobility. 
%Given the mobility is already a coarse-grained quantity from the solvent dynamics, it would be interesting to obtain the constrained dynamics by coarse-graining over the rapidly-varying mobility and forces at the same time as coarse-graining over the solvent. %Since there is a separation of lengthscales 
%since typically one thinks about averaging forces -- it could be that starting from the Brownian dynamics equations, in which the mobility matrix already coarse-grains the solvent degrees of freedom, and then further coarse-graining, is not the right order of operations -- it could be that one must coarse-grain over the rapidly-varying mobility and forces \emph{at the same time as} coarse-graining over the solvent. 
%It would be interesting to do do this latter calculation and see if it agrees with ours. 
%\mhc{this paragraph is too wordy and some of it may be better elsewhere, however the last point is something to think about. (just wanted to get down thought on a page)}\mhc{is david dean's comment relevant here?}

In this work, we have mostly considered the constrained equations with ``soft'' constraints -- meaning the constraints are approximations for very stiff forces. In some situations, e..g in statistics or in machine learning~\cite{leimkuhler2021better,schonle2025efficient}, one wishes to use ``hard'' constraints instead. In that case the extrinsic dynamics write exactly as in Eq.~\eqref{eq:ODP0}, except $U^{\rm eff} \rightarrow U^{\rm eff} - k_B T \log |C|$ now contains a geometrical correction. We report the detailed formula in Table~\ref{tab:formulas}. To see how this changes dynamics in practice, we should take an example where the constraint is ``curved''. In the example of 2D brownian walker on the parabola for instance, one can show that the effective dynamics say in $x$ are given by 
\begin{equation}
    \frac{dx}{dt} = \sqrt{2D(x)}\eta(t)
\end{equation}
where $D(x) = D/(1 + 4x^2/\lambda^2)$. Compared to Eq.~\eqref{eq:curvedline} in the soft constrained case, the hard constraints let no drift term arise. This is due to the fact that the constraint being hard, it is exact, and therefore small wiggles around the constraint, which can introduce more weight in curved regions, here do not arise. 

Taking a step back, allows us to comment on the potential advantages of methods like singular perturbation theory. Unlike Lagrange multiplier methods, which require guessing beforehand the right form of the equations, singular perturbation is a straightforward procedure that does not require any arbitrary guesses. 
One could thus export such a method to project other kinds of equations, starting with non overdamped or generalized Langevin equations, where particles exhibit exotic phenomena like resonances in constrained configurations experimentally~\cite{franosch2009persistent,franosch2011resonances}.

\begin{acknowledgments}

%The conception of this work was made possible thanks to numerous discussions with Aleksandar Donev. 

The authors acknowledge further fruitful discussions with David Richard, Marylou Gabri\'e and Eric Vanden-Eijnden. We are grateful to Eric Vanden-Eijnden for providing calculations involving Lagrange multipliers. 
M.H.C. acknowledges support from the Natural Sciences and Engineering Research Council of Canada (NSERC), RGPIN-2023-04449 / Cette recherche a \'{e}t\'{e} financ\'{e}e par le Conseil de recherches en sciences naturelles et en g\'{e}nie du Canada (CRSNG).
\end{acknowledgments}

\section*{Author Declarations}

\subsection*{Conflict of Interest}

The authors have no conflicts to disclose.

\subsection*{Author Contributions}

\textbf{Sophie Marbach}: Conceptualization (equal); Investigation (equal); Simulations (equal); Methodology (equal); Visualization (equal); Writing - original draft (equal); Writing - review \& editing (equal).
\textbf{Adam Carter}: Simulations (equal); Visualization (equal); 
\textbf{Miranda Holmes-Cerfon}: Conceptualization (equal); Investigation (equal); Simulations (equal); Methodology (equal); Visualization (equal); Writing - original draft (equal); Writing - review \& editing (equal).

\section*{Data Availability Statement}

The data that support the findings of this study are available within the article. 

\appendix

\section{Numerical simulations}
\label{app:numerics}

Numerical simulations of overdamped Langevin equations in Sec.~\ref{sec:tethered} and Sec.~\ref{sec:2dnondiag} are performed by integrating Langevin equations with the Euler-Maruyama method. For a degree of freedom say $x$, the diffusion coefficient in the $x$ direction is estimated as 
\[
D^{\rm est} = \frac{1}{N_{\rm msd}}\sum_{i=1}^{N_{\rm msd}} \frac{\Delta ((i+1)\delta t)-\Delta (i \,\delta t)}{2\delta t}.
\]
Here  $\delta$ is the time step for saving data, and we take $N_{\rm msd}=100$.  
The mean-squared displacements are defined as 
\begin{equation}
\label{eq:msd}
    \Delta (\tau) = \frac{1}{N_d-j}\sum_{k=1}^{N_d-j} (x(k \, \delta t+\tau) - x(k\, \delta t))^2,
\end{equation}
where $\tau=j\delta$, and $N_d$ is the total number of saved data points.

\section{Additional details for the physical examples}
\label{app:tethered}
In this appendix we provide details and calculus in particular for the tethered particle model in Sec.~\ref{sec:tethered}.

\paragraph{Case of angular constraint}

We start from the unconstrained system of a tether and a free rotating particle of Sec.~\ref{sec:tethered}b.
If the constraints are 
\begin{align*}
    &c_1(x,y,a\theta) = \frac{1}{a}\left(x^2 + (a+h -y)^2 - a^2 \right) \\
    &c_2(x,y,a\theta) = a\theta - a\beta = a\theta - a\arctan(x/(a+h-y)).
\end{align*}
The constraint matrix is thus 
\begin{align*}
    C = \begin{pmatrix}
        \frac{2x}{a} & -\frac{2(a+h -y)}{a} & 0 \\
        - a\frac{(a+h-y)}{(a+h-y)^2 + x^2} & - a\frac{x}{(a+h-y)^2 + x^2} & 1
    \end{pmatrix}
\end{align*}
and then 
\begin{align*}
    CC^T = \begin{pmatrix}
        4 \frac{(a+h-y)^2 + x^2}{a^2} & 0 \\
        0 & 1 + \frac{a^2}{(a+h-y)^2 + x^2}
    \end{pmatrix}
\end{align*}
and the pseudo-determinant $|C| = 2\sqrt{1 + \frac{(a+h-y)^2 + x^2}{a^2}}$.

One can evaluate the projected mobility  as 
\begin{equation*}
    M_P = \frac{D^{\rm eff}(x,y) }{k_B T}  \begin{pmatrix}
        \frac{(a+h-y)^2}{a^2} & \frac{(a+h-y)x}{a^2} & \frac{(a+h-y)}{a} \\
        \frac{(a+h-y)x}{a^2}  & \frac{x^2}{a^2} & \frac{x}{a}\\
        \frac{(a+h-y)}{a} & \frac{x}{a} & 1
    \end{pmatrix}
\end{equation*}
where $D^{\rm eff}(x,y) = \frac{a^2D_{\theta}D}{ (x^2 + (a+h-y)^2)D_{\theta} + D}$ is a form of weighted harmonic average of the diffusivities $D,a^2D_{\theta}$. This 
 has a Cholesky decomposition  
\begin{equation*}
    \sigma_P = \sqrt{\frac{D^{\rm eff}(x,y)}{k_B T}  } \begin{pmatrix}
        \frac{(a+h-y)}{a} & 0 & 0 \\
        \frac{x}{a} & 0 & 0 \\
        1 & 0 & 0 
    \end{pmatrix}
\end{equation*}
such that $\sigma_P \sigma_P^T = M_P$.

The constrained dynamics then have $\theta$ evolving as %evolution of the system through a single coordinate, $\theta$, which evolves as
\begin{align*}
    a\frac{d\theta}{dt} = &- \frac{D^{\rm eff}(x,y)}{k_B T} \left( \frac{(a + h - y)}{a} F_x  + \frac{x}{a} F_y \right) \\
    &+ \partial_x (M_P)_{x\theta} + \partial_y (M_P)_{y\theta} \\
    &+ \sqrt{2 k_B T D^{\rm eff}(x,y) } \eta_{\theta}.
\end{align*}
One can then check (not shown here for simplicity) that $\partial_x (M_P)_{x\theta} + \partial_y (M_P)_{y\theta} = 0$
and then we can use the fact that on the manifold, the constraints are satisfied to further simplify the equations of motion. In that case $D^{\rm eff}(x,y)|_{\mathcal{M}} = \frac{a^2D_{\theta}D}{ a^2 D_{\theta} + D} $ and with further angular formulas recover Eq.~\eqref{eq:angletethered}.

\paragraph{Sine of angular constraint}

Now we consider a constraint on the sine of the angle, as 
\begin{align*}
    &c_1(x,y,a\theta) = \frac{1}{a}(x^2 + (a+h -y)^2 - a^2)  \\
    &c_2(x,y,a\theta) = a\left(\sin \theta - x/\sqrt{x^2+(a+h-y)^2}\right)
\end{align*}
Note we have modified the constraints and variables so they all have dimensions of length. 
The constraint matrix is
\[
C = \begin{pmatrix}
\frac{2x}{a} & -\frac{2(a+h -y)}{a} & 0\\
-a\frac{(a+h-y)^2}{(x^2+(a+h-y)^2)^{3/2}} & 
-a\frac{x(a+h-y)}{(x^2+(a+h-y)^2)^{3/2}} & 
\cos\theta
\end{pmatrix}.
\]
We thus have
\[
CC^T = \begin{pmatrix}
    4 \frac{x^2+(a+h-y)^2}{a^2} & 0 \\
    0 & \frac{a^2(a+h-y)^2}{(x^2+(a+h-y)^2)^2} + \cos^2\theta
\end{pmatrix}.
\]
This has determinant $|CC^T| = 4\big(\frac{(a+h-y)^2}{x^2+(a+h-y)^2} + \frac{x^2+(a+h-y)^2}{a^2} \cos^2\theta\big)$, which, on the constraint set, equals $|CC^T| = 8\cos^2\theta$.

%Now assume
%\[
%M = \frac{1}{k_BT} \begin{pmatrix} D & & \\ & D & \\ &&D_\theta \end{pmatrix}.
%\]
%Some calculations to help with the projected mobility:
%\[
%CMC^T = \frac{1}{k_BT} \begin{pmatrix}
%    4D \frac{x^2+(a+h-y)^2}{a^2} & 0 \\
%    0 & D_\theta\left(\frac{a^2(a+h-y)^2}{(x^2+(a+h-y)^2)^2} + %\cos^2\theta\right)
%\end{pmatrix}.
%\]

\begin{widetext}
One can evaluate the projected mobility as 
\[
 M_P = \frac{D^{\rm eff}(x,y,\theta) }{k_B T}  \begin{pmatrix}
        \frac{a^2(x,y,\theta)}{a^2} & \frac{x}{a+h-y} \frac{a^2(x,y,\theta)}{a^2} & \frac{a(x,y,\theta)}{a} \\
       \frac{x}{a+h-y} \frac{a^2(x,y,\theta)}{a^2} & \frac{x^2}{(a+h-y)^2} \frac{a^2(x,y,\theta)}{a^2} & \frac{x}{a+h-y} \frac{a(x,y,\theta)}{a}\\
      \frac{a(x,y,\theta)}{a} & \frac{x}{a+h-y} \frac{a(x,y,\theta)}{a} & 1
    \end{pmatrix}
\]
\end{widetext}
where $D^{\rm eff}(x,y,\theta) = \frac{a^2D_{\theta}D (a+h-y)^2}{ (x^2 + (a+h-y)^2)^2 D_{\theta} \cos^2 \theta + D(a+h-y)^2}$ and $a(x,y,\theta) =  \sqrt{x^2 + (a+h-y)^2}\cos\theta$. 
This has a Cholesky decomposition  
\begin{equation*}
    \sigma_P = \sqrt{\frac{D^{\rm eff}(x,y,\theta)}{k_B T}  } \begin{pmatrix}
        \frac{a(x,y,\theta)}{a} & 0 & 0 \\
        \frac{a(x,y,\theta)}{a}\frac{a+h-y}{x} & 0 & 0 \\
        1 & 0 & 0 
    \end{pmatrix}
\end{equation*}
such that $\sigma_P \sigma_P^T = M_P$.

The constrained dynamics then have $\theta$ evolving as %evolution of the system through a single coordinate, $\theta$, which evolves as
\begin{align*}
    a\frac{d\theta}{dt} = &- \frac{D^{\rm eff}(x,y,\theta)}{k_B T} \left( \frac{a(x,y,\theta)}{a} F_x  + \frac{a(x,y,\theta)}{a} \frac{x}{a+h-y} F_y \right) \\
    &+ \partial_x (M_P)_{x\theta} + \partial_y (M_P)_{y\theta} + \partial_{a\theta} (M_P)_{\theta\theta} \\
    &+ \sqrt{2 k_B T D^{\rm eff}(x,y,\theta) } \eta_{\theta}.
\end{align*}
Before figuring out the drift terms it is useful to notice that we can now evaluate the different functions on the constraint manifold. In particular, $a(x,y,\theta) = a \cos \theta$ and $D^{\rm eff}(x,y,\theta) = \frac{a^2D_{\theta}D }{ a^2 D_{\theta} + D}$ simplifies to the same effective diffusion coefficient we had with the other constraints. We obtain
\begin{align*}
    a\frac{d\theta}{dt} = &- \frac{D^{\rm eff}}{k_B T} \left( \cos \theta F_x  + \sin \theta F_y \right) \\
    &+ V + \sqrt{2 k_B T D^{\rm eff} } \eta_{\theta}.
\end{align*}
where we bundled all drift terms in $V = \partial_x (M_P)_{x\theta} + \partial_y (M_P)_{y\theta} + \partial_{a\theta} (M_P)_{\theta\theta}$. We thus see that there is no difference in all the other terms except for $V$. Now let us calculate $V$ explicitly. Using the expression of $M_P$, taking the derivatives and then using the simplifications mentioned previously, we obtain
\[
V = \frac{ D^{\rm eff}}{a} \tan \theta.
\]%\mhc{I don't see why you divide by $a$ here -- is this consistent with $M_P\partial_\theta \log|C|$ ? }
This is consistent with the additional drift in intrinsic form, which is proportional to $- \partial_{\theta} \log |C| = 2 \tan \theta$. 

\section{Additional details for the toy models}
In this appendix we provide details and further calculus to investigate the toy models presented in Sec.~\ref{sec:toymodels}.

\subsection{Change of coordinates on the parabolic curve}
\label{app:changeparabolic}

A natural coordinate in the constrained space is $s$, the arc-length parametrization along the curve, 
$$s = \int_0^x \sqrt{1 + 4\tilde{x}^2/\lambda^2} \mathrm{d}\tilde{x}.$$ 
The other coordinate we will take to be the value of the constraint itself, $z = (y-x^2/\lambda)$, since this variable is associated with fluctuations on a small lengthscale. %\sophie{there is a question of making this orthogonal but then the jacobian etc are horrible}.\mhc{I don't think you can/don't need to -- it's orthogonal on the manifold and that's what matters. } 
%\mhc{I think you should use $s$ for $u$, and $z$ for $v$, and something else for $s$, to be consistent with the notation later in the paper}
The jacobian matrix is 
$$J = \pp{(s,z)}{(x,y)} = \begin{pmatrix}
    \partial_x s & \partial_y s  \\ \partial_x z & \partial_y z 
\end{pmatrix}  = \begin{pmatrix}
    \sqrt{1 + (2x/\lambda)^2}  & 0  \\ - 2x/\lambda  & 1
\end{pmatrix}.$$
Before making more progress it'll also be useful to get the inverse of the jacobian 
$$J^{-1} = \begin{pmatrix}
    \partial_s x & \partial_z x  \\ \partial_s y & \partial_z y 
\end{pmatrix} = \begin{pmatrix}
    1/\sqrt{1+(2x/\lambda)^2} & 0 \\ (2x/\lambda)/\sqrt{1+(2x/\lambda)^2} & 1
\end{pmatrix}$$
which importantly highlights that $\pp{x}{z} = 0$, which will simplify some calculations further down. 

By the chain rule 
\begin{align*}
\pp{}{x} &= \pp{s}{x} \pp{}{s} + \pp{z}{x} \pp{}{z} = \sqrt{1+(2x/\lambda)^2}\pp{}{s} - (2x/\lambda)\pp{}{z} \\
\pp{}{y} &= \pp{}{z}. 
\end{align*}

% $v = (y-x^2/\lambda)/\sqrt{1+4x^2/\lambda}$, where the additional rescaling of $\sqrt{1+4x^2/\lambda}$ comes from requiring that the basis on which these coordinates are defined is indeed an orthogonal basis. The jacobian matrix is $J = \pp{(u,v)}{(x,y)} = \begin{pmatrix}
%     \partial_x u & \partial_y u  \\ \partial_x v & \partial_y v 
% \end{pmatrix}$ 
% can be expressed as
% $$J = \begin{pmatrix}
%     \sqrt{1 + 4x^2/\lambda^2}  & 0  \\ - \frac{2 x}{\lambda} \frac{1}{\sqrt{1 + 4x^2/\lambda^2}}  - v \frac{4x/\lambda}{(1 + 4x^2/\lambda^2)} & \frac{1}{\sqrt{1 + 4x^2/\lambda^2}}
% \end{pmatrix}$$
% and the inverse
% % $J^{-1} = \begin{pmatrix}
% %     \partial_u x & \partial_v x  \\ \partial_u y & \partial_v y 
% % \end{pmatrix}$ 
% % for which we know $\partial_v y  = 1$ and $\partial_v x = -2x/\lambda$ and we are simply missing the $u$ components. Now there are several choices for $u$ and its not obvious which one should be the correct one. If I want the matrix to be as simple as possible I can take $\partial_u x = 1$ and $\partial_u y = 0$. Such that 
% % $$J^{-1} = \begin{pmatrix}
% %     1 & -2x/\lambda \\ 0 & 1 
% % \end{pmatrix}$$
% % and its inverse is
% % $$ J = \begin{pmatrix}
% %     1 & 2x/\lambda \\ 0 & 1 \end{pmatrix}$$. 
% By the chain rule $\pp{}{x} = \pp{u}{x} \pp{}{u} + \pp{v}{x} \pp{}{v} = \sqrt{1+s^2}\pp{}{u} - \frac{s}{\sqrt{1 + s^2}}\pp{}{v} - \frac{2s}{(1+s^2)}v \pp{}{v}$ and $\pp{}{y} = \pp{}{v} \frac{1}{\sqrt{1+s^2}}$.
Now we need the Laplacian operators. 
Clearly
$$\pp{^2 }{y^2} = \pp{^2}{z^2} $$ and then
\begin{align*}
\pp{^2}{x^2} =& \left(\sqrt{1+(2x/\lambda)^2}\pp{}{s} - (2x/\lambda)\pp{}{z} \right)^2 \\
= &\sqrt{1+(2x/\lambda)^2}\pp{}{s} \left( \sqrt{1+(2x/\lambda)^2}\pp{}{s}\right) \\
& - \sqrt{1+(2x/\lambda)^2}\pp{}{s} \left( (2x/\lambda)\pp{}{z} \right)  \\ 
&- (2x/\lambda)\pp{}{z} \left(  \sqrt{1+(2x/\lambda)^2}\pp{}{s}\right)  \\
&+ (2x/\lambda)\pp{}{z} \left( (2x/\lambda)\pp{}{z} \right) \\
=& \left( 1+(2x/\lambda)^2 \right) \pp{^2}{s^2} + \frac{\partial x}{\partial s}  \frac{4 x}{\lambda^2} \frac{\sqrt{1+(2x/\lambda)^2}}{\sqrt{1+(2x/\lambda)^2}} \pp{}{s} \\
&-   \frac{2x}{\lambda}\sqrt{1+(2x/\lambda)^2} \pp{^2}{s \partial z} - \frac{\partial x}{\partial s} \frac{2}{\lambda} \sqrt{1+(2x/\lambda)^2} \pp{}{z} \\
&-  \frac{2x}{\lambda}\sqrt{1+(2x/\lambda)^2} \pp{^2}{s \partial z} - \frac{\partial x}{\partial z}  \frac{4 x}{\lambda^2} \frac{(2x/\lambda)}{\sqrt{1+(2x/\lambda)^2}} \pp{}{s} \\
&+ \left( \frac{2x}{\lambda}\right)^2 \pp{^2}{z^2} + \frac{4x}{\lambda^2} \frac{\partial x}{\partial z}  \pp{}{z}.
\end{align*}
Now we make use of the inverse Laplacian formulas to simplify things out
\begin{align*}
\pp{^2}{x^2} =&\left( 1+(2x/\lambda)^2 \right) \pp{^2}{s^2} +  \frac{1}{\sqrt{1+(2x/\lambda)^2}} \frac{4 x}{\lambda^2}  \pp{}{s} \\
&-   \frac{2x}{\lambda}\sqrt{1+(2x/\lambda)^2} \pp{^2}{s \partial z} - \frac{2}{\lambda}  \pp{}{z} \\
&-  \frac{2x}{\lambda}\sqrt{1+(2x/\lambda)^2} \pp{^2}{s \partial z} + \left( \frac{2x}{\lambda}\right)^2 \pp{^2}{z^2}  \\
=& \left( \frac{2x}{\lambda}\right)^2 \pp{^2}{z^2} + \left(1+\left( \frac{2x}{\lambda}\right)^2\right) \pp{^2}{s^2} \\
&- \frac{4x}{\lambda} \sqrt{1 +\left( \frac{2x}{\lambda}\right)^2} \pp{^2}{s \partial z} \\
&- \frac{2}{\lambda} \pp{}{z} + \frac{4x}{\lambda^2} \frac{1}{\sqrt{1+\left( \frac{2x}{\lambda}\right)^2}} \pp{}{s} .   
\end{align*}
%. \mhc{yikes. I didn't check this laplacian. But, are you sure you accounted for the fact that $w=w(s)$?} \sophie{just checked, yes. I did}

To go back to the initial coordinate system, it is useful to derive further operators $\pp{}{s}  = \frac{1}{\sqrt{1+w^2}}\pp{}{x} + \frac{w}{\sqrt{1+w^2}} \pp{}{y}$ and $\pp{}{z} = \pp{}{y}$. Then we need the Laplacian, it is $\pp{^2}{s^2} = \frac{1}{1+w^2}\pp{^2}{x^2} + \frac{w^2}{1+w^2} \pp{^2}{y^2} + \frac{2w}{1+w^2} \pp{^2}{x \partial y} + \frac{2}{\lambda} \frac{1}{(1+w^2)^2}\pp{}{y} -\frac{4x}{\lambda^2} \frac{1}{(1+w^2)^2}\pp{}{x} $. 

\subsection{Calculation of $\mathcal{L}_1f_1$ for the parabolic curve}
\label{sec:l1f1}

We provide details for calculating $\mathcal{L}_1 f_1$ to obtain Eq.~\eqref{eq:1Dcurved} of the main text:
\begin{align*}
    \mathcal{L}_1 f_1 =& z \frac{2x}{\lambda} \sqrt{1+\left( \frac{2x}{\lambda}\right)^2} \pp{}{s} \left(\frac{\frac{2x}{\lambda}}{\sqrt{1+\left( \frac{2x}{\lambda}\right)^2}} z \pp{a}{s} \right) \\
    &- \frac{2L_s}{\lambda} \pp{}{z} \left(\frac{\frac{2x}{\lambda}}{\sqrt{1+\left( \frac{2x}{\lambda}\right)^2}} z \pp{a}{s} \right)  \\
    &- 2 \frac{2x}{\lambda} \sqrt{1 +\left( \frac{2x}{\lambda}\right)^2} \pp{^2}{s \partial z}  \left(\frac{\frac{2x}{\lambda}}{\sqrt{1+\left( \frac{2x}{\lambda}\right)^2}} z \pp{a}{s} \right)  \\ 
    = & (z^2-2) \left( \frac{2x}{\lambda}\right)^2 \pp{^2a}{s^2} \\
    &+ (z^2-2) \frac{2x}{\lambda} \sqrt{1+\left( \frac{2x}{\lambda}\right)^2}  \frac{2}{\lambda}\pp{x}{s} \frac{1}{\left( 1+\left( \frac{2x}{\lambda}\right)^2\right)^{3/2}} \pp{a}{s} \\
    & - \frac{2L_s}{\lambda} \frac{\frac{2x}{\lambda}}{\sqrt{1+\left( \frac{2x}{\lambda}\right)^2}}  \pp{a}{s}.
\end{align*}
And then we simplify things out and replace $\pp{x}{s}$
\begin{align*}
    \mathcal{L}_1 f_1 =& (z^2-2) \left( \frac{2x}{\lambda}\right)^2 \pp{^2a}{s^2} \\
    &+ (z^2-2) \frac{4x}{\lambda^2} \frac{1}{\left( 1+\left( \frac{2x}{\lambda}\right)^2\right)^{3/2}} \pp{a}{s} \\
    & - \frac{2L_s}{\lambda} \frac{\frac{2x}{\lambda}}{\sqrt{1+\left( \frac{2x}{\lambda}\right)^2}}  \pp{a}{s}.
\end{align*}
and finally carrying out the integration over $z$, as
\begin{align*}
    \langle \mathcal{L}_1 f_1, \pi_0 \rangle = &- \left( \frac{2x}{\lambda}\right)^2 \pp{^2a}{s^2} - \frac{4x}{\lambda^2} \frac{1}{\left( 1+\left( \frac{2x}{\lambda}\right)^2\right)^{3/2}} \pp{a}{s} \\
    & - \frac{2L_s}{\lambda} \frac{\frac{2x}{\lambda}}{\sqrt{1+\left( \frac{2x}{\lambda}\right)^2}}  \pp{a}{s}. 
\end{align*}

\subsection{Recovering the constrained equations on the parabolic curve using the friction route}

To obtain the constrained dynamics on the curved line (for $x$), \textit{e.g.} to find Eq.~\eqref{eq:curvedline}, 
an alternative derivation uses the projected friction $\Gamma_P$. In this case we need 
\begin{align*}
    P &= I -  C (C^T  C)^{-1} C  \\
  &= \frac{1}{1 + 4x^2/\lambda^2}\begin{pmatrix}
    1 & 2x/\lambda \\ 2x/\lambda & 4x^2/\lambda^2 \end{pmatrix}
    \end{align*}
and then $\Gamma = \frac{k_B T}{D} I$ such that $\Gamma_P = P \Gamma P = \frac{k_B T}{D} P $ and $\Gamma_P^{\dagger} = \frac{D}{k_B T} P$. We clearly see that $\Gamma_P^{\dagger} = M_P$ and so the square root of the mobility is the same. In our case the pseudo-determinant of $C$ is $|C| = |CC^T|^{1/2} = \sqrt{1 + 4x^2/\lambda^2}$. The terms that contribute to the dynamics, are according to Table~\ref{tab:formulas}, 
\begin{align*}
    \frac{dx}{dt} =& - (\Gamma_P^{\dagger})_{xx} \partial_x k_B T \log |C|  + \sqrt{2 k_B T \Gamma_P^{\dagger}}_{xx} \eta_x(t) \\
    &+ k_B T (P_{xx} \partial_x (\Gamma_P^{\dagger})_{xx} + P_{xy} \partial_x (\Gamma_P^{\dagger})_{yx} )
\end{align*}
which after some algebra, simplifies exactly to Eq.~\eqref{eq:curvedline}. And similarly for $y$.

\subsection{Derivation for non-diagonal projected mobility using singular perturbation theory}
\label{sec:OUextra}

Here we provide an alternative derivation for the projected dynamics in the case of a non-diagonal mobility, \textit{i.e.} to recover Eq.~\eqref{eq:projMobOff}. As a reminder, we consider a 2D random walker $(x(t),y(t))$ moving in a potential $U(x,y) = \half ky^2$, with a constant but non-diagonal mobility tensor 
\[
M = \begin{pmatrix} M_{11} & M_{12} \\  M_{12} & M_{22} \end{pmatrix}.
\]
The equations of motion are given in Eq.~\eqref{eq:2dnondiag}.
The generator for the dynamics is 
\begin{multline*}
\mathcal Lf = \\
-M_{12}ky\partial_x f - M_{22}ky\partial_y f + M_{11}\partial_{xx}f + 2M_{12}\partial_{xy}f + M_{22}\partial_{yy}f. 
\end{multline*}
Now, we assume that $k=\tilde k / \epsilon^2$ where $\epsilon \ll 1$, and we rescale $Y = y/\epsilon$. Under this rescaling the generator has the form
\[
\mathcal L = \frac{1}{\epsilon^2}\mathcal L_0 + \frac{1}{\epsilon}\mathcal L_1 + \mathcal L_2
\]
where
\begin{align*}
\mathcal L_0 &= -M_{22}\tilde k Y \partial_{Y} + M_{22}\partial_{YY}  \\
\mathcal L_1 &= -M_{12}\tilde k Y \partial_x + 2M_{12}\partial_{xY} \\
\mathcal L_2 &= M_{11}\partial_{xx}.
\end{align*}
Now we assume a solution to the backward equation $\partial_t f = \mathcal Lf$ of the form $f=f_0 + \epsilon f_1 + \epsilon^2 f_2 + \cdots$. The leading-order equation is $\mathcal L_0 f_0 = 0$ which implies, since $\mathcal L_0$ is only an operator on $Y$, that $f_0$ is constant in $Y$. Hence, $f_0 = a(x,t)$ for some function $a$. 

At next order, we must solve $\mathcal L_0 f_1 = -\mathcal L_1 f_0$, or 
\[
-M_{22}\tilde k Y \partial_{Y}f_1 + M_{22}\partial_{YY}f_1 = M_{12}\tilde k Y \partial_xa.
\]
The solution is 
\[
f_1(x,Y,t) = -\frac{M_{12}}{M_{22}}Y\partial_x a. 
\]
The next order equation is $\mathcal L_0 f_2 = \partial_t f_0 - \mathcal L_1 f_1 - \mathcal L_2 f_0$. This has a solution only if $\langle \pi(Y),(\partial_t f_0 - \mathcal L_1 f_1 - \mathcal L_2 f_0)\rangle = 0$, where $\pi$ solves $\mathcal L_0^*\pi = 0$. We find that 
\[
\pi(Y) = \frac{1}{\sqrt{2\pi\tilde k}} e^{-\half \tilde k Y^2}.
\]
We also have that 
\begin{multline*}
 \partial_t f_0 - \mathcal L_1 f_1 - \mathcal L_2 f_0 \\
 = \partial_t a - \frac{M_{12}^2}{M_{22}}\tilde k Y^2 \partial_{xx}a + 2\frac{M_{12}^2}{M_{22}}\partial_{xx}a - M_{11}\partial_{xx}a. 
\end{multline*}
Integrating this against $\pi$ and setting equation to 0 gives
\[
\partial_t a = \left(M_{11} - \frac{M_{12}^2}{M_{22}} \right) \partial_{xx}a.
\]
This is the backward equation for a process which diffuses in $x$, with effective diffusivity given indeed by Eq.~\eqref{eq:projMobOff}.

\section{Some useful formulas for changing variables }\label{sec:diffgeom}

%\mhc{perhaps present general formulas as well as 2d?}\\

We present these formulas for a two-dimensional change of variables, for ease of presentation, however the formulas hold more general (without any modification except to the number of components of each vector). 

Consider the change of variables $(x,y)\to (s,z)$. The Jacobian of the mapping is
\[
J = \frac{\partial(s,z)}{\partial(x,y)} = 
\begin{pmatrix}
    \pp{s}{x} & \pp{s}{y} \\
    \pp{z}{x} & \pp{z}{y}
\end{pmatrix}.
\]

\paragraph{Basis vectors.}
Let $e_x,e_y$ be the basis vectors in the original coordinates, and let $e_s,e_z$ be the basis vectors in the new coordinates. 
A basis vector may be thought of as representing a derivation -- $e_x$ is shorthand for $\pp{}{x}$, etc.  

From the chain rule, the basis vectors transform as 
\begin{align*}
    (e_s\;\; e_z) &= (e_x\;\; e_y)J^{-1} &\Leftrightarrow\quad& 
    \left(\pp{}{s} \;\; \pp{}{z}\right) = \left(\pp{}{x} \;\; \pp{}{y}\right) J^{-1}\\
    (e_x\;\; e_y) &= (e_s\;\; e_z)J &\Leftrightarrow\quad& 
    \left(\pp{}{x} \;\; \pp{}{y}\right) = \left(\pp{}{s} \;\; \pp{}{z}\right) J
\end{align*}

\paragraph{Tangent vectors.}
Any tangent vector, say $\v$, can be written in local coordinates as 
\[
\v = v^x e_x + v^ye_y = (e_x\;\;e_y)\begin{pmatrix} v^x\\v^y\end{pmatrix},
\]
where $(v^x,v^y)$ are the coordinates of the tangent vector in the original coordinate system. 
Similarly we have $\v = v^se_s + v^ze_z$ in the new coordinate system coordinate system. 
From the way bases transform, we have
\[
\v = (e_x \;\; e_y)\begin{pmatrix}v^x\\v^y\end{pmatrix}
= (e_s \;\; e_z)J\begin{pmatrix}v^x\\v^y\end{pmatrix}
\]
which shows that coordinates are related as 
\[
\begin{pmatrix}v^s\\v^z\end{pmatrix} = J\begin{pmatrix}v^x\\v^y\end{pmatrix}.
\]
That is, coordinates transform \emph{contravariantly}, i.e. with the inverse ($(J^{-1})^{-1}$) of how the bases transform. 

\paragraph{Metric tensor.}
The metric tensor in the new space is
\[
g = (J^{-1})^TJ^{-1} \quad\text{with inverse } \quad
g^{-1} = JJ^T.
\]
Recall that a metric tensor gives the inner product between vectors in the chosen coordinate system: if vectors with local coordinates $\v,\w$ in the original coordinate system have inner product $\v\cdot\w$, then in the new coordinate system, expressed in local coordinates $\v'=J\v$, $\w'=J\w$, the vectors have inner product $\langle \v',\w'\rangle_g = (\v')^Tg\w'$.

\paragraph{Gradient.}
We write $\nabla f = \grad f$ to mean the tangent vector equal to
\[
\grad f  = (e_x\;\;e_y)\begin{pmatrix} \pp{f}{x} \\ \pp{f}{y} \end{pmatrix}
\]
in our original coordinate system. 
Converting both the bases and the partial derivatives gives
\[
\grad f = (e_s\;\;e_z)JJ^T
\begin{pmatrix} \pp{f}{s}\\\pp{f}{z}\end{pmatrix}
= (e_s\;\;e_z)g^{-1}\begin{pmatrix} \pp{f}{s}\\\pp{f}{z}\end{pmatrix}.
\]
Therefore, the coordinates of $\grad f$ in the new basis are 
\[
g^{-1}\begin{pmatrix} \partial f / \partial s\\\partial f / \partial z\end{pmatrix} = g^{-1}D_{\s}f,
\]
where we write $D_\s = \begin{pmatrix} \partial f / \partial s\\ \partial f / \partial z\end{pmatrix}$ for the collection of directional derivatives. 

In general, a gradient on an $n$-dimensional manifold $\mathcal M$ with metric tensor $g$ is computed as 
\[
\gradM f = g^{-1} Df,
\]
where $Df = (\partial_1 f, \ldots,\partial_n f)^T$ is the collection of partial derivatives with respect to the $n$ local coordinate variables.

\paragraph{Volume element.}
The volume element in the new coordinates is $\sqrt{|g|}dsdz$, where $|g|$ is the determinant of $g$.

\paragraph{Divergence.}
The divergence of a field of tangent vectors $\v(s,z) = v^s(s,z)e_s + v^z(s,z)e_z$ is 
\[
\div\v = \frac{1}{\sqrt{|g|}}\left(\partial_s(\sqrt{|g|}v^s) + \partial_z(\sqrt{|g|} v^z) \right).
\]

In general, the divergence of a vector field $\v$ on an $n$-dimensional manifold $\mathcal M$ with metric tensor $g$ is computed as
\[
\divM \v = \sum_{i=1}^n \frac{1}{\sqrt{|g|}}\partial_i(\sqrt{|g|}v^i),
\]
where $v^i$ is the $i$th component of $\v$ with respect to the unnormalized coordinate bases. %\mhc{need more info here about bases, or in above section on basis vectors}

\paragraph{Mobility tensor.}
The mobility tensor $M$ is a matrix representation of a linear operator, which takes a tangent vector $\v$ and produces a tangent vector $\w$. Suppose $\v=(v^x,v_y), \w=(w^x,w^y)$ are written in the original local coordinates, and let $\v' = J\v, \w'=J\w$ be the local coordinates in the new coordinate system. Then, we have
\[
M\v =\w \quad\Rightarrow\quad MJ^{-1}\v' = J^{-1}\w'
\quad\Rightarrow\quad JMJ^{-1}\v' = \w'. 
\]
Therefore the matrix representation of the mobility tensor in the new coordinate system is
\[
M' = JMJ^{-1}.
\]

\section{Properties of the constrained dynamics}\label{sec:proofs}

In this section we show some of the mathematical properties of a constrained diffusion Eq.~\eqref{eq:ODP0} with mobility tensor satisfying assumption (A), rewritten here for ease of reference:
\begin{equation*}
    \dd{\x}{t} = -M_P\nabla U^{\rm eff} + k_BT\nabla\cdot  M_P + \sqrt{2k_BT}M^{1/2}_P \eta(t).
\end{equation*}
with assumption
\begin{itemize}
    \item[(A)] $M_P\nabla c_i(\x) = 0$ (the 0-vector) for $i=1,\ldots,m$, for  $\x$ in a neighbourhood of $ \mathcal M$. 
\end{itemize}
We will show: 
\begin{enumerate}[(i),nosep]
    \item The dynamics Eq.~\eqref{eq:ODP0} preserves the constraints, \textit{i.e.} $dc_i(\x_t)/dt = 0$. 
    \item The stationary distribution is Eq.~\eqref{eq:pidelta}; 
    \item The dynamics satisfies detailed balance. 
\end{enumerate}\medskip
Thus, Eq.~\eqref{eq:ODP0} has all the desired properties of a constrained reversible diffusion.

\paragraph{(i): The dynamics Eq.~\eqref{eq:ODP0} preserves the constraints.}
This follows from a direction calculation: Using It\^{o}'s formula, we have 
\[
\frac{dc_i(\x_t)}{dt} = \mathcal L c_i + \sqrt{2k_BT}(\nabla c_i)^TM^{1/2}_P \eta(t),
\]
where 
\[
\mathcal L f = k_BTe^{\frac{U^{\rm eff}}{k_BT}}\nabla\cdot (e^{-\frac{U^{\rm eff}}{k_BT}}M_P\nabla f).
\]
We have that $\mathcal Lc_i=0$ on $\mathcal M_P$ since $M_P\nabla c_i=0$ on $\mathcal M$.
We also have that $(\nabla c_i)^TM^{1/2}_P=0$, because $M_P$ and $(M^{1/2}_P)^T$ have the same null space, as shown in the following fact:
\begin{itemize}
    \item For any symmetric matrix $A$, the matrices $(A^{1/2})^T$ and $A$ have the same null space.\smallskip
    
    Proof:  clearly $\mbox{null}(A^{1/2})^T\subset \mbox{null } A$. Suppose the inclusion doesn't go the other way. Then there exists a vector $\v\neq 0$ such that $A\v=0$ but $(A^{1/2})^T\v\neq 0$. But $|(A^{1/2})^T\v|^2 = \v^TA\v = 0$, so $\v=0$, a contradiction.
\end{itemize}
 
Therefore $dc_i(\x_t)/dt=0$.

\medskip

To show properties (ii) and (iii), we first change variables in the generator, as in Section \ref{sec:extrinsic}.  We obtain Eq.~\eqref{eq:Lu_s} on $\mathcal M$, repeated here for ease of reference: 
\begin{multline*}
    \mathcal Lf = k_BT|C|e^{\frac{U^{\rm eff}}{k_BT}}\divM\left( |C|^{-1}e^{-\frac{U^{\rm eff}}{k_BT}}M^{\s}_PD_{\s}f  \right).
\end{multline*}
Here
$M^{\s}_P \equiv (JM_PJ^T)_{\s\s}$. 
%
% It can also be written using intrinsic gradients as 
% \[
% \mathcal Lf = k_BT\frac{e^{\frac{U}{k_BT}}}{|\nabla \c|^{-1}}\divM\left( |\nabla \c|^{-1}e^{-\frac{U}{k_BT}}M^{\s}_Pg_{\s}\gradM f  \right).
% \]
From this, we can compute the adjoint $\mathcal L^*$ on $\mathcal M$: 
\begin{equation}\label{eq:Lstar_s}
\mathcal L^* p = 
k_BT\divM\left( 
|C|^{-1}e^{-\frac{U^{\rm eff}}{k_BT}}
M^{\s}_PD_{\s}(|C|e^{\frac{U^{\rm eff}}{k_BT}}p)
\right).
\end{equation}

\paragraph{(ii): The stationary distribution is Eq.~\eqref{eq:pidelta}.}

Showing $\pi$ in Eq.~\eqref{eq:pidelta} is a stationary distribution is typically done by showing that $\mathcal L^*\pi=0$. However, because $\pi$ contains a delta-function, we cannot compute $\mathcal L^*\pi$. Instead, we use the weak form of the equation $\mathcal L^*\pi=0$, and ask that $\langle \pi, \mathcal Lf\rangle = 0$ for all functions $f$ in the domain of the generator (say all bounded, twice differentiable functions $f$). To compute the inner product we will make use of the coarea formula $\delta(\c(\x)) = |C|^{-1}\dsigmaM$ , 
% \begin{equation}
%     \delta(\c(\x))d\x = |\nabla \c|^{-1}\dsigmaM,
% \end{equation}
where $\dsigmaM$ is the natural volume measure on $\mathcal M$, induced from the ambient space by restriction. In local coordinates, $\dsigmaM = |g_{\s}|^{1/2}ds_1ds_2\cdots ds_d$.
We compute, %using the coarea formula Eq.~\eqref{eq:coarea} and the expression Eq.~\eqref{eq:Lu_scoords} for the generator valid near $\mathcal M$, 
\begin{align*}
    \langle \pi, \mathcal Lf\rangle &= Z^{-1}\int \delta(\c(\x))e^{-\frac{U^{\rm eff}(\x)}{k_BT}}\mathcal Lf(\x) d\x\\ 
&= Z^{-1}\int k_BT \divM\left( |C|^{-1}e^{-\frac{U^{\rm eff}}{k_BT}}M^{\s}_PD_{s}f  \right)\dsigmaM\\
&=0,
\end{align*}
by the divergence theorem on $\mathcal M$. 

% This calculation additionally shows that the locally stationary distribution in intrinsic coordinates, i.e. with respect to the natural surface measure $d\sigma_{\mathcal M}$, is
% \begin{equation}
%     \pi_{\s}=|\nabla_{\x}\c|^{-1}e^{-\frac{V}{k_BT}}.
% \end{equation}

\paragraph{(iii): The dynamics are reversible.}

To show the diffusion is reversible, we show that there is no flux in steady-state \cite{holmesbook}. From Eq.~\eqref{eq:Lstar_s}, the flux is
\[
\j=
-k_BT
|C|^{-1}e^{-\frac{U^{\rm eff}}{k_BT}}
M^{\s}_PD_{\s}(|C|e^{\frac{U^{\rm eff}}{k_BT}}p).
\]
This clearly vanishes for $\pi_{\s}\propto |C|^{-1}e^{-\frac{U^{\rm eff}}{k_BT}}$, which is the stationary distribution in local coordinates.

\section{Some additional details about the projected friciton and mobility}

\subsection{Identity $P_MM_PP_M^T = P_MM_P$}\label{sec:PMidentity}

We now show that $P_MM_PP_M^T = P_MM_P$. First, define
\[
\QM = I-\PM. 
\]
Notice that 
\begin{equation*}
\PM = M\PM^TM^{-1}
\quad\Leftrightarrow\quad
\QM = M\QM^TM^{-1}.
\end{equation*}
Hence, showing the desired identity, is equivalent to showing that 
\[
\QM M\QM^T = \QM M.
\]
We focus on this latter identity. 
We have
\begin{align*}
    \QM M\QM^T &= MC(C^TMC)^{-1}C^TMC(C^TMC)^{-1}C^TM\\
    &= MC(C^TMC)^{-1}C^TM\\
    &= \QM M. 
\end{align*}

% \begin{equation}\label{eq:PPT}
% \PM = M\PM^TM^{-1}
% \quad\Leftrightarrow\quad
% \QM = M\QM^TM^{-1}.
% \end{equation}

% We show that 
% \begin{equation}
%     \PM M\PM^T = \PM M.
% \end{equation}
% Notice that this is equivalent to showing that $\QM M\QM^T = \QM M$, after applying identity Eq.~\eqref{eq:PPT} (recall $\QM = I-\PM = MC(C^TMC)^{-1}C^T$). 
% Since the algebra is slightly simpler for the latter we show this instead. We have
% \begin{align*}
%     \QM M\QM^T &= MC(C^TMC)^{-1}C^TMC(C^TMC)^{-1}C^TM\\
%     &= MC(C^TMC)^{-1}C^TM\\
%     &= \QM M. 
% \end{align*}

\subsection{Showing the equivalence between the projected mobility and the inverse of the projected friction, $M_P \equiv \Gamma_P^{\dagger}$.}\label{app:equivalenceMobFriction}

%\sophie{I haven't re-read this yet but it's not the trickiest part and I can see it to be true.} 
We wish to show that $M_P=P_MM$ is the pseudoinverse of $\Gamma_P = P\Gamma P$. 
%To show this it is enough to show that $M_P  = P_M M$ satisfies all the properties required to be the pseudo-inverse of $\Gamma_P$. 
The pseudo inverse of a matrix $A \in \mathbb{R}^{n\times m}$ is the unique matrix $A^{\dagger}\in \mathbb{R}^{m\times n}$ that satisfies (1) $A A^{\dagger} A = A$,  (2) $A^{\dagger} A  A^{\dagger} = A^{\dagger}$, (3)$ (A A^{\dagger})^{T} = A A^{\dagger}$ and (4) $ ( A^{\dagger} A)^{T} = A^{\dagger} A$. 

We recall that \begin{align*}
& M_P = M - M C^T(C M C^T)^{-1}C M\\ 
& \Gamma_P = (I - C^T(CC^T)^{-1}C ) M^{-1} (I - C^T(CC^T)^{-1}C)
\end{align*}
and we will now show the different properties. 

\begin{itemize}
\item Let's first show $(M_P\Gamma_P)^T = M_P\Gamma_P$.
We have
\begin{widetext}
\begin{align*}
M_P\Gamma_P &= (M - M C^T(C M C^T)^{-1}C M)( M^{-1}  - C^T(CC^T)^{-1}CM^{-1}  ) (I - C^T(CC^T)^{-1}C) \\
& = (I - M C^T(C M C^T)^{-1}C - M C^T(CC^T)^{-1}CM^{-1} + M C^T(C M C^T)^{-1}C M C^T(CC^T)^{-1}CM^{-1} )(I - C^T(CC^T)^{-1}C)  \\
& = (I - M C^T(C M C^T)^{-1}C - M C^T(CC^T)^{-1}CM^{-1} + M C^T(CC^T)^{-1}CM^{-1} )(I - C^T(CC^T)^{-1}C)  \\
& = (I - M C^T(C M C^T)^{-1}C )(I - C^T(CC^T)^{-1}C)  \\
& = I - M C^T(C M C^T)^{-1}C  - C^T(CC^T)^{-1}C + M C^T(C M C^T)^{-1}C C^T(CC^T)^{-1}C \\
& =  I - M C^T(C M C^T)^{-1}C  - C^T(CC^T)^{-1}C + M C^T(C M C^T)^{-1}C \\
& =  I - C^T(CC^T)^{-1}C  = P
\end{align*}
\end{widetext}
which obviously remains unchanged with respect to the transpose operation, by the construction of $P$. 
\item Similarly one can show $(\Gamma_PM_P)^T = \Gamma_PM_P = P$.
\item Now let's calculate $M_P \Gamma_P M_P$ knowing that $M_P \Gamma_P = P$ as we showed above. 
\begin{align*}
M_P \Gamma_P M_P  =& ( I - C^T(CC^T)^{-1}C ) M_P \\
=& M - C^T (CC^T)^{-1}C M - M C^T (CM C^T)^{-1}C M  \\
&+  C^T (C C^T)^{-1}C M  C^T (CM C^T)^{-1}C M  \\
=& M - C^T (CC^T)^{-1}C M - M C^T (CM C^T)^{-1}C M \\
&+  C^T (C C^T)^{-1} C M  \\
=& M - M C^T (CM C^T)^{-1}C M   \\
=& M_P
\end{align*}
so that $M_P \Gamma_P M_P  = M_P$ is verified. Notice that this also shows that $PM_P = M_P$. 
\item Finally $\Gamma_P M_P \Gamma_P = P \Gamma_P = P P \Gamma P = P \Gamma P = \Gamma_P$. 
\end{itemize}

We indeed verified properties (1-4) and so $\Gamma_P^{\dagger} = M_P$.

\subsection{Divergence of the mobility gives a constraint-dependent drift term}
\label{sec:divM}

%\mhc{I rewrote this -- please check}

We wish to show identity \eqref{eq:divM}, which, written component-wise for the $i$th component, is
\begin{equation}
   \partial_j M_{P,ij} =  -M_{P,ij} \partial_j \log |C| + P_{kj} \partial_j M_{P,ki} .
    \label{eq:comp}
\end{equation}
Here we use Einstein notation to imply summation over repeated indices. 

We first rewrite the divergence term, using the fact that $PM_P = M_P$ as we showed in Appendix~\ref{app:equivalenceMobFriction}, and that $M_P$ is symmetric: 
\begin{align*}
\partial_j M_{P,ij} &= \partial_jM_{P,ji}
= \partial_j(P_{jk}M_{P,ki})\\
&= (\partial_jP_{jk})M_{P,ki} + P_{jk}(\partial_jM_{P,ki}).
\end{align*}
Since $P$ is symmetric, we have that $P_{jk}\partial_jM_{P,ki} = P_{kj} \partial_j M_{P,ki}$, so \eqref{eq:comp} is shown once we show that 
\begin{equation}\label{eq:logdet}
(\partial_jP_{jk})M_{P,ki} = -M_{P,ij} \partial_j \log |C|.
\end{equation}
In vector notation, $M_P\nabla\cdot P = -M_P\nabla \log |C|$ (where we used that $M_P$ is symmetric). 

We show \eqref{eq:logdet} by expanding the derivatives on each side of the identity. Recalling that $P = I - C^T (CC^T)^{-1}C$, the left-hand side is 
\begin{align*}
    M_{P,ij}\partial_kP_{kj}
    &= -M_{P,ij}\partial_k\left( C_{uk}(CC^T)^{-1}_{uv}C_{vj}\right)\\
    &= -M_{P,ij}\partial_k(C_{uk}(CC^T)^{-1}_{uv})C_{vj} - M_{P,ij}C_{uk}(CC^T)^{-1}_{uv}\partial_kC_{vj}\\
    &= - M_{P,ij}C_{uk}(CC^T)^{-1}_{uv}\partial_kC_{vj}
\end{align*}
where the last step follows because $M_{P,ij}C_{vj} = 0$, since $M_PC^T = 0$ in a neighbourhood of $\mathcal M$. We can further rewrite this using the fact that $\partial_k C_{vj} = \partial^2_{kj}c_v = \partial_j C_{vk}$, to obtain
\begin{equation}\label{eq:tmp}
M_{P,ij}\partial_kP_{kj} = - M_{P,ij}(CC^T)^{-1}_{uv}C_{uk}\partial_jC_{vk}.
\end{equation}

Now we expand the right-hand side of \eqref{eq:logdet}, using Jacobi's formula for the gradient of the determinant of an invertible matrix $A$:
\[
\partial_j|A| = |A|\mbox{Tr}\left(A^{-1}\partial_j A\right). 
\]
Applying this identity to $\partial_j\log|C| = \partial_j\log |CC^T|^{1/2}$ gives that
\begin{align*}
\partial_j\log|C| &= \frac{1}{2}\mbox{Tr}\left((CC^T)^{-1}\partial_j(CC^T) \right)\\
&= \frac{1}{2} (CC^T)^{-1}_{uv}\partial_j(CC^T)_{uv}\\
&= \frac{1}{2} (CC^T)^{-1}_{uv}\left( (\partial_jC_{uk})C_{vk} + C_{uk}\partial_jC_{vk}\right)\\
&= (CC^T)^{-1}_{uv}C_{uk}\partial_jC_{vk}
\end{align*}
where the last step follows because the two sums are identical. 
Thus, we have that
\[
M_{P,ij}\partial_j\log |C| = M_{P,ij}(CC^T)^{-1}_{uv}C_{uk}\partial_jC_{vk},
\]
which is equal to the expression for $M_{P,ij}\partial_kP_{kj}$ that we derived in \eqref{eq:tmp}. We have shown \eqref{eq:logdet} and hence we have shown identity \eqref{eq:comp}. 

Notice that in this derivation, the only properties we required of $M_P$ were (i) it is symmetric, (ii) $M_PC^T=0$ in a neighbourhood of $\mathcal M$, and (iii) $PM_P=M_P$. Hence, results \eqref{eq:comp}, \eqref{eq:logdet} hold for other forms of projected mobility that satisfy these properties -- we do not have to use the specific projection $P_M$. 

\medskip

As a side remark, we observe that \eqref{eq:logdet} is also proven in Ciccotti et al~\cite{ciccotti2008projection}, Lemma A.2, by multiplying both sides of their Eq.~(A.4) by $M_P$ and using the fact that $M_PP = M_P$. 

\medskip

\section*{References}
\bibliography{Constraints}

@article{papanicolaou1975asymptotic,
  title={Asymptotic analysis of deterministic and stochastic equations with rapidly varying components},
  author={Papanicolaou, George C and Kohler, Werner},
  journal={Communications in Mathematical Physics},
  volume={45},
  number={3},
  pages={217--232},
  year={1975},
  publisher={Springer}
}

@article{papanicolaou1976some,
  title={Some probabilistic problems and methods in singular perturbations},
  author={Papanicolaou, George C},
  journal={The Rocky Mountain Journal of Mathematics},
  pages={653--674},
  year={1976},
  publisher={JSTOR}
}

@article{kurtz1973limit,
  title={A limit theorem for perturbed operator semigroups with applications to random evolutions},
  author={Kurtz, Thomas G},
  journal={Journal of Functional Analysis},
  volume={12},
  number={1},
  pages={55--67},
  year={1973},
  publisher={Elsevier}
}

@inproceedings{kurtz2005averaging,
  title={Averaging for martingale problems and stochastic approximation},
  author={Kurtz, Thomas G},
  booktitle={Applied Stochastic Analysis: Proceedings of a US-French Workshop, Rutgers University, New Brunswick, NJ, April 29--May 2, 1991},
  pages={186--209},
  year={2005},
  organization={Springer}
}

@article{katzenberger1991,
  title={Solutions of a stochastic differential equation forced onto a manifold by a large drift},
  author={G. S. Katzenberger},
  journal={Annals of Probability},
  volume={19},
  number={4},
  pages={1587-1628},
  year={1991}
}

@article{funaki1993,
  title={Degenerative convergence of diffusion process toward a submanifold by strong drift},
  author={Funaki, Tadahisa and Nagai, Hideo},
  journal={Stochastics: An International Journal of Probability and Stochastic Processes},
  volume={44},
  number={1-2},
  pages={1--25},
  year={1993},
  publisher={Taylor \& Francis}
}

@book{arnold,
  title={Mathematical aspects of classical and celestial mechanics},
  author={Arnold, Vladimir I. and Kozlov, Valery V. and Neishtadt, Anatoly I. },
  volume={3},
  year={2006},
  publisher={Springer}
}

@book{landau,
  title={Mechanics: Volume 1},
  author={Landau, Lev Davidovich and Lifshitz, Evgenii Mikhailovich},
  volume={1},
  edition={Third},
  year={1976},
  publisher={Butterworth-Heinemann}
}

@article{rubin1957motion,
  title={Motion under a strong constraining force},
  author={Rubin, Hanan and Ungar, Peter},
  journal={Communications on pure and applied mathematics},
  volume={10},
  number={1},
  pages={65--87},
  year={1957},
  publisher={Wiley Online Library}
}

@article{maraner1995complete,
  title={A complete perturbative expansion for quantum mechanics with constraints},
  author={Maraner, P},
  journal={Journal of Physics A: Mathematical and General},
  volume={28},
  number={10},
  pages={2939--2951},
  year={1995}
}

@article{froese2001realizing,
  title={Realizing holonomic constraints in classical and quantum mechanics},
  author={Froese, Richard and Herbst, Ira},
  journal={Communications in Mathematical Physics},
  volume={220},
  number={3},
  pages={489--535},
  year={2001},
  publisher={Springer}
}

@article{sakai2017influenza,
  title={Influenza A virus hemagglutinin and neuraminidase act as novel motile machinery},
  author={Sakai, Tatsuya and Nishimura, Shin I and Naito, Tadasuke and Saito, Mineki},
  journal={Scientific reports},
  volume={7},
  number={1},
  pages={1--11},
  year={2017},
  publisher={Nature Publishing Group}
}

@article{leimkuhler2021better,
  title={Better training using weight-constrained stochastic dynamics},
  author={Leimkuhler, Benedict and Vlaar, Tiffany and Pouchon, Timoth{\'e}e and Storkey, Amos},
  journal={arXiv preprint arXiv:2106.10704},
  year={2021}
}

@article{Leimkuhler.2016, 
year = {2016}, 
title = {Efficient molecular dynamics using geodesic integration and solvent–solute splitting}, 
author = {Leimkuhler, Benedict and Matthews, Charles}, 
journal = {Proceedings of the Royal Society A: Mathematical, Physical and Engineering Sciences}, 
issn = {1364-5021}, 
doi = {10.1098/rspa.2016.0138}, 
pages = {20160138 -- 22}, 
number = {2189}, 
volume = {472}
}

@article{mckinley2012asymptotic,
  title={Asymptotic analysis of microtubule-based transport by multiple identical molecular motors},
  author={McKinley, Scott A and Athreya, Avanti and Fricks, John and Kramer, Peter R},
  journal={Journal of theoretical biology},
  volume={305},
  pages={54--69},
  year={2012},
  publisher={Elsevier}
}

@article{lee2018modeling,
  title={Modeling the relative dynamics of DNA-coated colloids},
  author={Lee-Thorp, James P and Holmes-Cerfon, Miranda},
  journal={Soft matter},
  volume={14},
  number={40},
  pages={8147--8159},
  year={2018},
  publisher={Royal Society of Chemistry}
}

@article{fogelson2019transport,
  title={Transport facilitated by rapid binding to elastic tethers},
  author={Fogelson, Ben and Keener, James P},
  journal={SIAM Journal on Applied Mathematics},
  volume={79},
  number={4},
  pages={1405--1422},
  year={2019},
  publisher={SIAM}
}

@article{andersen1983rattle,
  title={Rattle: A “velocity” version of the shake algorithm for molecular dynamics calculations},
  author={Andersen, Hans C},
  journal={Journal of computational Physics},
  volume={52},
  number={1},
  pages={24--34},
  year={1983},
  publisher={Elsevier}
}

@article{krautler2001fast,
  title={A fast SHAKE algorithm to solve distance constraint equations for small molecules in molecular dynamics simulations},
  author={Kr{\"a}utler, Vincent and Van Gunsteren, Wilfred F and H{\"u}nenberger, Philippe H},
  journal={Journal of computational chemistry},
  volume={22},
  number={5},
  pages={501--508},
  year={2001},
  publisher={Wiley Online Library}
}

@article{grier1997optical,
  title={Optical tweezers in colloid and interface science},
  author={Grier, David G},
  journal={Current opinion in colloid \& interface science},
  volume={2},
  number={3},
  pages={264--270},
  year={1997},
  publisher={Elsevier}
}

@book{lekkerkerker2024colloids,
  title={Colloids and the depletion interaction},
  author={Lekkerkerker, Henk NW and Tuinier, Remco and Vis, Mark},
  year={2024},
  publisher={Springer Nature}
}

@article{jana2019translational,
  title={Translational and rotational dynamics of colloidal particles interacting through reacting linkers},
  author={Jana, Pritam Kumar and Mognetti, Bortolo Matteo},
  journal={Physical Review E},
  volume={100},
  number={6},
  pages={060601},
  year={2019},
  publisher={APS}
}

@article{fogelson2018enhanced,
  title={Enhanced nucleocytoplasmic transport due to competition for elastic binding sites},
  author={Fogelson, Ben and Keener, James P},
  journal={Biophysical journal},
  volume={115},
  number={1},
  pages={108--116},
  year={2018},
  publisher={Elsevier}
}

@article{bressloff2013stochastic,
  title={Stochastic models of intracellular transport},
  author={Bressloff, Paul C and Newby, Jay M},
  journal={Reviews of Modern Physics},
  volume={85},
  number={1},
  pages={135},
  year={2013},
  publisher={APS}
}

@article{leibler1991dynamics,
  title={Dynamics of reversible networks},
  author={Leibler, Ludwik and Rubinstein, Michael and Colby, Ralph H},
  journal={Macromolecules},
  volume={24},
  number={16},
  pages={4701--4707},
  year={1991},
  publisher={ACS Publications}
}

@article{zwanzig1992diffusion,
  title={Diffusion past an entropy barrier},
  author={Zwanzig, Robert},
  journal={The Journal of Physical Chemistry},
  volume={96},
  number={10},
  pages={3926--3930},
  year={1992},
  publisher={ACS Publications}
}

@article{marbach2022mass,
  title={Mass changes the diffusion coefficient of particles with ligand-receptor contacts in the overdamped limit},
  author={Marbach, Sophie and Holmes-Cerfon, Miranda},
  journal={Physical Review Letters},
  volume={129},
  number={4},
  pages={048003},
  year={2022},
  publisher={APS}
}

@article{kallus2017,
  title={Free energy of singular sticky-sphere clusters},
  author={Kallus, Yoav and Holmes-Cerfon, Miranda},
  journal={Physical Review E},
  volume={95},
  number={2},
  pages={022130},
  year={2017}
}

@article{rubi2019entropic,
  title={Entropic diffusion in confined soft-matter and biological systems},
  author={Rubi, J Miguel},
  journal={EPL (Europhysics Letters)},
  volume={127},
  number={1},
  pages={10001},
  year={2019},
  publisher={IOP Publishing}
}

@book{jacobs_diffusion_1967,
	address = {Berlin, Heidelberg},
	title = {Diffusion {Processes}},
	isbn = {978-3-642-86416-2 978-3-642-86414-8},
	url = {http://link.springer.com/10.1007/978-3-642-86414-8},
	urldate = {2022-08-25},
	publisher = {Springer Berlin Heidelberg},
	author = {Jacobs, M. H.},
	year = {1967},
	doi = {10.1007/978-3-642-86414-8},
}

@article{kalinay_corrections_2006,
	title = {Corrections to the {Fick}-{Jacobs} equation},
	volume = {74},
	issn = {1539-3755, 1550-2376},
	url = {https://link.aps.org/doi/10.1103/PhysRevE.74.041203},
	doi = {10.1103/PhysRevE.74.041203},
	number = {4},
	urldate = {2022-08-25},
	journal = {Physical Review E},
	author = {Kalinay, P. and Percus, J. K.},
	month = oct,
	year = {2006},
	pages = {041203},
}

@article{perkins1992hydrodynamic,
  title={Hydrodynamic interaction of a spherical particle with a planar boundary: II. Hard wall},
  author={Perkins, GS and Jones, RB},
  journal={Physica A: Statistical Mechanics and its Applications},
  volume={189},
  number={3-4},
  pages={447--477},
  year={1992},
  publisher={Elsevier}
}

@article{fatkullin2010reduced,
  title={Reduced dynamics of stochastically perturbed gradient flows},
  author={Fatkullin, Ibrahim and Kovacic, Gregor and Vanden-Eijnden, Eric},
  year={2010}
}

@article{wang2015crystallization,
  title={Crystallization of DNA-coated colloids},
  author={Wang, Yu and Wang, Yufeng and Zheng, Xiaolong and Ducrot, {\'E}tienne and Yodh, Jeremy S and Weck, Marcus and Pine, David J},
  journal={Nature communications},
  volume={6},
  number={1},
  pages={1--8},
  year={2015},
  publisher={Nature Publishing Group}
}

@article{feng2013specificity,
  title={Specificity, flexibility and valence of DNA bonds guide emulsion architecture},
  author={Feng, Lang and Pontani, Lea-Laetitia and Dreyfus, R{\'e}mi and Chaikin, Paul and Brujic, Jasna},
  journal={Soft Matter},
  volume={9},
  number={41},
  pages={9816--9823},
  year={2013},
  publisher={Royal Society of Chemistry}
}

@article{ciccotti2008projection,
  title={Projection of diffusions on submanifolds: Application to mean force computation},
  author={Ciccotti, Giovanni and Lelievre, Tony and Vanden-Eijnden, Eric},
  journal={Communications on Pure and Applied Mathematics: A Journal Issued by the Courant Institute of Mathematical Sciences},
  volume={61},
  number={3},
  pages={371--408},
  year={2008},
  publisher={Wiley Online Library}
}

@article{gehrels2022programming,
  title={Programming Directed Motion with DNA-Grafted Particles},
  author={Gehrels, Emily W and Rogers, W Benjamin and Zeravcic, Zorana and Manoharan, Vinothan N},
  journal={ACS nano},
  year={2022},
  publisher={ACS Publications}
}

@article{waszkiewicz2025trimer,
  title={The trimer paradox: The effect of stiff constraints on equilibrium distributions in overdamped dynamics},
  author={Waszkiewicz, Radost and Lisicki, Maciej},
  journal={The Journal of Chemical Physics},
  volume={162},
  number={18},
  year={2025},
  publisher={AIP Publishing}
}

@article{Ciccotti.2005, 
year = {2005}, 
title = {{Blue Moon Sampling, Vectorial Reaction Coordinates, and Unbiased Constrained Dynamics}}, 
author = {Ciccotti, Giovanni and Kapral, Raymond and Vanden-Eijnden, Eric}, 
journal = {ChemPhysChem}, 
issn = {1439-4235}, 
doi = {10.1002/cphc.200400669}, 
url = {http://doi.wiley.com/10.1002/cphc.200400669}, 
pages = {1809 -- 1814}, 
number = {9}, 
volume = {6}, 
month = {09}
}

@article{lelievre2012langevin,
  title={Langevin dynamics with constraints and computation of free energy differences},
  author={Lelievre, Tony and Rousset, Mathias and Stoltz, Gabriel},
  journal={Mathematics of computation},
  volume={81},
  number={280},
  pages={2071--2125},
  year={2012}
}

@article{schonle2025efficient,
  title={Efficient Monte-Carlo sampling of metastable systems using non-local collective variable updates},
  author={Sch{\"o}nle, Christoph and Carbone, Davide and Gabri{\'e}, Marylou and Leli{\`e}vre, Tony and Stoltz, Gabriel},
  journal={arXiv preprint arXiv:2512.16812},
  year={2025}
}

@article{mammen1998polyvalent,
  title={Polyvalent interactions in biological systems: implications for design and use of multivalent ligands and inhibitors},
  author={Mammen, Mathai and Choi, Seok-Ki and Whitesides, George M},
  journal={Angewandte Chemie International Edition},
  volume={37},
  number={20},
  pages={2754--2794},
  year={1998},
  publisher={Wiley Online Library}
}

@inproceedings{alon2002rolling,
  title={From rolling to arrest on blood vessels: leukocyte tap dancing on endothelial integrin ligands and chemokines at sub-second contacts},
  author={Alon, Ronen and Feigelson, Sara},
  booktitle={Seminars in immunology},
  volume={14},
  pages={93--104},
  year={2002},
  organization={Elsevier}
}

@article{korn2008dynamic,
  title={Dynamic states of cells adhering in shear flow: from slipping to rolling},
  author={Korn, CB and Schwarz, US},
  journal={Physical review E},
  volume={77},
  number={4},
  pages={041904},
  year={2008},
  publisher={APS}
}

@article{muller2019mobility,
  title={Mobility-based quantification of multivalent virus-receptor interactions: New insights into influenza A virus binding mode},
  author={M\"{u}ller, Matthias and Lauster, Daniel and Wildenauer, Helen HK and Herrmann, Andreas and Block, Stephan},
  journal={Nano letters},
  volume={19},
  number={3},
  pages={1875--1882},
  year={2019},
  publisher={ACS Publications}
}

@article{perry2015two,
  title={Two-dimensional clusters of colloidal spheres: Ground states, excited states, and structural rearrangements},
  author={Perry, Rebecca W and Holmes-Cerfon, Miranda C and Brenner, Michael P and Manoharan, Vinothan N},
  journal={Physical review letters},
  volume={114},
  number={22},
  pages={228301},
  year={2015},
  publisher={APS}
}

@article{sakai2018unique,
  title={Unique directional motility of influenza C virus controlled by its filamentous morphology and short-range motions},
  author={Sakai, Tatsuya and Takagi, Hiroaki and Muraki, Yasushi and Saito, Mineki},
  journal={Journal of virology},
  volume={92},
  number={2},
  pages={e01522--17},
  year={2018},
  publisher={Am Soc Microbiol}
}

@article{ley2007getting,
  title={Getting to the site of inflammation: the leukocyte adhesion cascade updated},
  author={Ley, Klaus and Laudanna, Carlo and Cybulsky, Myron I and Nourshargh, Sussan},
  journal={Nature Reviews Immunology},
  volume={7},
  number={9},
  pages={678--689},
  year={2007},
  publisher={Nature Publishing Group}
}

@article{varilly2012general,
  title={A general theory of DNA-mediated and other valence-limited colloidal interactions},
  author={Varilly, Patrick and Angioletti-Uberti, Stefano and Mognetti, Bortolo M and Frenkel, Daan},
  journal={The Journal of chemical physics},
  volume={137},
  number={9},
  pages={094108},
  year={2012},
  publisher={American Institute of Physics}
}

@article{mirkin1996dna,
  title={A DNA-based method for rationally assembling nanoparticles into macroscopic materials},
  author={Mirkin, Chad A and Letsinger, Robert L and Mucic, Robert C and Storhoff, James J},
  journal={Nature},
  volume={382},
  number={6592},
  pages={607--609},
  year={1996},
  publisher={Nature Publishing Group}
}

@article{levitz2025probing,
  title={Probing particle dynamics in a fully opaque porous network using X-ray differential dynamic radiography (XDDR)},
  author={Levitz, P and Michot, L and Malikova, N and Scheel, M and Weitkamp, T},
  journal={Soft Matter},
  volume={21},
  number={16},
  pages={3067--3079},
  year={2025},
  publisher={Royal Society of Chemistry}
}

@article{thorneywork2020direct,
  title={Direct detection of molecular intermediates from first-passage times},
  author={Thorneywork, Alice L and Gladrow, Jannes and Qing, Yujia and Rico-Pasto, Marc and Ritort, Felix and Bayley, Hagan and Kolomeisky, Anatoly B and Keyser, Ulrich F},
  journal={Science advances},
  volume={6},
  number={18},
  pages={eaaz4642},
  year={2020},
  publisher={American Association for the Advancement of Science}
}

@article{franosch2011resonances,
  title={Resonances arising from hydrodynamic memory in Brownian motion},
  author={Franosch, Thomas and Grimm, Matthias and Belushkin, Maxim and Mor, Flavio M and Foffi, Giuseppe and Forr{\'o}, L{\'a}szl{\'o} and Jeney, Sylvia},
  journal={Nature},
  volume={478},
  number={7367},
  pages={85--88},
  year={2011},
  publisher={Nature Publishing Group UK London}
}

@article{franosch2009persistent,
  title={Persistent correlation of constrained colloidal motion},
  author={Franosch, Thomas and Jeney, Sylvia},
  journal={Physical Review E—Statistical, Nonlinear, and Soft Matter Physics},
  volume={79},
  number={3},
  pages={031402},
  year={2009},
  publisher={APS}
}

@article{schallamach1963theory,
  title={A theory of dynamic rubber friction},
  author={Schallamach, A},
  journal={Wear},
  volume={6},
  number={5},
  pages={375--382},
  year={1963},
  publisher={Elsevier}
}

@article{fish2025libmobility,
  title={libMobility: A Python library for hydrodynamics at the Smoluchowski level},
  author={Fish, Ryker and Carter, Adam and Diez-Silva, Pablo and Delgado-Buscalioni, Rafael and Pelaez, Raul P and Sprinkle, Brennan},
  journal={arXiv preprint arXiv:2510.02135},
  year={2025}
}

@article{cui2022comprehensive,
  title={Comprehensive view of microscopic interactions between DNA-coated colloids},
  author={Cui, Fan and Marbach, Sophie and Zheng, Jeana Aojie and Holmes-Cerfon, Miranda and Pine, David J},
  journal={Nature communications},
  volume={13},
  number={1},
  pages={2304},
  year={2022},
  publisher={Nature Publishing Group UK London}
}

@article{rogers2011direct,
  title={Direct measurements of DNA-mediated colloidal interactions and their quantitative modeling},
  author={Rogers, W Benjamin and Crocker, John C},
  journal={Proceedings of the National Academy of Sciences},
  volume={108},
  number={38},
  pages={15687--15692},
  year={2011},
  publisher={National Academy of Sciences}
}

@article{sprinkle2020driven,
  title={Driven dynamics in dense suspensions of microrollers},
  author={Sprinkle, Brennan and Van Der Wee, Ernest B and Luo, Yixiang and Driscoll, Michelle M and Donev, Aleksandar},
  journal={Soft Matter},
  volume={16},
  number={34},
  pages={7982--8001},
  year={2020},
  publisher={Royal Society of Chemistry}
}

@article{gompper20252025,
  title={The 2025 motile active matter roadmap},
  author={Gompper, Gerhard and Stone, Howard A and Kurzthaler, Christina and Saintillan, David and Peruani, Fernado and Fedosov, Dmitry A and Auth, Thorsten and Cottin-Bizonne, Cecile and Ybert, Christophe and Cl{\'e}ment, Eric and others},
  journal={Journal of Physics: Condensed Matter},
  volume={37},
  number={14},
  pages={143501},
  year={2025},
  publisher={IOP Publishing}
}

@book{faxen1921einwirkung,
  title={Einwirkung der Gefasswande auf den Widerstand gegen die Bewegung einer kleinen Kugel in einer zahen Flussigkeit},
  author={Fax{\'e}n, H.},
  year={1921},
  publisher={F{\"o}rf.}
}

@article{reguera_kinetic_2001,
	title = {Kinetic equations for diffusion in the presence of entropic barriers},
	volume = {64},
	issn = {1063-651X, 1095-3787},
	url = {https://link.aps.org/doi/10.1103/PhysRevE.64.061106},
	doi = {10.1103/PhysRevE.64.061106},
	number = {6},
	urldate = {2022-08-25},
	journal = {Physical Review E},
	author = {Reguera, D. and Rubí, J. M.},
	month = nov,
	year = {2001},
	pages = {061106},
}

@article{sprik1998free,
  title={Free energy from constrained molecular dynamics},
  author={Sprik, Michiel and Ciccotti, Giovanni},
  journal={The Journal of chemical physics},
  volume={109},
  number={18},
  pages={7737--7744},
  year={1998},
  publisher={American Institute of Physics}
}

@article{marbach2022nanocaterpillar,
  title={The nanocaterpillar's random walk: diffusion with ligand--receptor contacts},
  author={Marbach, Sophie and Zheng, Jeana Aojie and Holmes-Cerfon, Miranda},
  journal={Soft Matter},
  volume={18},
  number={16},
  pages={3130--3146},
  year={2022},
  publisher={Royal Society of Chemistry}
}

@article{Kampen.1981, 
year = {1981}, 
title = {{Statistical mechanics of trimers}}, 
author = {Kampen, N G van}, 
journal = {Applied Scientific Research. An International Journal on Thermal, Mechanical, and Electromagnetic Phenomena in Continua}, 
doi = {10.1007/bf00382618}, 
url = {http://link.springer.com/10.1007/BF00382618}, 
pages = {67 -- 75}, 
number = {1-2}, 
volume = {37}, 
month = {00}
}

@article{fixman1978simulation,
  title={Simulation of polymer dynamics. I. General theory},
  author={Fixman, Marshall},
  journal={The Journal of Chemical Physics},
  volume={69},
  number={4},
  pages={1527--1537},
  year={1978},
  publisher={American Institute of Physics}
}

@article{ottinger1994brownian,
  title={Brownian dynamics of rigid polymer chains with hydrodynamic interactions},
  author={{\"O}ttinger, Hans Christian},
  journal={Physical Review E},
  volume={50},
  number={4},
  pages={2696},
  year={1994},
  publisher={APS}
}

@incollection{morse2003Theory,
  title = {Theory of Constrained Brownian Motion},
  booktitle = {Advances in Chemical Physics},
  author = {Morse, David C.},
  year = {2003},
  pages = {65--189},
  publisher = {John Wiley \& Sons, Ltd},
  doi = {10.1002/0471484237.ch2},
  chapter = {2},
  isbn = {978-0-471-48423-3},
  langid = {english}
}

@article{mannattil2022Thermal,
  title = {Thermal Fluctuations of Singular Bar-Joint Mechanisms},
  author = {Mannattil, Manu and Schwarz, J. M. and Santangelo, Christian D.},
  year = {2022},
  journal = {Physical Review Letters},
  volume = {128},
  number = {20},
  pages = {208005},
  issn = {0031-9007, 1079-7114},
  doi = {10.1103/PhysRevLett.128.208005},
  langid = {english}
}

@article{maxian2023bending,
  title={Bending fluctuations in semiflexible, inextensible, slender filaments in Stokes flow: Toward a spectral discretization},
  author={Maxian, Ondrej and Sprinkle, Brennan and Donev, Aleksandar},
  journal={The Journal of Chemical Physics},
  volume={158},
  number={15},
  year={2023},
  publisher={AIP Publishing}
}

@article{maxian2024simulation,
  title={A simulation platform for slender, semiflexible, and inextensible fibers with Brownian hydrodynamics and steric repulsion},
  author={Maxian, Ondrej and Donev, Aleksandar},
  journal={Physics of Fluids},
  volume={36},
  number={12},
  year={2024},
  publisher={AIP Publishing}
}

@article{hinch1994Brownian,
  title = {Brownian motion with stiff bonds and rigid constraints},
  author = {Hinch, E. J.},
  year = {1994},
  journal = {Journal of Fluid Mechanics},
  volume = {271},
  pages = {219--234},
  publisher = {Cambridge University Press},
  issn = {1469-7645, 0022-1120},
  doi = {10.1017/S0022112094001746},
  langid = {english}
}

@book{Landau.1976, 
year = {1976}, 
title = {Mechanics}, 
author = {Landau, Lev Davidovich and Lifshitz, E M}, 
series = {Butterworth-Heinemann}, 
publisher = {Butterworth-Heinemann}
}

@article{holmes2016stochastic,
  title={Stochastic disks that roll},
  author={Holmes-Cerfon, Miranda},
  journal={Physical Review E},
  volume={94},
  number={5},
  pages={052112},
  year={2016},
  publisher={APS}
}

@book{holmesbook,
  title={Applied Stochastic Analysis},
  author={Holmes-Cerfon, Miranda},
  volume={33},
  year={2024},
  publisher={American Mathematical Society, Courant Institute of Mathematical Sciences}
}

@article{HolmesCerfon:2013jw,
author = {Holmes-Cerfon, Miranda and Gortler, Steven J and Brenner, Michael P},
title = {{A geometrical approach to computing free-energy landscapes from short-ranged potentials}},
journal = {Proceedings of the National Academy of Sciences},
year = {2013},
volume = {110},
number = {1},
pages = {E5--E14}
}

@article{Sharma.2021, 
year = {2021}, 
title = {{NonReversible Sampling Schemes on Submanifolds}}, 
author = {Sharma, Upanshu and Zhang, Wei}, 
journal = {SIAM Journal on Numerical Analysis}, 
issn = {0036-1429}, 
doi = {10.1137/20m1378752}, 
pages = {2989--3031}, 
number = {6}, 
volume = {59}
}

@article{brenner1961slow,
	author = {Brenner, Howard},
	journal = {Chemical Engineering Science},
	number = {3-4},
	pages = {242--251},
	title = {The slow motion of a sphere through a viscous fluid towards a plane surface},
	volume = {16},
	year = {1961}}

@article{maxian2021integral,
  title={Integral-based spectral method for inextensible slender fibers in Stokes flow},
  author={Maxian, Ondrej and Mogilner, Alex and Donev, Aleksandar},
  journal={Physical Review Fluids},
  volume={6},
  number={1},
  pages={014102},
  year={2021},
  publisher={APS}
}

@article{funkenbusch2024approaches,
  title={Approaches for fast Brownian dynamics simulation with constraints},
  author={Funkenbusch, William T and Silmore, Kevin S and Swan, James W},
  journal={Journal of Computational Physics},
  volume={509},
  pages={113043},
  year={2024},
  publisher={Elsevier}
}

@article{ottinger2015preservation,
  title={Preservation of thermodynamic structure in model reduction},
  author={{\"O}ttinger, Hans Christian},
  journal={Physical Review E},
  volume={91},
  number={3},
  pages={032147},
  year={2015},
  publisher={APS}
}

@article{melio2024soft,
  title={Soft and stiff normal modes in floppy colloidal square lattices},
  author={Melio, Julio and Henkes, Silke E and Kraft, Daniela J},
  journal={Physical Review Letters},
  volume={132},
  number={7},
  pages={078202},
  year={2024},
  publisher={APS}
}

@article{delmotte2025modeling,
  title={Modeling complex particle suspensions: Perspectives on the rigid multiblob method},
  author={Delmotte, Blaise and Usabiaga, Florencio Balboa},
  journal={Physical Review Fluids},
  volume={10},
  number={10},
  pages={100701},
  year={2025},
  publisher={APS}
}

@book{pavliotis2008multiscale,
  title={Multiscale methods: averaging and homogenization},
  author={Pavliotis, Grigoris and Stuart, Andrew},
  year={2008},
  publisher={Springer Science \& Business Media}
}

@article{rogers2016using,
  title={Using DNA to program the self-assembly of colloidal nanoparticles and microparticles},
  author={Rogers, W Benjamin and Shih, William M and Manoharan, Vinothan N},
  journal={Nature Reviews Materials},
  volume={1},
  number={3},
  pages={1--14},
  year={2016},
  publisher={Nature Publishing Group}
}

@article{macfarlane2011nanoparticle,
  title={Nanoparticle superlattice engineering with DNA},
  author={Macfarlane, Robert J and Lee, Byeongdu and Jones, Matthew R and Harris, Nadine and Schatz, George C and Mirkin, Chad A},
  journal={science},
  volume={334},
  number={6053},
  pages={204--208},
  year={2011},
  publisher={American Association for the Advancement of Science}
}

@article{mao2015mechanical,
  title={Mechanical instability at finite temperature},
  author={Mao, Xiaoming and Souslov, Anton and Mendoza, Carlos I and Lubensky, Tom C},
  journal={Nature communications},
  volume={6},
  number={1},
  pages={5968},
  year={2015},
  publisher={Nature Publishing Group UK London}
}

@article{meng2010free,
  title={The free-energy landscape of clusters of attractive hard spheres},
  author={Meng, Guangnan and Arkus, Natalie and Brenner, Michael P and Manoharan, Vinothan N},
  journal={Science},
  volume={327},
  number={5965},
  pages={560--563},
  year={2010},
  publisher={American Association for the Advancement of Science}
}
\end{document}